\documentclass[fleqn,usenatbib]{mnras}

\usepackage{newtxtext,newtxmath}
\usepackage[T1]{fontenc}

\DeclareRobustCommand{\VAN}[3]{#2}
\let\VANthebibliography\thebibliography
\def\thebibliography{\DeclareRobustCommand{\VAN}[3]{##3}\VANthebibliography}

\usepackage{graphicx}	% Including figure files
\usepackage{amsmath}	% Advanced maths commands
\title[The $z\simeq7$ rest-frame UV LF from VIDEO]{The bright end of the galaxy luminosity function at $\boldsymbol{z}\simeq\boldsymbol{7}$ from the VISTA VIDEO survey}

\author[R. G. Varadaraj et al.]{
R. G. Varadaraj,$^{1}$\thanks{rohan.varadaraj@physics.ox.ac.uk}
R. A. A. Bowler,$^{2}$
M. J. Jarvis,$^{1,3}$
N. J. Adams,$^{2}$
and B. H{\"a}u{\ss}ler$^{4}$
\\
$^{1}$Sub-department of Astrophysics, University of Oxford, Denys Wilkinson Building, Keble Road, Oxford, OX1 2DL, UK\\
$^{2}$Jodrell Bank Centre for Astrophysics, University of Manchester, Oxford Road, Manchester, M13 9PL, UK\\
$^{3}$Department of Physics, University of the Western Cape, Bellville 7535, South Africa\\
$^{4}$ European Southern Observatory, Alonso de Cordova 3107, Vitacura, Santiago, Chile
}

\date{Accepted XXX. Received YYY; in original form ZZZ}

\pubyear{2023}

\begin{document}
\label{firstpage}
\pagerange{\pageref{firstpage}--\pageref{lastpage}}
\maketitle

\begin{abstract}
    We have conducted a search for $z\simeq7$ Lyman break galaxies over 8.2 square degrees of near-infrared imaging from the VISTA Deep Extragalactic Observations (VIDEO) survey in the \textit{XMM-Newton} - Large Scale Structure (XMM-LSS) and the Extended \textit{Chandra} Deep Field South (ECDF-S) fields.
    Candidate galaxies were selected from a full photometric redshift analysis down to a $Y+J$ depth of 25.3 ($5\sigma$), utilizing deep auxiliary optical and~\emph{Spitzer}/IRAC data to remove brown dwarf and red interloper galaxy contaminants.
    Our final sample consists of 28 candidate galaxies at $6.5\le z \le7.5$ with $-23.5 \le M_{\mathrm{UV}} \le -21.6$.
    We derive stellar masses of $9.1 \le \mathrm{log}_{10}(M_{\star}/M_{\sun}) \le 10.9$ for the sample, suggesting that these candidates represent some of the most massive galaxies known at this epoch. 
    We measure the rest-frame UV luminosity function (LF) at $z\simeq7$, confirming previous findings of a gradual decline in number density at the bright-end ($M_{\mathrm{UV}} < -22$) that is well described by a double-power law (DPL).
    We show that quasar contamination in this magnitude range is expected to be minimal, in contrast to conclusions from recent pure-parallel~\emph{Hubble} studies.
    Our results are up to a factor of ten lower than previous determinations from optical-only ground-based studies at $M_{\rm UV} \lesssim - 23$.
    We find that the inclusion of $YJHK_{s}$ photometry is vital for removing brown-dwarf contaminants, and $z \simeq 7$ samples based on red-optical data alone could be highly contaminated ($\gtrsim 50$ per cent).
    In comparison with other robust $z > 5$ samples, our results further support little evolution in the very bright-end of the rest-frame UV LF from $z = 5$--$10$, potentially signalling a lack of mass quenching and/or dust obscuration in the most massive galaxies in the first Gyr.
\end{abstract}

\begin{keywords}
galaxies: formation -- galaxies: evolution -- galaxies: high-redshift
\end{keywords}

%%%%%%%%%%%%%%%%%%%%%%%%%%%%%%%%%%%%%%%%%%%%%%%%%%

%%%%%%%%%%%%%%%%% BODY OF PAPER %%%%%%%%%%%%%%%%%%

\section{Introduction}

Observing the formation and evolution of galaxies at very high-redshifts is vital for understanding the build-up of early structures and all subsequent galaxy evolution to the present day. 
The Lyman-break technique has been used successfully for over three decades to search for high-redshift galaxies \citep[][]{Guhathakurta90, Steidel96} via the redshifted Lyman-$\alpha$ break at $\lambda_{\mathrm{rest}} = 1216$\AA, by utilizing the strong spectral break that appears at the wavelength of the Lyman-$\alpha$ line due to absorption by the IGM along the line of sight.
The selection of high-redshift $(z\ge7)$ Lyman-break galaxies (LBGs) has been revolutionised thanks to the unparalleled near-infrared capabilities of the \textit{James Webb Space Telescope (JWST)}. 
Early results have identified and spectroscopically confirmed galaxies in the range $z =10.6$--$13.1$~\citep[e.g.][]{curtislake22, bunker23, haro23} and rest-frame optical spectroscopy has revealed the gas phase properties for the first time ~\citep[e.g.][]{fujimoto22spec, cameron23, curti23}.
One surprising result from the first samples of galaxy candidates found by \textit{JWST} has been the discovery of an unexpected number of luminous ($M_{\mathrm{UV}} \sim -21$) sources at $z>8$ \citep[e.g.][]{naidu22, finkelstein22} that have challenged models of galaxy formation \citep{ferrara22}.

Galaxies form and reside within dark matter halos, and the luminosity function (LF, number density as a function of brightness) provides a connection to the dark-matter halo-mass function and early galaxy formation since galaxy formation efficiency has a strong dependency on halo mass \citep[e.g.][]{vale06, behroozi13, wechsler18}. 
This makes the LF a key measurement in galaxy evolution. 
As at $z\gtrsim6$ LBGs are typically selected and studied in the near-infrared (NIR), which provides access to the rest-frame UV emission from young stars, many studies have derived the rest-frame UV LF (hereafter UV LF) from photometric samples at $z>4$ \citep[e.g.][]{mclure2013, finkelstein15, stefanon19, bouwens21}{}{}.
The shape of the UV LF gives insight into feedback properties within galaxies and ionising properties of galaxies in the early Universe \citep[e.g.][]{benson03, McLure2010, bradley12, Bowler15}. A common form used to fit the rest-frame UV LF in the local Universe is the Schechter function \citep{Schechter76}, $\phi(L)dL = \phi^{*}(L/L^{*})^{\alpha}e^{-L/L^{*}}d(L/L^{*})$, where $\alpha$ is the faint end slope, $L^{*}$ is the characteristic luminosity and $\phi^{*}$ is the density normalisation. 
The faint end follows a power law and a rapid decline is seen in the bright end, necessitating an exponential cut-off beyond the knee at $L^{*}$. 
This suggests the growth of the brightest, most massive galaxies is quenched in lower redshift LBG samples \citep[e.g.][]{burg10, stevans18, adams22}{}{}, since the dark matter halo mass function has a shallower high-mass slope and is more akin to a double power law. 
Probing the bright end of the rest-frame UV LF requires galaxy searches to be conducted in degree-scale surveys, since the number density of bright objects declines rapidly. For example, LBGs at $z\simeq7$ have a surface density of a few per square degree at $M_{\mathrm{UV}} < - 22$ \citep[][]{Bowler14}{}{}. 
The widest survey fields imaged by the \textit{Hubble Space Telescope (HST)} only cover $1136 \ \mathrm{arcmin}^{2}$ \citep[e.g.][]{bouwens21}, proving only weak constraints on the LF at $M_{\mathrm{UV}} \lesssim -21$.
Prior to the launch of space missions capable of imaging a wide area (e.g.~\textit{Euclid} and~\textit{Roman}), ground-based NIR surveys provide the only insight into the number density of the very bright galaxy population at $z > 6$.
The first determinations of the $z\simeq7$ UV LF with degree-scale surveys showed an excess of bright galaxies compared to the commonly used Schechter function, with a double power law (DPL) fit preferred \citep{Bowler2012, Bowler14}.
The DPL form has also been identified in optical-only studies from the Hyper-Suprime Cam GOLDRUSH program \citep{ono18, harikane22}.
This bright end excess appears to continue out to $z\simeq10$ \citep{stefanon19, Bowler_2020}. 
Additionally, there is evidence that whatever mechanism causes the excess in bright galaxies at $z\simeq7$ has a rapid onset, with a clear steepening of the bright-end slope found to occur in the 400 Myrs between $z=7$ and $z=5$ \citep[][]{Bowler15, adams22}{}{}

Interestingly, many simulations do not predict \emph{intrinsic} UV LFs with a DPL shape at $z>5$, with a steeper bright-end only appearing after the addition of significant dust obscuration (e.g. \citet{Cai2014}, see compilation in \citet{Bowler15}).
While at lower redshifts the steep decline in luminous/massive galaxies \citep[e.g.][]{yang09}{}{} can be attributed to mass quenching \citep[][]{peng10}{}{}, attenuation or a lack of attenuation is predicted to be the more significant effect in shaping the UV LF at high redshifts \citep[e.g.][]{Cai2014, vijayan21}.
Additionally, as active galactic nucei (AGN) begin to grow and turn on in the early Universe, energetic feedback from their accretion disc may suppress star formation in massive galaxies \citep[e.g.][]{bower12, dave19, lovell22}. 
While quasars at $z\simeq7$ are extremely rare \citep[e.g.][]{mortlock11, banados18, wang21}{}{}, the quasar LF appears to rapidly increase to lower redshifts, which leads to the faint end of the AGN LF being comparable to the bright end around $M_{\mathrm{UV}} \simeq -23$ at $z\simeq5$ \citep[][]{canodiaz11, adams22, harikane22}. 
By determining the functional form of the luminosity function at different redshifts, the onset of potential quenching and/or dust obscuration can be investigated. Comparison of the bright end of the galaxy UV LF with simulations, plus follow-up observations with \textit{JWST} and ALMA, can then uncover the dominant mechanism responsible for the shape of the UV LF.

A key step in searching for $z\simeq7$ galaxies is the removal of red, dusty galaxies at $z\sim1-2$ and cool Galactic M, L and T dwarf stars \citep[e.g.][]{Stanway2008, Bowler15} which have photometry that can mimic a Lyman break. % in the $Z$ or $y$ band. 
The selection of robust samples requires the use of deep optical, NIR and \textit{Spitzer} Infrared Array Camera (IRAC) photometry via a colour-colour selection or spectral energy distribution (SED) fitting \citep[e.g.][]{Bowler14, rebels}{}{}. 
In this work we conduct a search for $z\simeq7$ galaxies within the VISTA VIDEO XMM-LSS and ECDF-S fields.
Crucially, these fields are covered by a range of deep multi-wavelength data including the optical, red optical and NIR filters extending to $2.2\,\mu{\rm m}$ (and up to $5\,\mu{\rm m}$ with~\emph{Spitzer}/IRAC data) allowing the robust removal of contaminant populations.
This work comprises the largest ground-based optical+NIR search for $z\simeq7$ galaxies to date, providing the most robust constraints on the bright-end of the rest-frame UV LF prior to the launch of wide-area space missions.

This paper is structured as follows. 
In Section \ref{sec:data} we outline the multi-wavelength datasets, and in Section~ \ref{sec:selection} we present the selection process of our $z\simeq7$ LBG sample. 
We present our candidate $z\simeq7$ galaxies and their properties in Section~\ref{finalsample}, and in Section \ref{sec:uvlf} we measure the rest-frame UV LF at $z\simeq7$ and present our results. We discuss contamination and potential contribution by AGN in Section \ref{sec:discussion}. 
We then present our conclusions in Section \ref{sec:conclusion}. 
In the Appendix we present SED fitting and postage stamps of all candidate objects. 
We also present an alternate calculation of the UV LF at $z\simeq7$ using a more inclusive sample.
All magnitudes are reported in the AB system \citep[][]{oke1983}. We assume a standard $\Lambda\mathrm{CDM}$ cosmology, $\mathrm{H}_{0}=70 \ \mathrm{km \ s^{-1} \ Mpc^{-1}}, \Omega_{\mathrm{M}}=0.3, \Omega_{\Lambda} = 0.7$. 

\section{Data}
\label{sec:data}

In this study we use multi-wavelength data across two fields, XMM-LSS and ECDF-S. These fields are selected for their coverage by the VISTA Deep Extragalactic Observations (VIDEO) survey in the near-infrared $YJHK_{s}$ bands \citep{Jarvis2013}, in addition to having overlapping deep optical data. We use the final data release of the VISTA VIDEO survey, which is now publicly available\footnote{http://www-astro.physics.ox.ac.uk/\textasciitilde video/public/Home}. The VIDEO data in the ELAIS-S1 field is not included due to a current lack of deep optical imaging, particularly in the red optical bands around the expected position of the Lyman break. The total  overlapping optical and NIR area used in this work covers $8.22 \ \mathrm{deg}^{2}$. The field footprints are shown in Fig. \ref{fig:cdfs}. The optical and NIR bands are used for SED fitting, and \textit{Spitzer}/Infrared Array Camera (IRAC) imaging in the mid-infrared is used to remove low-redshift dusty interlopers (see Section \ref{sec:contaminants}). 

\subsection{XMM-LSS}
\label{sec:XMM} % used for referring to this section from elsewhere

The VIDEO data are split into three VISTA pointings, or tiles (XMM1, 2, 3), of $1.5 \ \mathrm{deg}^{2}$ each as shown in Fig. \ref{fig:cdfs}. Optical data are taken from the Data Release $3$ of the Hyper Suprime-Cam Subaru Strategic Program \citep[HSC-SSP,][]{AiharaDR3}. XMM1 contains an `ultradeep' HSC pointing with longer exposure than the `deep' pointings. The DR3 data provide an increase in 5$\sigma$ depth of $\sim0.1-0.7$ mag on DR2 \citep[][]{AiharaDR2}. HSC-SSP also provides two narrow bands, NB816 and NB921, near the expected position of the Lyman break at $z\simeq7$. This can aid in reducing errors on photometric redshifts. In addition, imaging from the \textit{Spitzer} Extragalactic Representative Volume Survey \citep[SERVS, ][]{SERVS} is used along with the deeper Spitzer Extended Deep Survey \citep[SEDS,][]{ashby13} where available. The area containing overlapping VIDEO and HSC data is 4.33 $\ \mathrm{deg}^{2}$, accounting for masking of bright stars and artefacts such as stellar ghosts. 

\subsection{ECDF-S}
\label{sec:CDFS}
\begin{figure}
    \includegraphics[width=\columnwidth]{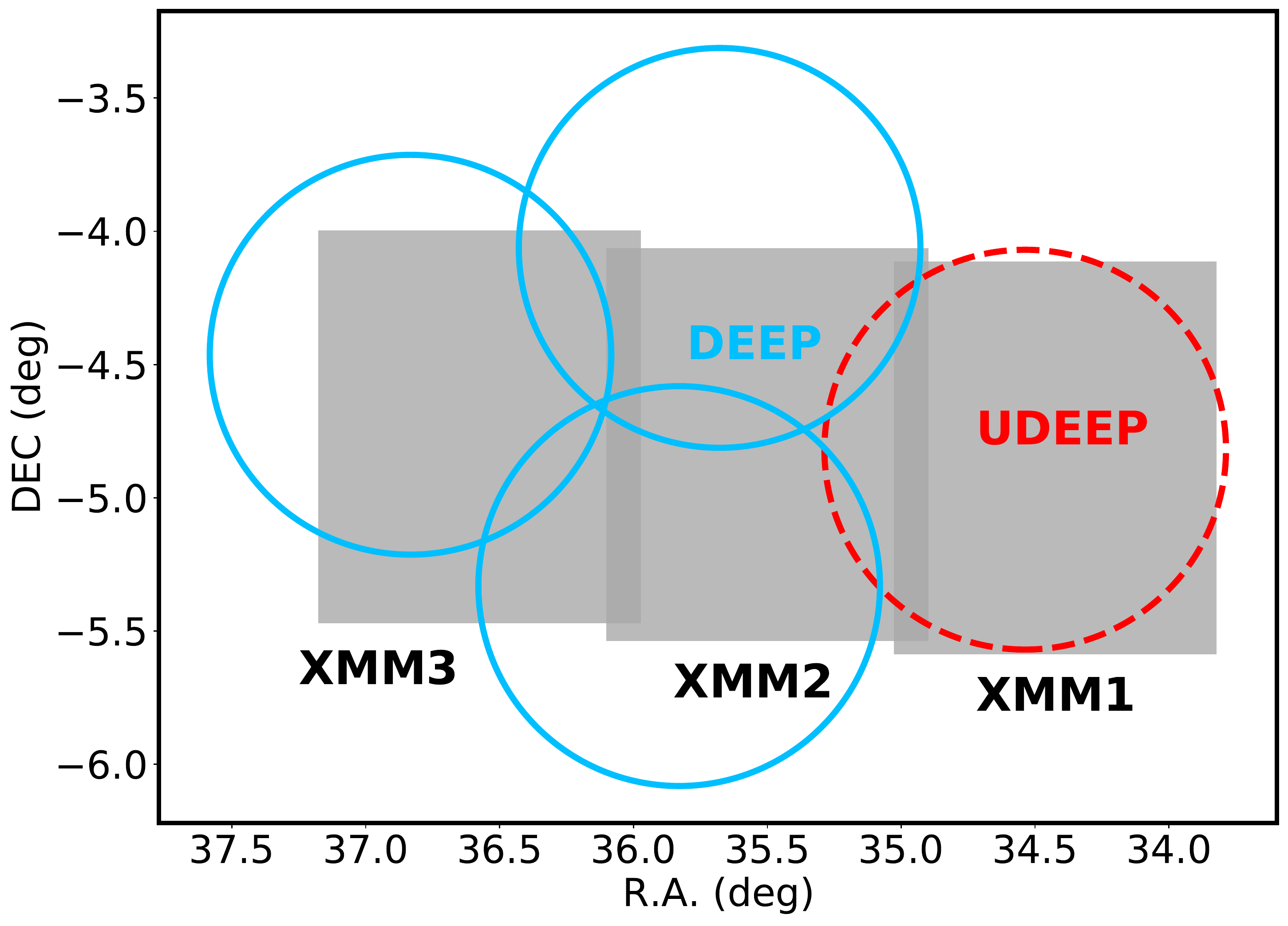}
    \includegraphics[width=\columnwidth]{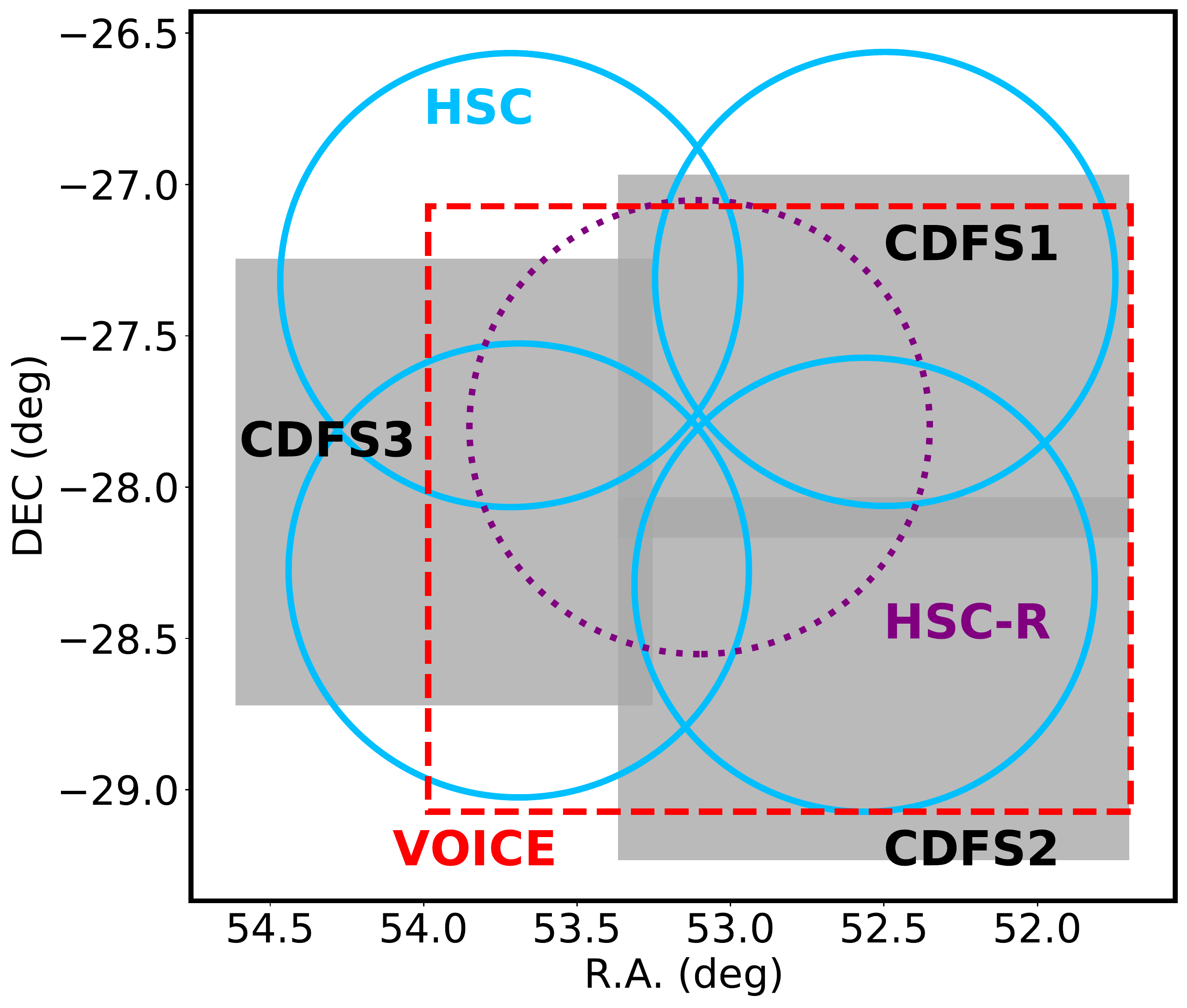}

    \caption{XMM-LSS footprint (top) and ECDF-S footprint (bottom). The VIDEO tiles are shown in grey and the HSC pointings in blue circles. In ECDF-S, the VOICE pointing is shown as a dashed red square. The central HSC-$R$ pointing is shown as a dotted purple circle. In XMM-LSS, the ultradeep HSC pointing is shown as a dashed red circle.}
    \label{fig:cdfs}
\end{figure}

Similarly to XMM-LSS, the VIDEO images in ECDF-S are split into three tiles (CDFS1, 2, 3) of $1.5 \ \mathrm{deg}^{2}$ each as shown in Fig. \ref{fig:cdfs}. Optical data is comprised of ancillary HSC data compiled by \citet{NiCDFS}, in the form of four pointings in \textit{G, I} and \textit{Z} and a central single pointing of the \textit{R}-band. The HSC data are a combination of different observation conditions, leading to poor seeing in the $I$- and $Z$-bands. The full width at half maximum (FWHM) of the point spread function (PSF) varies by 0.1 arcsec across the field in these bands (see Section \ref{sec:catalogue_creation}). Furthermore, there is poor coverage in the $R$-band. We therefore complement the HSC data with optical data from VST Optical Imaging of the CDFS and ES1 fields \citep[VOICE, ][]{VOICE}, covering $4 \ \mathrm{deg}^{2}$. \textit{Spitzer}/IRAC imaging is taken from the Cosmic Dawn Survey \citep[CDS, ][]{CDS}. The area containing overlapping VIDEO, HSC and VOICE data is 3.89 $\ \mathrm{deg}^{2}$.

\subsection{Image processing and depths}

The astrometry of the VIDEO data is matched to GAIA \citep{GaiaDR2}. 
If not already matched to GAIA, all other auxiliary data was shifted into this frame using \textsc{Scamp} \citep{scamp}, and the pixel-scale was matched using \textsc{Swarp} \citep{swarp}. The $5\sigma$ depths were computed across the images to determine flux uncertainties. Circular apertures with a 2 arcsec diameter were placed on empty regions of the image. For the \emph{Spitzer}/IRAC data we use a 2.8 arcsec diameter aperture to account for the poorer resolution. The \textsc{Segmentation} map produced by \textsc{SExtractor} \citep{sextractor} was used to avoid foreground objects. To measure the standard deviation of the fluxes, the median absolute deviation (MAD) was used. This is more robust to outliers than Gaussian fitting, and is given by $\mathrm{MAD} = \mathrm{median(|\mathrm{flux} - \mathrm{median}|)}$, with $\sigma = 1.4826\times\mathrm{MAD}$. Local depth maps in each image were produced by splitting the image into local regions with a size determined by taking the closest 300 apertures to each point. The global depths were taken as the mode of these local depths and are reported in Table~\ref{tab:Depths}.

\begin{table}
	\centering
	\caption{The $5\sigma$ limiting magnitudes for each band used in this work, in each VISTA VIDEO tile (see Fig. \ref{fig:cdfs}). The local depths were measured by placing 2 arcsec diameter circular apertures on empty regions of the images. The depth quoted here is the mode of these local depths. IRAC depths were measured in 2.8 arcsec diameter circular apertures to account for the poorer resolution. The first section shows bands from VST. The second section shows bands from HSC. The third section shows bands from VISTA. The final section shows bands from \textit{Spitzer}/IRAC.}
	\label{tab:Depths}
	\begin{tabular}{lcccccr} % four columns, alignment for each
		\hline
        Filter & XMM1 & XMM2 & XMM3 & CDFS1 & CDFS2 & CDFS3   \\[1ex]
        \hline
         
        \textit{u} & - & - & - & 25.5 & 25.6 & 25.4 \\[1ex]
          
        \textit{g} & - & - & - & 26.0 & 25.9 & 26.0 \\[1ex]
           
        \textit{r} & - & - & - & 26.0 & 26.0 & 26.0 \\[1ex]
            
        \textit{i} & - & - & - & 24.6 & 24.8 & 24.7 \\[1ex]
        
        \hline
        
         \textit{G} & 27.6 & 27.1 & 27.1 & 26.5 & 26.2 & 26.4 \\[1ex] 
        
         \textit{R} & 27.1 & 26.6 & 26.5 & 25.5 & 25.5 & 25.5 \\[1ex]
        
         \textit{I} & 26.9 & 26.2 & 26.2 & 25.6 & 25.5 & 25.5 \\[1ex]
         
         NB0816 & 26.0 & 25.4 & 25.4 & - & - & - \\[1ex]
        
         \textit{Z} & 26.5 & 25.9 & 25.9 & 25.1 & 24.8 & 24.8 \\[1ex]
         
        NB0921 & 26.0 & 25.3 & 25.4 & - & - & - \\[1ex]
        
         \textit{y} & 25.6 & 24.7 & 24.7 & - & - & - \\[1ex] 
         
         \hline
         
         \textit{Y} & 25.2 & 25.1 & 25.2 & 25.2 & 25.1 & 25.3 \\[1ex] 
         
         \textit{J} & 24.7 & 24.7 & 24.7 & 24.6 & 24.9 & 24.6 \\[1ex]
         
         \textit{Y+J} & 25.3 & 25.3 & 25.3 & 25.3 & 25.3 & 25.2 \\[1ex]
         
         \textit{H} & 24.2 & 24.3 & 24.3 & 24.1 & 24.2 & 24.1 \\[1ex]
         
         $K_{s}$ & 23.8 & 23.9 & 23.9 & 23.8 & 23.7 & 23.7 \\[1ex]
         
         \hline
         
         3.6$\micron$ & 24.3 & 24.3 & 24.4 & 25.0 & 24.9 & 24.9 \\[1ex]
         
         4.5$\micron$ & 24.1 & 24.1 & 24.1 & 24.8 & 24.8 & 24.8 \\[1ex]
         
         \hline
	\end{tabular}
\end{table}

\subsection{Catalogue creation}
\label{sec:catalogue_creation}

At $z\simeq7$, the Lyman break will cause objects to drop out of either the $Z$, $y$ or $Y$ bands. The VIDEO-$Y$ band is deeper than the HSC-$y$ band (see Table \ref{tab:Depths}), hence we produced fiducial catalogues  by running \textsc{SExtractor} in dual image mode on a $Y+J$ stacked image (parameters used are provided in Table \ref{tab:sextractor_params}). This probes the rest-frame UV ($\sim 1500$\AA) at $z\simeq7$.  The photometry was performed in 2 arcsec diameter apertures, which encloses 70-80\% of the total flux of a point source. This provides high signal-to-noise whilst balancing against the need for a large aperture correction. For the \textit{Spitzer}/IRAC photometry, 2.8 arcsec diameter apertures were used to account for the broader PSF.

\begin{table}
    \centering
    \begin{tabular}{cc}
        \hline
         Parameter & Value  \\
         \hline
         DETECT\_MINAREA & 3 \\ 
         DETECT\_THRESH & 1.4 \\
         ANALYSIS\_THRESH & 1.4 \\
         DEBLEND\_NTHRESH & 32 \\
         DEBLEND\_MINCONT & 0.0001 \\
         BACK\_SIZE & 64 \\
         BACK\_FILTERSIZE & 9 \\
         BACKPHOTO\_TYPE & LOCAL \\
         BACKPHOTO\_THICK & 24 \\
         \hline
         
    \end{tabular}
    \caption{Values of parameters used when running \textsc{SExtractor} in dual image mode.}
    \label{tab:sextractor_params}
\end{table}

Since fluxes are measured in fixed apertures, the measurement must be corrected to a total flux by accounting for light beyond the aperture using the shape of the PSF. 
The FWHM of the $Y$-band imaging is 0.8 arcsec. However, as mentioned in Section \ref{sec:CDFS} for ECDF-S the FWHM of the PSF varies markedly (by up to 0.1 arcsec) across the field for the $I-$ and $Z-$bands.
We thus used \textsc{PSFEx} \citep{psfex} to generate a local PSF model to aperture correct our photometry by splitting each tile into a 10$\times$10 grid.
For our analysis we only considered regions where VOICE, HSC and VIDEO data overlap (see Fig. \ref{fig:cdfs}). Bright stars and regions of low signal-to-noise at the edges of the VISTA VIDEO tiles are masked.

\section{Galaxy selection}
\label{sec:selection}

The selection of robust $z\simeq7$ candidates required several steps, which we detail in this section and present an overview of in Table \ref{tab:colour_cut}.

\subsection{Initial selection}
\label{sec:initial selection}

The $Y+J$ selected catalogues contained $1,814,513$ and $1,846,027$ objects in XMM-LSS and ECDF-S respectively. Objects without overlapping VOICE+HSC+VIDEO imaging are removed, shown as the `remove missing data' step in Table \ref{tab:colour_cut}. Bright objects were selected by requiring detections at $>5\sigma$ significance in $Y+J$ for XMM-LSS. In the ECDF-S data, a $5\sigma$ cut was insufficient because $Z$ is $\sim1$ mag shallower than $Y$ (see Table \ref{tab:Depths}), making it challenging to select drop-out objects. We therefore applied a brighter $8\sigma$ cut in ECDF-S, corresponding to $\sim0.5$ mag brighter than the $5\sigma$ cut in XMM-LSS.

We then require $<2\sigma$ significance (i.e. a non-detection) in bands bluewards of and including $i/I$. In ECDF-S the HSC $R$-band is not used for this due to its poor coverage and because $r$ is deeper by $0.5$mag. Similarly, the $i$-band is not used for this since HSC-$I$ is $\gtrsim0.7$mag deeper (see Table \ref{tab:Depths}). The $u-$band is also not used due to its significantly shallower depth. Note that all of these bands are later included for SED fitting in Section \ref{sed_preamble}. The results of these steps on reducing the sample are shown in Table \ref{tab:colour_cut}. 

\begin{table}
    \centering
	\caption{The results of our selection in XMM-LSS (top) and ECDF-S (bottom). The left column describes the selection step and the right column shows the number of objects remaining after this step. SED fitting is used to find candidates with photometric redshifts $z>6$ and remove low-redshift interlopers, and visual checks are used to remove artefacts. The objects remaining after the removal of brown dwarfs is the final number of candidate galaxies in each field.}
	\label{tab:colour_cut}
	\begin{tabular}{lcr} % four columns, alignment for each
		\hline
        Selection step & Objects remaining\\[1ex]%&
        \hline
        \multicolumn{2}{c}{\textbf{XMM-LSS}}  \\[1ex]
        Initial catalogue & 1814513  \\%& 100\% \\[1ex]
        Remove missing data & 1707431  \\%& 5.9\% \\[1ex]
        $5\sigma$ $Y+J$ cut & 807190  \\%& 52.7\% \\[1ex]
        $<2\sigma$ in $I$ & 11821  \\%& 98.5\% \\[1ex]
        $<2\sigma$ in $R$ & 3398  \\ %& 71.3\% \\[1ex]
        $<2\sigma$ in $G$ & 3217  \\%& 5.3\% \\[1ex] 
        SED fitting: $z>6$ & 307 \\ 
        Visual selection & 121 \\
        Dusty $z\sim1-2$ galaxies & 97\\
        Brown dwarf removal & 22\\
        \hline
        %Final candidates & 22 \\
        %\hline
        \multicolumn{2}{c}{\textbf{ECDF-S}} \\
        Initial catalogue & 1846027  \\%& 100\% \\[1ex]
        Remove missing data & 1291653  \\%& 30.0\% \\[1ex]
        $8\sigma$ $Y+J$ cut & 428695  \\%& 67.0\% \\[1ex]
        $<2\sigma$ in $I$ & 2859 \\% & 99.3\% \\[1ex]
        $<2\sigma$ in $r$ & 1504  \\%& 47.4\% \\[1ex]
        $<2\sigma$ in $g$ & 1427  \\%& 5.1\% \\[1ex]
        $<2\sigma$ in $G$ & 1390  \\%& 2.6\% \\[1ex]
        SED fitting: $z>6$ & 186 \\ 
        Visual selection & 70 \\
        Dusty $z\sim1-2$ galaxies & 61\\
        Brown dwarf removal & 6\\
        \hline
        %Final candidates & 6 \\
        %\hline
	\end{tabular}
\end{table}

\begin{table*}
\caption{The photometry of sources in our final $z\simeq7$ sample. The top section shows objects in XMM-LSS, and the bottom section shows objects in ECDF-S. Objects are ordered by their photometric redshift. The first column shows the object ID, and the next two columns show the coordinates of the candidate. The remaining columns show the photometry in the bands available in each field. We require a $<2\sigma$ detections bluewards of and including the $i/I$-bands, so we only present bands redwards of this. The photometry is measured in a 2.0 arcsec diameter circular aperture apart from the \textit{Spitzer}/IRAC bands where 2.8 arcsec diameter apertures are used to account for the broader PSF. The photometry is corrected to a total flux assuming a point-source correction. If the measured flux is $<2\sigma$ we present an upper limit. Objects with confused \textit{Spitzer}/IRAC photometry are marked with an asterisk.}
\label{tab:photometry}
\begin{tabular}{lccccccccccc}
\hline
ID & RA & DEC & Z & NB921 & y & Y & J & H & Ks & [3.6] & [4.5] \\[1ex]
\hline
VIDEO\_z7\_1$^{*}$ & 02:22:56.06 & -05:28:00.37 & $26.3^{+0.5}_{-0.3}$ & $24.9^{+0.2}_{-0.2}$ & $25.3^{+0.6}_{-0.4}$ & $24.7^{+0.2}_{-0.2}$ & $24.4^{+0.2}_{-0.2}$ & $24.9^{+0.7}_{-0.4}$ & $23.7^{+0.2}_{-0.2}$ & $22.2^{+0.1}_{-0.1}$ & $22.0^{+0.1}_{-0.1}$ \\[1ex]
VIDEO\_z7\_2 & 02:25:51.03 & -04:03:16.98 & $26.3^{+0.7}_{-0.4}$ & $25.5^{+0.4}_{-0.3}$ & $25.1^{+0.7}_{-0.4}$ & $24.5^{+0.2}_{-0.2}$ & $24.5^{+0.3}_{-0.2}$ & $24.3^{+0.4}_{-0.3}$ & >$24.1$ & $23.4^{+0.2}_{-0.2}$ & $23.7^{+0.3}_{-0.2}$ \\[1ex]
VIDEO\_z7\_3 & 02:26:39.93 & -04:01:09.51 & $26.3^{+0.5}_{-0.4}$ & $24.9^{+0.2}_{-0.1}$ & $24.6^{+0.3}_{-0.2}$ & $24.4^{+0.2}_{-0.1}$ & $24.0^{+0.2}_{-0.2}$ & $24.4^{+0.4}_{-0.3}$ & >$24.2$ & $23.3^{+0.2}_{-0.2}$ & $23.8^{+0.2}_{-0.2}$ \\[1ex]
VIDEO\_z7\_4 & 02:27:38.62 & -04:51:36.34 & >$26.2$ & $25.2^{+0.2}_{-0.2}$ & >$24.6$ & $24.7^{+0.2}_{-0.2}$ & $25.0^{+0.5}_{-0.4}$ & $24.9^{+0.7}_{-0.4}$ & >$24.2$ & $23.5^{+0.2}_{-0.2}$ & $23.5^{+0.2}_{-0.2}$ \\[1ex]
VIDEO\_z7\_5$^{*}$ & 02:26:30.02 & -04:20:32.23 & >$26.2$ & $25.9^{+0.5}_{-0.3}$ & $24.8^{+0.4}_{-0.3}$ & $24.4^{+0.2}_{-0.2}$ & $24.4^{+0.3}_{-0.2}$ & $24.2^{+0.4}_{-0.3}$ & $24.2^{+0.5}_{-0.3}$ & $21.8^{+0.2}_{-0.2}$ & $21.8^{+0.2}_{-0.2}$ \\[1ex]
VIDEO\_z7\_6 & 02:16:32.44 & -05:30:05.82 & >$25.9$ & $24.9^{+0.2}_{-0.2}$ & $24.6^{+0.2}_{-0.2}$ & $24.7^{+0.2}_{-0.2}$ & >$24.6$ & >$24.1$ & $24.3^{+0.5}_{-0.4}$ & $24.5^{+0.3}_{-0.2}$ & >$25.1$ \\[1ex]
VIDEO\_z7\_7 & 02:27:30.78 & -04:25:12.11 & $26.4^{+0.4}_{-0.3}$ & $24.7^{+0.1}_{-0.1}$ & $24.6^{+0.3}_{-0.2}$ & $24.5^{+0.1}_{-0.1}$ & $25.2^{+0.5}_{-0.3}$ & $24.4^{+0.4}_{-0.3}$ & >$23.9$ & $23.8^{+0.3}_{-0.2}$ & $24.2^{+0.6}_{-0.4}$ \\[1ex]
VIDEO\_z7\_8 & 02:22:05.84 & -04:54:10.85 & >$26.0$ & >$25.5$ & $25.0^{+0.6}_{-0.4}$ & $24.6^{+0.2}_{-0.2}$ & $24.8^{+0.4}_{-0.3}$ & >$24.4$ & $24.4^{+0.5}_{-0.4}$ & $23.9^{+0.3}_{-0.2}$ & >$24.4$ \\[1ex]
VIDEO\_z7\_9 & 02:23:47.30 & -05:21:25.49 & $25.4^{+0.2}_{-0.2}$ & $25.0^{+0.2}_{-0.2}$ & $24.4^{+0.2}_{-0.2}$ & $24.5^{+0.1}_{-0.1}$ & $24.3^{+0.2}_{-0.2}$ & $24.3^{+0.3}_{-0.3}$ & $24.3^{+0.5}_{-0.3}$ & $22.9^{+0.1}_{-0.1}$ & $23.2^{+0.1}_{-0.1}$ \\[1ex]
VIDEO\_z7\_10$^{*}$ & 02:16:36.51 & -04:54:50.62 & >$26.8$ & >$26.4$ & $24.9^{+0.2}_{-0.2}$ & $24.8^{+0.3}_{-0.2}$ & $24.6^{+0.3}_{-0.2}$ & $24.7^{+0.6}_{-0.4}$ & $24.3^{+0.6}_{-0.4}$ & $22.1^{+0.2}_{-0.2}$ & $22.2^{+0.2}_{-0.2}$ \\[1ex]
VIDEO\_z7\_11 & 02:24:59.99 & -04:43:54.27 & >$26.6$ & >$26.0$ & $24.7^{+0.3}_{-0.2}$ & $24.7^{+0.3}_{-0.2}$ & $24.9^{+0.4}_{-0.3}$ & $24.7^{+0.4}_{-0.3}$ & >$24.2$ & $23.6^{+0.2}_{-0.2}$ & >$24.3$ \\[1ex]
VIDEO\_z7\_12$^{*}$ & 02:17:00.07 & -04:12:26.78 & >$26.6$ & >$26.6$ & $25.1^{+0.2}_{-0.2}$ & $24.8^{+0.2}_{-0.2}$ & $25.3^{+0.6}_{-0.4}$ & >$24.7$ & >$23.8$ & $23.6^{+0.2}_{-0.2}$ & $24.4^{+0.3}_{-0.2}$ \\[1ex]
VIDEO\_z7\_13 & 02:25:17.59 & -04:38:54.33 & >$26.2$ & >$25.4$ & $24.8^{+0.4}_{-0.3}$ & $24.7^{+0.2}_{-0.1}$ & $24.7^{+0.3}_{-0.2}$ & $24.6^{+0.4}_{-0.3}$ & $24.7^{+0.6}_{-0.4}$ & >$24.0$ & $24.0^{+0.4}_{-0.3}$ \\[1ex]
VIDEO\_z7\_14 & 02:27:37.26 & -04:10:52.04 & >$26.6$ & >$25.8$ & $25.2^{+0.5}_{-0.3}$ & $24.8^{+0.2}_{-0.2}$ & $25.2^{+0.6}_{-0.4}$ & $25.0^{+0.7}_{-0.4}$ & $24.7^{+0.7}_{-0.4}$ & >$24.6$ & >$24.2$ \\[1ex]
VIDEO\_z7\_15 & 02:21:29.47 & -05:15:27.24 & >$26.5$ & >$25.7$ & $25.5^{+0.6}_{-0.4}$ & $25.2^{+0.3}_{-0.2}$ & $25.3^{+0.4}_{-0.3}$ & >$24.5$ & >$24.2$ & >$24.1$ & >$23.8$ \\[1ex]
VIDEO\_z7\_16 & 02:22:27.88 & -05:13:41.40 & >$26.5$ & >$25.9$ & $25.5^{+0.7}_{-0.4}$ & $24.9^{+0.2}_{-0.2}$ & >$25.1$ & $25.1^{+0.6}_{-0.4}$ & >$24.6$ & >$24.6$ & >$24.3$ \\[1ex]
VIDEO\_z7\_17 & 02:25:30.47 & -05:18:01.46 & >$26.6$ & >$25.5$ & $25.3^{+0.7}_{-0.4}$ & $24.9^{+0.2}_{-0.2}$ & $25.2^{+0.4}_{-0.3}$ & >$24.6$ & >$24.2$ & >$24.0$ & >$24.5$ \\[1ex]
VIDEO\_z7\_18 & 02:21:16.56 & -04:33:20.28 & >$26.7$ & >$25.8$ & $25.2^{+0.3}_{-0.2}$ & $24.4^{+0.1}_{-0.1}$ & $24.6^{+0.2}_{-0.2}$ & $24.3^{+0.3}_{-0.2}$ & $24.6^{+0.6}_{-0.4}$ & $22.6^{+0.2}_{-0.2}$ & $23.1^{+0.2}_{-0.2}$ \\[1ex]
VIDEO\_z7\_19 & 02:18:06.79 & -04:25:30.56 & >$27.1$ & >$26.5$ & $25.3^{+0.2}_{-0.2}$ & $24.7^{+0.2}_{-0.2}$ & $25.0^{+0.4}_{-0.3}$ & >$24.7$ & >$24.3$ & $24.9^{+0.4}_{-0.3}$ & >$24.3$ \\[1ex]
VIDEO\_z7\_20 & 02:15:31.44 & -05:09:07.95 & >$26.6$ & >$26.4$ & $25.1^{+0.3}_{-0.2}$ & $24.6^{+0.2}_{-0.2}$ & $24.7^{+0.4}_{-0.3}$ & $24.6^{+0.7}_{-0.4}$ & >$24.0$ & $24.1^{+0.2}_{-0.2}$ & $25.0^{+0.5}_{-0.3}$ \\[1ex]
VIDEO\_z7\_21$^{*}$ & 02:26:46.18 & -04:59:53.51 & >$26.4$ & >$25.7$ & >$24.5$ & $24.6^{+0.2}_{-0.1}$ & $24.5^{+0.3}_{-0.2}$ & $24.5^{+0.4}_{-0.3}$ & $24.4^{+0.6}_{-0.4}$ & $24.1^{+0.4}_{-0.3}$ & $23.4^{+0.2}_{-0.2}$ \\[1ex]
VIDEO\_z7\_22$^{*}$ & 02:16:25.10 & -04:57:38.56 & >$27.2$ & >$26.2$ & $26.5^{+0.7}_{-0.4}$ & $24.7^{+0.2}_{-0.2}$ & $24.1^{+0.2}_{-0.1}$ & $24.1^{+0.2}_{-0.2}$ & $23.6^{+0.3}_{-0.2}$ & $23.1^{+0.2}_{-0.2}$ & $22.9^{+0.2}_{-0.2}$ \\[1ex]
\hline
VIDEO\_z7\_23$^{*}$ & 03:30:17.50 & -28:14:20.71 & >$25.1$ & - & - & $24.3^{+0.1}_{-0.1}$ & $24.4^{+0.2}_{-0.2}$ & $24.5^{+0.4}_{-0.3}$ & $23.7^{+0.3}_{-0.2}$ & $22.7^{+0.2}_{-0.2}$ & $22.5^{+0.2}_{-0.2}$ \\[1ex]
VIDEO\_z7\_24$^{*}$ & 03:35:19.41 & -27:49:32.15 & >$25.3$ & - & - & $24.3^{+0.1}_{-0.1}$ & $24.6^{+0.3}_{-0.2}$ & $24.9^{+0.7}_{-0.4}$ & >$24.0$ & >$24.8$ & $24.8^{+0.5}_{-0.3}$ \\[1ex]
VIDEO\_z7\_25 & 03:34:44.08 & -28:02:51.72 & >$25.5$ & - & - & $24.2^{+0.1}_{-0.1}$ & $23.7^{+0.1}_{-0.1}$ & $23.2^{+0.1}_{-0.1}$ & $22.8^{+0.1}_{-0.1}$ & $22.5^{+0.2}_{-0.2}$ & $22.4^{+0.2}_{-0.2}$ \\[1ex]
VIDEO\_z7\_26 & 03:28:29.28 & -27:59:28.38 & >$25.5$ & - & - & $24.5^{+0.1}_{-0.1}$ & $24.2^{+0.2}_{-0.1}$ & $23.5^{+0.1}_{-0.1}$ & $23.1^{+0.1}_{-0.1}$ & $22.4^{+0.2}_{-0.2}$ & $22.1^{+0.2}_{-0.2}$ \\[1ex]
VIDEO\_z7\_27 & 03:30:42.79 & -27:17:30.39 & >$25.6$ & - & - & $24.2^{+0.1}_{-0.1}$ & $24.3^{+0.2}_{-0.2}$ & $24.7^{+0.6}_{-0.4}$ & $23.4^{+0.2}_{-0.2}$ & $23.8^{+0.2}_{-0.2}$ & $24.3^{+0.2}_{-0.2}$ \\[1ex]
VIDEO\_z7\_28 & 03:34:29.33 & -28:13:00.47 & >$25.1$ & - & - & $24.1^{+0.1}_{-0.1}$ & $23.5^{+0.1}_{-0.1}$ & $24.4^{+0.5}_{-0.3}$ & $24.2^{+0.4}_{-0.3}$ & $23.7^{+0.2}_{-0.2}$ & $22.4^{+0.2}_{-0.2}$ \\[1ex]

\hline
\end{tabular}
\end{table*}

\subsection{SED fitting and photometric redshifts}
\label{sed_preamble}

We make use of all available bands to select Lyman break galaxies at $z\simeq7$ by using a SED fitting analysis. SED fitting is done using \textsc{LePhare} \citep{arnouts99, ilbert06}, which works by minimising $\chi^{2}$ to find the best-fitting redshift and galaxy templates. The redshift was allowed to vary between $z=0-9$. \citet{bruzualcharlot} stellar population models were used with metallicities of $Z = [0.2, 0.4, 1.0] \ Z_{\odot}$. The star-formation histories explored were constant, instantaneous bursts and exponentially declining with timescales in the range $\tau = 0.05-10$ Gyr. Stellar population ages could range from 10 Myr to 13.8 Gyr, limited by the age of the Universe at a given redshift. Dust reddening was prescribed by the \citet{calzetti} dust law, with attenuation in the range $0.0 \le A_{V} \le 4.0$ to allow for very dusty low-redshift interlopers. A \citet{chabrier} initial mass function is assumed, and absorption by the IGM was applied according to \citet{madau95}. \textit{Spitzer}/IRAC photometry was not used initially to determine the photometric redshifts of objects due to the larger uncertainties and high rates of confusion due to a larger PSF. Instead, we included the IRAC photometry (when unconfused) for a separate check for low-redshift interlopers (see Section \ref{sec:contaminants}). 

Lyman-$\alpha$ emission can increase the photometric redshift of objects by $\Delta z \sim 0.5$ \citep{Bowler14} due to the addition of flux to the broadband. This is accounted for by simultaneously fitting \citet{bruzualcharlot} templates which have lines of equivalent widths $0$ \AA $ \ \le \mathrm{EW}_{0} \le 240$ \AA \ added . To do this, the continuum level was measured from the mean flux between 1250\AA \ and 1300\AA. The photometric redshift constraints from this Lyman-$\alpha$ analysis are stronger in XMM-LSS due to many overlapping filters around the expected position of the Lyman break, and due to the availability of narrowband filters that can very precisely pick out excess flux due to an emission line. It is highly degenerate in ECDF-S due to the gap between $Z$ and $Y$.

Candidates were first required to have their best fitting solution at $z>6$. The fits then had to be sufficiently good, defined as $\chi^{2} < 11.3$ ($12.9$) for XMM-LSS (ECDF-S). These values correspond to $2\sigma$ significance given 5 (6) degrees of freedom. We also required that the high-redshift solution was better than the low-redshift solution by a threshold $\Delta\chi^{2} > 4$, which corresponds to a $2\sigma$ significance.
This removes $2910$ objects in XMM-LSS and $1204$ in ECDF-S, leaving $307$ and $186$ objects respectively. 

\subsection{Visual selection}
\label{sec:vischeck}

We conducted a visual selection of the remaining objects to remove artefacts that appear as good high-redshift candidates in the photometry, primarily in the form of crosstalk in VIDEO images and diffraction spikes. Crosstalk is an artefact produced in the readout from the VIRCAM instrument on VISTA that produces ghost images at multiples of 128 pixels in the native scale from bright stars \citep[][]{bowler17}. Optical stacked postage stamp images are created to check for low-level optical flux that would indicate a low-redshift galaxy. The bands used for the optical stack are required to have similar $5\sigma$ depths, leading to a $GRI$ stack in XMM-LSS and a $grGI$ stack in ECDF-S. Single-band detections are also removed. This step removes $186$ objects in XMM-LSS and $116$ in ECDF-S, leaving $121$ and $70$ objects respectively.

Two objects in XMM-LSS (VIDEO\_z7\_8 and VIDEO\_z7\_12) appeared to have low-level flux in their optical stacks. This optical flux is offset from the position of the object in the $Y+J$ detection filter by $>1$ arcsec, suggesting the optical flux originates from a foreground object and hence we retain them.
One object (ID 1610530, RA 02:27:12.04, DEC -04:32:05.05) appeared to be remarkably bright at $M_{\mathrm{UV}} \simeq -24.4$, but a marginal detection at $3.6\micron$ around $2$ mag fainter than in $K_{s}$ and a non-detection at $4.5\micron$ resulted in very unusual photometry that passed our selection criterion. For such a rest-UV bright high-redshift object, we would expect much brighter \textit{Spitzer}/IRAC flux. We checked for this object in $J$-band imaging from the overlapping WIRCam Deep Survey \citep[WIRDS,][]{WIRDS}, where it was not present. However, a fainter object closely separated to the south is visible. This object is thus likely either a VIDEO artefact or a transient, supported by the faint $3.6\micron$ detection. The SED fitting and postage stamp images of this object are presented in Appendix \ref{sec: 1610530_stamp}.

\subsection{Removing contaminants}
\label{sec:contaminants}

Dusty galaxies at $z\sim1-2$ with Balmer breaks masquerading as Lyman breaks form one class of contaminant. These galaxies tend to have much redder continuum slopes than galaxies at $z\simeq7$, as observed in the NIR. We use SED fitting with the \textit{Spitzer}/IRAC bands to test for dusty low-redshift galaxies. If the SED fitting has a low-redshift solution that is preferred to the high-redshift solution when IRAC is included, the object is rejected if the IRAC photometry is unconfused. This step removes $24$ objects in XMM-LSS and $9$ objects in ECDF-S, leaving $97$ and $61$ objects respectively.

M, L and T brown dwarf (BD) stars comprise a major contaminant in high-redshift searches around a magnitude $m_{\mathrm{AB}} \sim 25$ where their number density in extragalactic surveys peaks \citep[][]{ryan11}{}{}. They have intrinsically red optical to NIR colours with heavy molecular spectral absorption complexes \citep[e.g.][]{Cushing08, Marley21}. The removal of BDs must be done carefully since peaks in their SEDs can line up with the ground-based $YJHK_{s}$ filters, mimicking a flat NIR colour. We fit our photometry with stellar templates taken from the SpeX prism library \citep{spex}. The $u$, $G, g$, $R$ and $r$ bands are excluded from the SED fitting since the templates do not contain any information at these wavelengths. The $\chi^{2}_{\mathrm{BD}}$ values are then compared to $\chi^{2}_{\mathrm{galaxy}}$ values, also fitted without these filters. In \citet{Bowler15}, a simulation of brown dwarfs in the Milky Way shows that removing objects with good brown dwarf fits at $\chi^{2}_{\mathrm{BD}} <1 0$ decreases the contamination rate to essentially zero. We therefore retain those high-redshift candidates which have $\chi^{2}_{\mathrm{BD}} > 10$.
This step removes around 80\% of objects (see Table \ref{tab:colour_cut}) compared to around 20\% in \citet{Bowler15}. We use a similar model to \citet{Bowler15} to compute the expected number of brown dwarfs in our fields, assuming a Galactic scale-height of 300 pc. 
%They suggest that the largest source at $z\simeq7$ is L- and T-type stars. 
Considering magnitudes down to the $5\sigma \ (8\sigma)$ depths in XMM-LSS (CDFS), we expect 480 and 220 dwarf stars respectively. Restricting this to the most common stellar types found by our initial SED fitting, we expect 146 and 76 dwarf stars respectively. We do not predict the exact number of dwarf stars because the Galactic scale-height is highly uncertain, with estimates varying between 300 pc \citep[][]{ryan17}{}{} and 400 pc \citep[][]{sorahana19}{}{}. The differences can thus be attributed to a higher LBG number density at the fainter magnitudes and lower redshift probed by their work ($-22.7 \le M_{\mathrm{UV}} \le -20.5$ at $z=6$), coupled with higher brown dwarf densities at brighter magnitudes probed by this work.

\citet{Bowler15} state that keeping objects with $\chi^{2}_{\mathrm{BD}} > 10 \ $ likely removes some genuine high-redshift galaxies, and they account for this with a completeness correction. We note that many of our candidates with $\chi^{2}_{\mathrm{BD}} < 10$ have a significantly better fitting high-redshift galaxy template. In an attempt to be complete to genuine $z\simeq7$ galaxies that may be removed from our primary robust sample due to being fit well as BDs, we conducted a more inclusive BD cut which retains these potential $z\simeq7$ galaxies, and is described in Appendix \ref{sec:inclusive calculation}.

%\subsection{Lyman-\texorpdfstring{$\alpha$}{alpha} emission}
%\label{sec:lya}
%The redshift range for our $z\simeq7$ UV LF is $6.5 < z < 7.5$, used to avoid objects being detected in bluer bands, i.e. $i/I$, and to ensure objects are detected at least in $J$. Lyman-$\alpha$ emission can increase the photometric redshift of objects by $\Delta z \sim 0.5$ \citep{Bowler14} due to the addition of flux to the broadband. To account for this, Lyman-$\alpha$ lines with equivalent widths $0 \le \mathrm{EW}_{0} \le 240$\AA \ are added to the \citet{bruzualcharlot} templates. To do this, the continuum level was measured from the mean flux between 1250\AA \ and 1300\AA. Candidates with $z_{\mathrm{Ly}\alpha} > 6.5$ were then retained. The photometric redshift constraints from this analysis are stronger in XMM-LSS due to many overlapping filters around the expected position of the Lyman break, and due to the availability of narrowband filters that can very precisely pick out excess flux due to an emission line. It is highly degenerate in ECDF-S due to the gap between $Z$ and $Y$. 

\subsection{SED fitting with \textsc{BAGPIPES}}
\label{sec:bagpipes fitting}

We repeat SED fitting on the final sample with Bayesian Analysis of Galaxies for Physical Inference and Parameter EStimation \citep[\textsc{BAGPIPES},][]{BAGPIPES} to compare photometric redshifts with \textsc{LePhare} and determine physical parameters such as stellar masses. The advantage of \textsc{BAGPIPES} is the implementation of grids of nebular emission line models based on input stellar population models. We used a fiducial constant star formation history (SFH) model. The time since star formation switched on varies between the start of the Universe and the redshift being considered. We adopt a uniform prior on the redshift, $0\le z \le 9$, and the ionization parameter, $-4 \le \mathrm{log}U \le -2$. The metallicity is fixed at $Z = 0.2Z_{\sun}$. Fixing of the metallicity at a value $Z < Z_{\sun}$ is motivated by recent spectroscopic measurements at $z\gtrsim7$ that suggest galaxies tend to be metal-poor \citep[e.g.][]{curtislake22, langeroodi22, matthee22, fujimoto22spec}. We also restrict the dust attenuation to the range $0 \le A_{V} \le 0.5$, limiting it to the largest attenuation we measure with \textsc{LePhare} (VIDEO\_z7\_25 and VIDEO\_z7\_26, shown in Table \ref{XMM_final}). If we do not restrict the range of $A_{V}$, large values occur for the objects with the most massive stellar masses: $A_{V} \gtrsim 1$ when  $\mathrm{log}_{10}(M/M_{\sun}) \gtrsim 11$. This is because dust extinction is degenerate with age (since we have fixed metallicity), so large values are a consequence of the constant SFH model rather than being a realistic estimate of the dust emission \citep[e.g.][]{inami22, ferrara22}.

\section{The final sample}
\label{finalsample}

% Define aliases for reference in sample table
\defcitealias{endsley2021}{E21}
\defcitealias{rebels}{B22}

\begin{table*}
    
    \caption{The 28 LBG candidates in XMM-LSS (top) and ECDF-S (bottom), ordered by photometric redshift as in Table \ref{tab:photometry}. The first column shows the object ID. The next four columns show properties of the high-redshift solution: photometric redshift, absolute magnitude in a top-hat filter at $1500$\AA \ with width $100$\AA, dust attenuation, and $\chi^{2}$ value as derived by \textsc{LePhare}. The next two columns show the secondary photometric redshift and $\chi^{2}$. The following two columns show the stellar type for fitting to standard MLT brown dwarf spectra, and $\chi^{2}_{\mathrm{BD}}$. The next two columns show the equivalent width for objects that have a best-fitting SED with Lyman-$\alpha$ emission. The final three columns show properties derived from SED fitting with \textsc{BAGPIPES}, namely the photometric redshift $z$, dust attenuation and stellar mass. Objects with IRAC non-detections have upper limits for their stellar masses. Object IDs marked with an asterisk have confused IRAC photometry, meaning stellar mass estimates may be overestimated. Object IDs marked with superscripts  1 and 2 have been previously identified in \citet{bouwens21} and \citet{endsley2021} respectively.} 
    \label{XMM_final}
    \begin{tabular}{lcccccccccccc}
    
        \hline
        ID & $z$ & $M_{\mathrm{UV}}$ & $A_{V}$ & ${\chi}^{2}$ & $z_{\mathrm{gal2}}$ & ${\chi}^{2}_{\mathrm{gal2}}$ & Stellar & ${\chi}^{2}_{\mathrm{BD}}$ & $\mathrm{EW_{0}}$  & $z$ & $A_{V}$ / mag & $\mathrm{log}_{10}(M_{\star}/M_{\sun})$\\

        & & / mag &/ mag & & & & Type   & &/ \ \AA  & (\textsc{BAGPIPES}) & (\textsc{BAGPIPES}) & (\textsc{BAGPIPES})  \\[1ex]
        \hline
        VIDEO\_z7\_1$^{*}$ &  $6.50^{+0.07}_{-0.06}$ & $-22.2\pm0.1$ & 0.2 & 11.9 & 1.3 & 18.4 & M8 & 18.2 & 40 &  $6.43^{+0.05}_{-0.06}$ & $0.41^{+0.07}_{-0.13}$ & $10.31^{+0.17}_{-0.32}$\\[1ex]
        VIDEO\_z7\_2 &  $6.52^{+0.08}_{-0.10}$ & $-22.3\pm0.2$ & 0.0 & 11.3 & 1.2 & 36.8 & T3 & 15.8 & & $6.53^{+0.03}_{-0.04}$ & $0.32^{+0.12}_{-0.15}$ & $10.20^{+0.15}_{-0.18}$ \\[1ex]
        VIDEO\_z7\_3 &  $6.53^{+0.02}_{-0.02}$ & $-22.5\pm0.1$ & 0.2 & 5.4 & 1.3 & 42.8 & M8 & 27.7 & &$6.49^{+0.03}_{-0.05}$ & $0.27^{+0.15}_{-0.17}$ & $10.17^{+0.24}_{-0.33}$ \\[1ex]
        VIDEO\_z7\_4 &  $6.53^{+0.03}_{-0.13}$ & $-22.0\pm0.2$ & 0.2 & 7.1 & 1.3 & 31.6 & T3 & 23.2 &&  $6.48^{+0.04}_{-0.06}$ & $0.42^{+0.06}_{-0.10}$ & $10.28^{+0.11}_{-0.15}$\\[1ex]
        VIDEO\_z7\_5 &  $6.57^{+0.04}_{-0.06}$ & $-22.4\pm0.1$ & 0.0 & 6.3 & 1.5 & 21.9 & T3 & 12.5& &$6.57^{+0.25}_{-0.02}$ & $0.32^{+0.14}_{-0.18}$ & $10.24^{+0.21}_{-0.28}$\\[1ex]
        VIDEO\_z7\_6$^{1}$ &  $6.58^{+0.03}_{-0.12}$ & $-22.0\pm0.1$& 0.0 & 7.9 & 1.3 & 46.2 & M8 & 31.0 &  50  & $6.46^{+0.04}_{-0.07}$ & $0.05^{+0.07}_{-0.03}$ & $9.12^{+0.29}_{-0.26}$  \\[1ex]
        VIDEO\_z7\_7 &  $6.58^{+0.02}_{-0.04}$  & $-22.1\pm0.1$ & 0.0 & 8.0 & 1.3 & 74.9 & M8 & 65.5 &  50  & $6.44^{+0.03}_{-0.04}$ & $0.17^{+0.17}_{-0.12}$ & $10.03^{+0.22}_{-0.37}$\\[1ex]
        VIDEO\_z7\_8 &  $6.64^{+0.31}_{-0.08}$ & $-22.1\pm0.2$ & 0.0 & 4.4 & 1.45 & 16.5 & M6 & 14.2 & & $6.73^{+0.18}_{-0.13}$ & $0.22^{+0.18}_{-0.15}$ & $10.04^{+0.23}_{-0.34}$\\[1ex]
        VIDEO\_z7\_9 &  $6.64^{+0.01}_{-0.22}$ & $-22.5\pm0.2$ & 0.0 & 4.1 & 1.25 & 23.5 & M8 & 17.7 & 240  & $6.30^{+0.05}_{-0.05}$ & $0.47^{+0.02}_{-0.04}$ & $10.55^{+0.05}_{-0.08}$\\[1ex]
        VIDEO\_z7\_10$^{2*}$ &  $6.69^{+0.12}_{-0.07}$ & $-22.2\pm0.2$ & 0.2 & 3.4 & 1.35 & 40.4 & T4 & 25.2 &  &$6.74^{+0.10}_{-0.10}$ & $0.46^{+0.03}_{-0.07}$ & $10.51^{+0.09}_{-0.13}$ \\[1ex]
        VIDEO\_z7\_11 &  $6.70^{+0.12}_{-0.06}$ & $-22.1\pm0.2$ & 0.0 & 2.7 & 1.5 & 26.9 & T8 & 20.4 &  &$6.70^{+0.11}_{-0.07}$ & $0.18^{+0.17}_{-0.13}$ & $9.93^{+0.25}_{-0.34}$\\[1ex]
        VIDEO\_z7\_12$^{*}$ &  $6.74^{+0.12}_{-0.10}$ & $-21.8\pm0.2$ & 0.0 & 8.6 & 1.5 & 30.0 & M6 & 33.9& & $6.68^{+0.10}_{-0.06}$ & $0.23^{+0.16}_{-0.13}$ & $9.94^{+0.16}_{-0.18}$\\[1ex]
        VIDEO\_z7\_13 & $6.79^{+0.22}_{-0.14}$ & $-22.3\pm0.2$ & 0.2 & 8.6 & 1.5 & 54.0 & T8 & 25.9& & $6.83^{+0.15}_{-0.17}$ & $0.13^{+0.16}_{-0.09}$ & $9.80^{+0.24}_{-0.34}$\\[1ex]
        VIDEO\_z7\_14 & $6.80^{+0.19}_{-0.15}$ & $-21.9\pm0.2$ & 0.0 & 7.8 & 1.5 & 28.2 & T8 & 18.9& &$6.79^{+0.13}_{-0.13}$ & $0.25^{+0.17}_{-0.17}$ & $<9.90$\\[1ex]
        VIDEO\_z7\_15 & $6.83^{+0.19}_{-0.18}$ & $-21.6\pm0.2$ & 0.0 & 4.7 & 1.5 & 24.4 & T8 & 12.7&& $6.77^{+0.19}_{-0.13}$ & $0.18^{+0.20}_{-0.14}$ & $<9.70$\\[1ex]
        VIDEO\_z7\_16 & $6.89^{+0.15}_{-0.20}$ & $-21.7\pm0.2$ & 0.0 & 8.0 & 1.5 & 31.8 & T8 & 23.5& &$6.79^{+0.14}_{-0.14}$ & $0.10^{+0.13}_{-0.07}$ & $<9.32$\\[1ex]
        VIDEO\_z7\_17 & $6.91^{+0.22}_{-0.24}$ & $-21.8\pm0.2$ & 0.0 & 7.8 & 1.15 & 35.1 & T8 & 16.9& &$6.79^{+0.15}_{-0.12}$ & $0.12^{+0.13}_{-0.09}$ & $<9.35$\\[1ex]
        VIDEO\_z7\_18 & $6.94^{+0.12}_{-0.10}$ & $-22.4\pm0.2$ & 0.0 & 6.3 & 1.5 & 45.4 & T3 & 23.6 & &$6.86^{+0.11}_{-0.11}$ & $0.44^{+0.04}_{-0.09}$ & $10.52^{+0.10}_{-0.14}$\\[1ex]
        VIDEO\_z7\_19 & $6.95^{+0.08}_{-0.09}$ & $-22.0\pm0.2$ & 0.0 & 9.4 & 1.25 & 88.2 & T8 & 35.9& &$6.76^{+0.11}_{-0.09}$ & $0.08^{+0.11}_{-0.05}$ & $9.40^{+0.28}_{-0.32}$\\[1ex]
        VIDEO\_z7\_20 & $6.99^{+0.09}_{-0.18}$ & $-22.3\pm0.2$ & 0.0 & 3.7 & 1.45 & 38.7 & T8 & 16.0 & &$6.78^{+0.10}_{-0.11}$ & $0.14^{+0.15}_{-0.09}$ & $9.55^{+0.24}_{-0.35}$\\[1ex]
        VIDEO\_z7\_21$^{1*}$ & $7.00^{+0.18}_{-0.31}$ & $-22.5\pm0.2$ & 0.1 & 4.5 & 1.5 & 41.8 & T8 & 17.5&  & $7.00^{+0.10}_{-0.15}$ & $0.25^{+0.14}_{-0.14}$ & $10.18^{+0.14}_{-0.19}$\\ [1ex]
        VIDEO\_z7\_22$^{1*}$ & $7.33^{+0.06}_{-0.07}$ & $-22.9\pm0.1$ & 0.0 & 11.4 & 1.6 & 33.6 & T1 & 14.5 &  & $7.29^{+0.06}_{-0.06}$ & $0.40^{+0.07}_{-0.10}$ & $10.59^{+0.11}_{-0.13}$\\[1ex]
        \hline
        VIDEO\_z7\_23$^{*}$ & $6.52^{+0.16}_{-0.13}$ & $-22.5\pm0.2$ & 0.0 & 6.7 & 1.45 & 12.2 & M6 & 14.7& &$6.49^{+0.19}_{-0.12}$ & $0.45^{+0.03}_{-0.07}$ & $10.61^{+0.09}_{-0.13}$\\[1ex]
        VIDEO\_z7\_24$^{*}$ & $6.57^{+0.2}_{-0.08}$ & $-22.4\pm0.2$ & 0.0 & 8.3 & 1.45 & 17.7 & M7 & 19.7& &$6.71^{+0.25}_{-0.21}$ & $0.04^{+0.06}_{-0.03}$ & $9.18^{+0.22}_{-0.20}$ \\[1ex]
        VIDEO\_z7\_25 & $6.59^{+0.44}_{-0.15}$ & $-22.9\pm0.3$ & 0.5 & 4.0 & 1.25 & 14.6 & L8 & 13.5& &$7.31^{+0.05}_{-0.05}$ & $0.46^{+0.03}_{-0.06}$ & $10.92^{+0.08}_{-0.14}$\\[1ex]
        VIDEO\_z7\_26 & $6.67^{+0.28}_{-0.32}$ & $-22.6\pm0.3$ & 0.5 & 4.0 & 1.5 & 8.7 & L8 & 14.7& &$7.30^{+0.05}_{-0.06}$ & $0.48^{+0.02}_{-0.03}$ & $10.86^{+0.06}_{-0.09}$\\[1ex]
        VIDEO\_z7\_27 & $6.67^{+0.24}_{-0.12}$ & $-22.6\pm0.2$ & 0.0 & 5.6 & 1.45 & 29.4 & M8 & 29.7& &$6.79^{+0.15}_{-0.12}$ & $0.20^{+0.18}_{-0.14}$ & $9.70^{+0.21}_{-0.31}$\\[1ex]
        VIDEO\_z7\_28 & $7.38^{+0.03}_{-0.06}$ & $-23.5\pm0.1$& 0.0 & 9.9 & 1.5 & 77.7 & T8 & 11.1& &$7.24^{+0.06}_{-0.09}$ & $0.09^{+0.14}_{-0.07}$ & $9.93^{+0.26}_{-0.50}$\\[1ex]

        \hline
    \end{tabular}
\end{table*}

 \begin{figure}
    \includegraphics[width=\columnwidth]{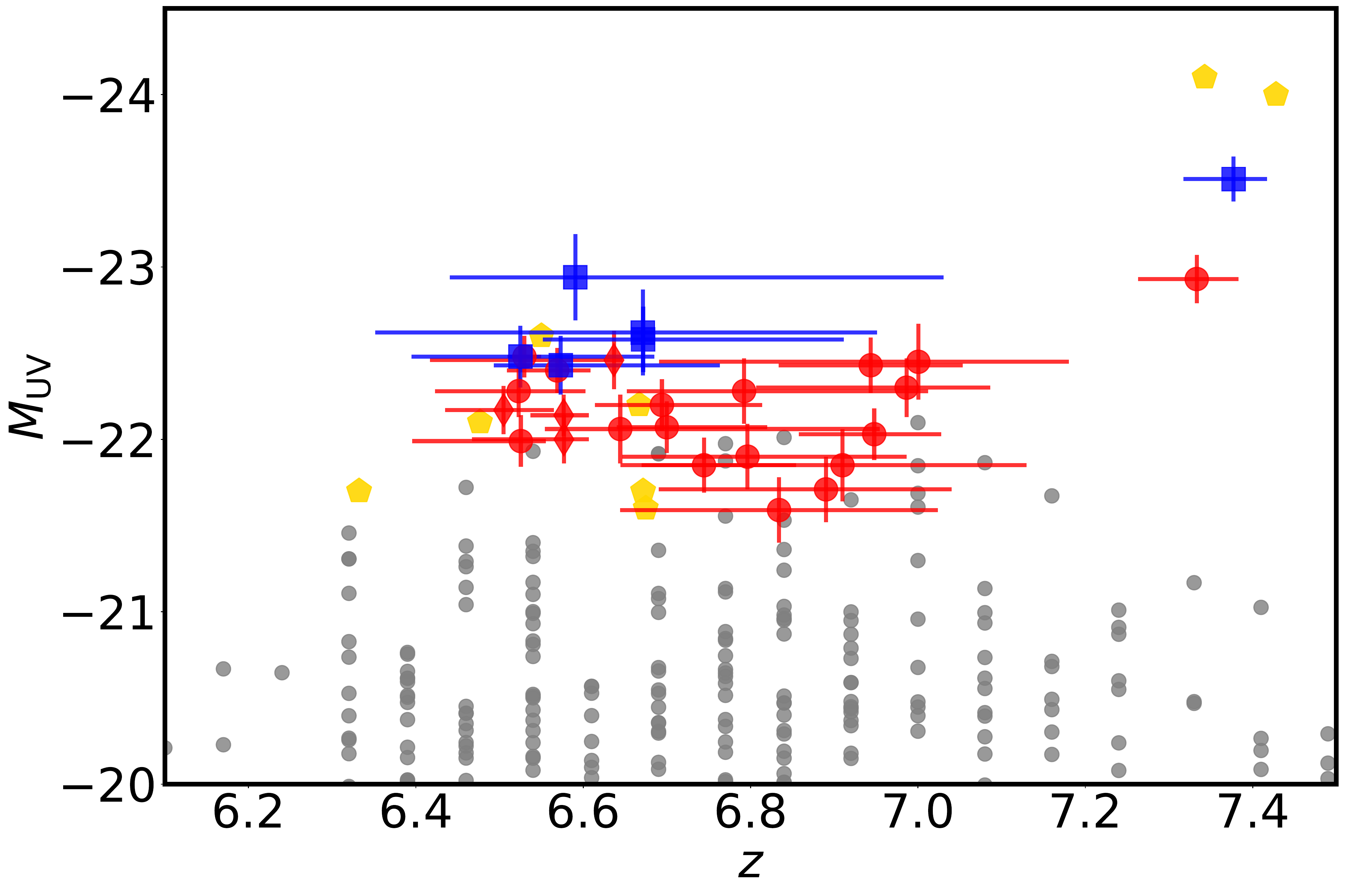}
    \caption{Rest-frame absolute UV magnitudes of the 28 LBGs in our sample against their photometric redshifts as measured by \textsc{LePhare}. Candidates from XMM-LSS are shown as red circles, and those with a Lyman-$\alpha$ best-fitting SED are shown as red diamonds. Candidates from ECDF-S are shown as blue squares. The grey circles are candidate galaxies from \citet{bouwens21}, which uses \textit{Hubble} data. The yellow pentagons show the high-redshift solution of SED fitting on objects from \citet{harikane22} crossmatched with our catalogues.}
    \label{fig:Muv vs z}
\end{figure}

The final sample consists of 28 objects, with photometry presented in Table \ref{tab:photometry}. The results of SED fitting and properties of objects (e.g. photometric redshift, $M_{\mathrm{UV}}$, dust attenuation) are shown in Table \ref{XMM_final}. Of the 22 primary candidates in XMM-LSS, 18 are newly identified. Five are within the ultradeep HSC pointing, and four have a best-fitting Lyman-$\alpha$ template (and have $z_{\mathrm{No \ Ly}\alpha}<6.5$). The best-fitting rest-frame equivalent widths for these objects are noted in Table \ref{XMM_final}.
In ECDF-S, all candidates are presented here for the first time. One object in ECDF-S had a best-fitting Lyman-$\alpha$ template, but was rejected due to having a good low-redshift fit. Therefore, all Ly$\alpha$ objects lie in XMM-LSS.
The absolute UV magnitudes, $M_{\mathrm{UV}}$, are computed by placing a top-hat filter at 1500\AA \ with width 100\AA \ on the rest-frame best-fitting SED. The sample has a mean redshift $\Bar{z} = 6.74$, and spans nearly 2 magnitudes in the rest-UV, $-23.5 \le M_{\mathrm{UV}} \le -21.6$. 
The distribution of photometric redshifts against UV absolute magnitude, split by field, is shown in Fig. \ref{fig:Muv vs z}. 
Candidates in ECDF-S are brighter on average than those in XMM-LSS simply by nature of the different selection thresholds that we imposed because of the shallower optical data available in this field. 
Uncertainties on the photometric redshifts tend to be lower in XMM-LSS compared to ECDF-S due to the proximity and overlap between $Z$, NB921, $y$ and $Y$ around the Lyman break. 
The grey points in Fig. \ref{fig:Muv vs z} are candidate galaxies from \citet{bouwens21}, which uses 1136 square arcminutes of \textit{Hubble} imaging to derive the rest-UV LF at $z\simeq7$. Their brightest candidate is $M_{\mathrm{UV}}\simeq-22.1$. Only 10 of our candidates are fainter than this, showing that these ground-based data better probe the bright-end of the rest-UV LF.

The SED fitting and stamps of a candidate galaxy from each field is shown in Fig. \ref{fig:candidates}. 
The SED fitting and stamps for all candidates in this work are shown in Appendix \ref{sec:merged}. 
The [3.6] and [4.5] \textit{Spitzer}/IRAC bands can be contaminated by nebular emission lines at $z > 6.5$ (primarily H$\mathrm{\alpha}$, H$\mathrm{\beta}$, \textsc{[O\,ii]} and \textsc{[O\,iii]}). The [3.6] - [4.5] colour will accordingly change with redshift as these lines move through the filters \citep[][]{smit2014, smit2015}. By accounting for nebular emission during SED fitting, we can obtain more precise photometric redshifts. An interesting example is the brightest object in our sample, VIDEO\_z7\_28, which has a red IRAC colour of $\mathrm{[3.6]} - \mathrm{[4.5]} = 1.3\pm0.4$ and a photometric redshift of $z=7.38^{+0.03}_{-0.06}$. \citet{Bowler_2020} extend the expected IRAC colours derived by \citet{smit2014} out to higher redshifts, predicting that galaxies with strong H$\beta+\mathrm{\textsc{[O\,iii]}}$ emission ($\mathrm{EW_{0}} = 500 - 2000$\AA) can have colours as high as $\mathrm{[3.6]} - \mathrm{[4.5]} \simeq 1.0$ at $z\simeq7.5$, in agreement with our photometry for this object.
As we discuss later in Section \ref{sec:previous work}, we show that several previous $z\sim7$ candidates at around $z=7.4$ are likely to be BDs. VIDEO\_z7\_28 passes our selection criterion with the BD fit showing $\chi^{2}_{\mathrm{BD}} = 11.1$. It is slightly extended in the data with a FWHM of 1.5 arcsec in comparison to the PSF of the $Y$-band. However, we cannot fully rule out a BD solution for this source, and further follow-up will be required to confirm this extremely luminous LBG.

\begin{figure*}
     \centering

         \includegraphics[width=\textwidth]{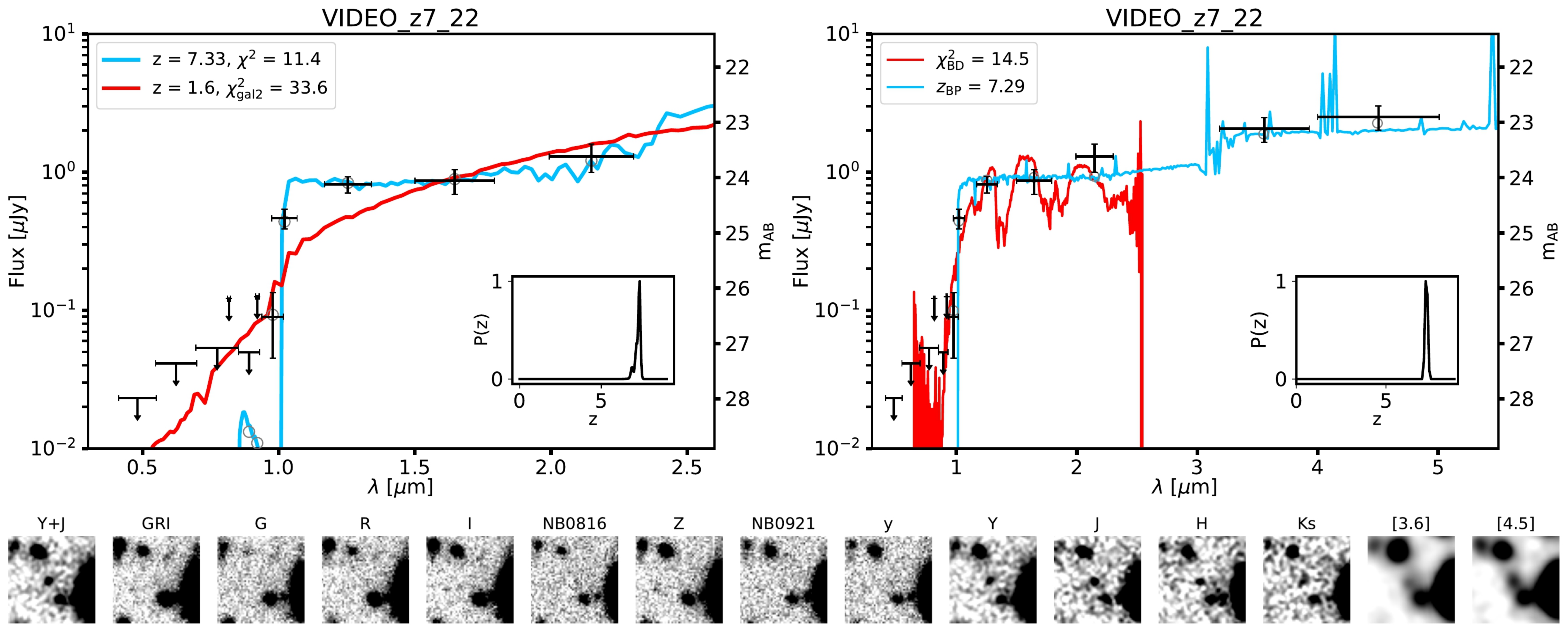}
         \includegraphics[width=\textwidth]{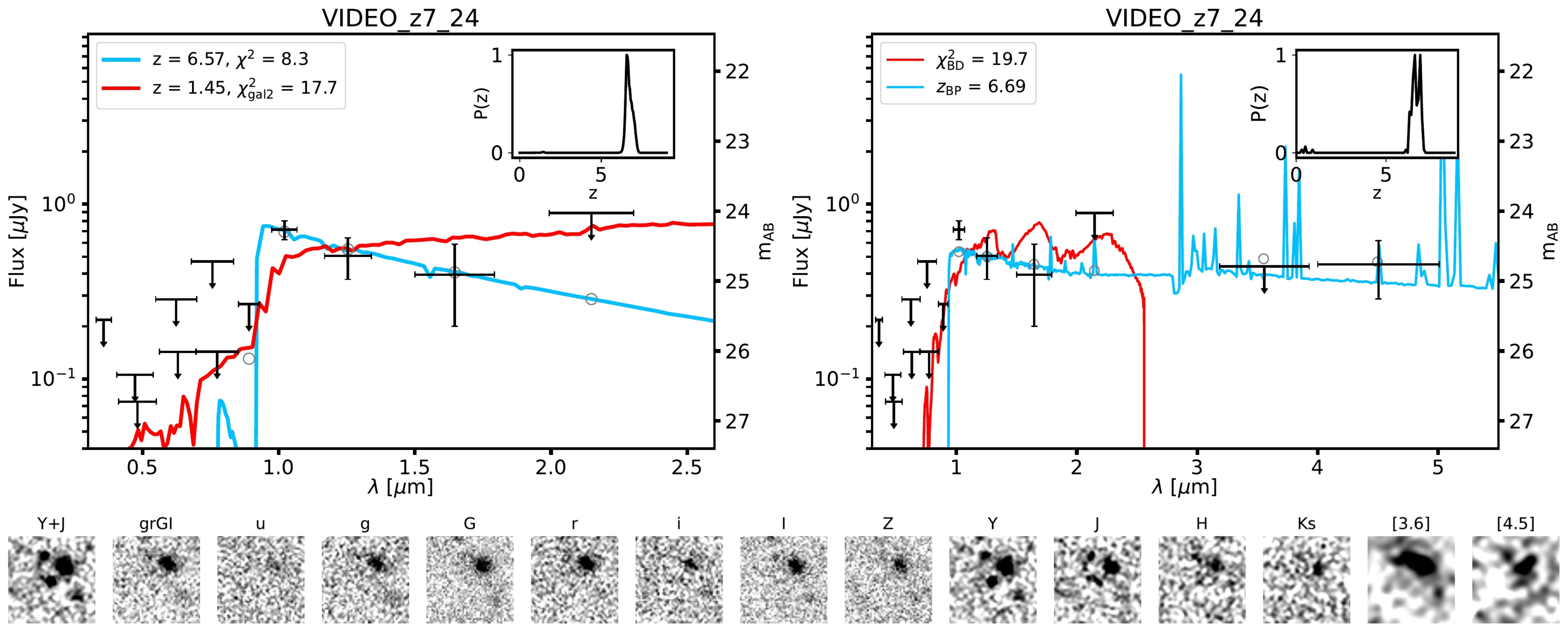}

        \caption{Two example candidate galaxies in XMM-LSS (top object, VIDEO\_z7\_22) and ECDF-S (bottom object, VIDEO\_z7\_24). For each object, the black points in both plots are the measured photometry, with non-detections replaced by $2\sigma$ upper limits. \textbf{Top Left:} SED fitting with \textsc{LePhare}. The blue curve shows the best-fitting high-redshift solution, and grey circles are its model photometry. The red curve shows the low-redshift solution. The legend shows the redshift and $\chi^{2}$ value of each solution. The wavelength range of this plot is limited to $\lambda\simeq2.5\micron$ to focus on the optical and NIR photometry around the Lyman break. The inset black curve shows the redshift probability distribution. \textbf{Top Right:} best-fitting BD template (red) along with the best-fit \textsc{BAGPIPES} galaxy fit. The chi-squared value for the brown dawrf fit, $\chi^{2}_{\mathrm{BD}}$, is computed without IRAC and without bands bluewards of $i/I$. The grey circles show the model photometry of the \textsc{BAGPIPES} solution.  The legend shows $\chi^{2}_{\mathrm{BD}}$ and the redshift of the BAGPIPES solution, $z_{\mathrm{BP}}$. The wavelength range of this plot extends out to $\lambda>5\micron$ to show the \textit{Spitzer}/IRAC photometry and the nebular emission component of the BAGPIPES fit. The inset black curve is the redshift probability found by \textsc{BAGPIPES}. \textbf{Bottom:} 10 arcsec $\times$ 10 arcsec postage stamps of the object in the filters used.}
        \label{fig:candidates}
\end{figure*}

\subsection{Comparison to previous work}
\label{sec:previous work}

The Reionisation Era Bright Emission Line Survey \citep[REBELS, ][]{rebels} is a search for bright $z>6.5$ galaxies in XMM-LSS and COSMOS, with the aim of following up these galaxies with the Atacama Large Millimeter/submillimeter Array (ALMA). 
Within XMM-LSS they make use of VIDEO \citep{Jarvis2013}, the UKIDSS Ultra Deep Survey \citep[UDS,][]{uds}, and optical data from HSC-SSP DR2 and the Canada-France Hawaii Telescope Legacy Survey (CFHTLS) D1 field \citep{cfhtls}. 
Of the 40 REBELS objects, 15 are in XMM-LSS and we recover three out of four that are bright enough to meet our $ 5\sigma \ Y+J$ cut, namely REBELS-01 (VIDEO\_z7\_22), REBELS-02 (VIDEO\_z7\_6) and REBELS-14 (VIDEO\_z7\_21). REBELS-10 is not recovered due to blending with a foreground object.
VIDEO\_z7\_21 and VIDEO\_z7\_22 are spectroscopically confirmed at $z=7.084$ and $z=7.177$ respectively, in agreement with our photometric redshifts. 
The difference in objects selected is due to deeper UDS data used by REBELS. By making use of the UDS data they are able to select fainter candidates, and their selection also identifies candidates at $z>7.5$.

\begin{figure*}
    \centering
    \includegraphics[width=\textwidth]{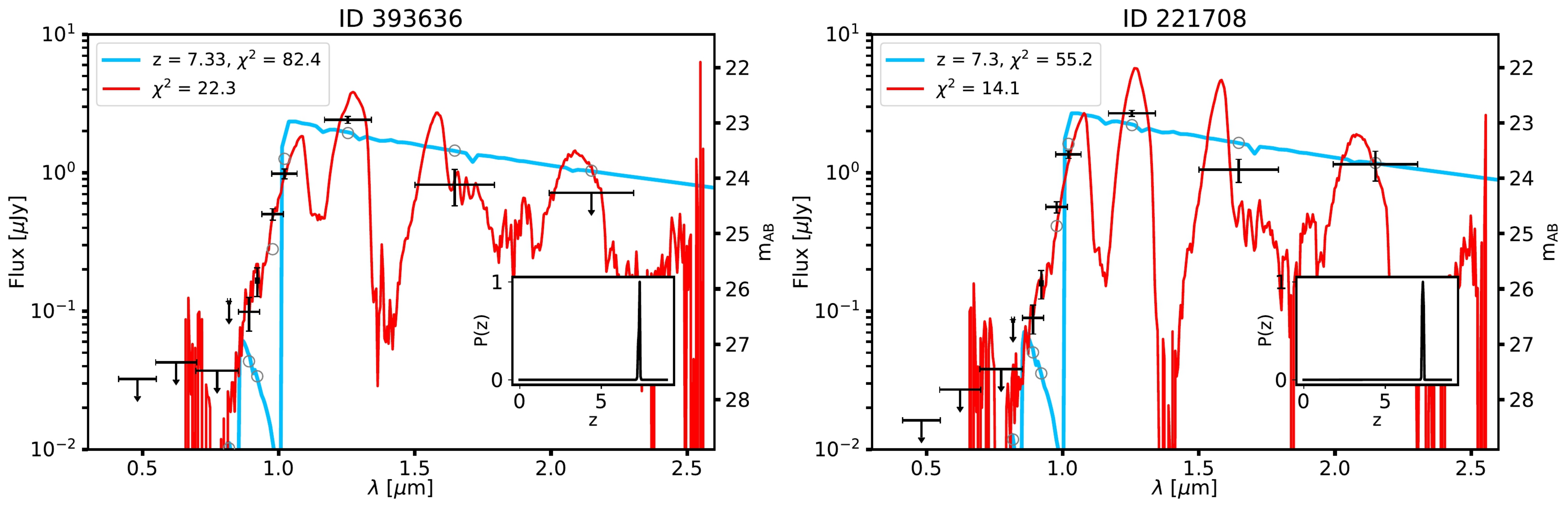}
    \includegraphics[width=\textwidth]{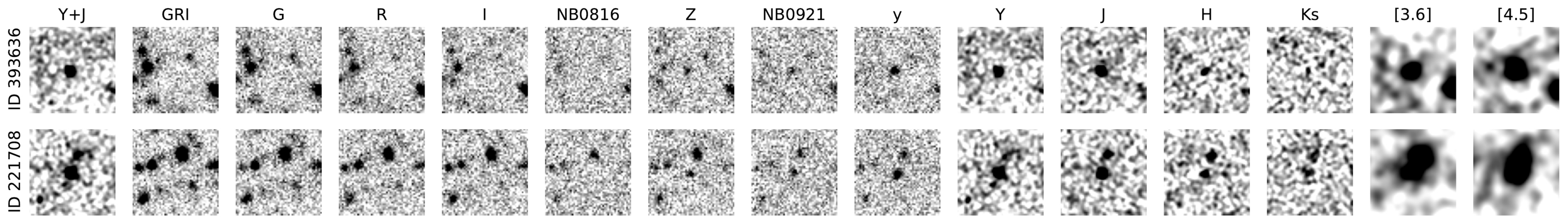}
    \caption{\textbf{Top:} SED fitting of the two brightest GOLDRUSH objects in XMM-LSS that overlap with our selection, ID 393636 (left) and ID 221708 (right). The black points are the measured photometry, with non-detections replaced by $2\sigma$ upper limits. The dwarf star SED is shown in red, and the high-redshift solution in blue. The grey circles are the model photometry for the high-redshift solution. Inset is the redshift probability distribution for the galaxy fit. The legend shows the redshift of the galaxy template and $\chi^{2}$ values. Both objects have preferred brown-dwarf fits and poor galaxy fits: the rising slope in the blue end of the brown dwarf SED and the drop in flux at $1.6\micron$ are captured by the optical and $YJHK_{s}$ bands respectively. \textbf{Bottom:} 10 arcsec $\times$ 10 arcsec postage stamps each object. From left to right we have: the detection filter $Y+J$, the optical stack used to check for low-level flux in each field, and the filters available for objects in each field ordered by increasing wavelength, as in Fig. \ref{fig:candidates}.}
    \label{fig:goldrush_dwarfs}
\end{figure*}

Great Optically Luminous Dropout Research Using HSC \citep[GOLDRUSH, ][]{harikane22}{}{} uses HSC-SSP DR2 data ($GRIZy$) to search for LBGs at $z=4-7$. 
This survey covers $20.2 \ \mathrm{deg}^{2}$ to depths comparable to the `deep' and `ultradeep' regions of XMM-LSS outlined in Section \ref{sec:XMM}. 
There are eight objects that overlap with our XMM-LSS data, all in the `ultradeep' HSC pointing.
Three out of eight of these objects are recovered in our $5\sigma \ Y+J$ cut.
%02h18m11.3856s -05d29m56.5539s
%02h17m00.9493s -04d40m35.1399s
%02h19m40.776s -05d13m53.7585s
We find that the two brightest objects recovered at $5\sigma$ in $Y+J$ have preferred brown dwarf fits, with poor galaxy fits. 
The SED fits for these objects are shown in Fig. \ref{fig:goldrush_dwarfs}. These objects have coordinates (RA 02:18:11.39 DEC -05:29:56.55) and (RA  02:17:00.95 DEC -04:40:35.14).
The third object (RA 02:19:40.78 DEC -05:13:53.76)\footnote{These have IDs 37484833881991283, 37485130234754131 and 37484567594038420 respectively in the \citet{harikane22} catalogue.} recovered at $5\sigma$ in $Y+J$ appears to have low-level optical flux, best seen in a $GRI$ stack, due to confusion with a nearby foreground object. 
Our SED fitting is also used on the remaining five objects that do not satisfy the $5\sigma \ Y+J$ cut. 
Of these, two objects have $z_{\mathrm{phot}} < 6.5$, and three objects have $z_{\mathrm{phot}} > 6.5$.
This appears to give a fiducial contamination rate of $\sim50$ per cent, but because we find brown dwarf fits for their brightest objects, it may be the case that their brightest bins are more significantly contaminated.
We show the photometric redshifts of the $z\ge6$ solutions of SED fitting and resulting absolute UV magnitudes $M_{\mathrm{UV}}$ for these objects in Fig. \ref{fig:Muv vs z}. The high-redshift solution for the two brown dwarfs places them at $z\simeq7.4$ with extremely bright absolute UV magnitudes of $M_{\mathrm{UV}} \simeq -24$. The remainder of the objects are fainter, similar to the bulk of our sample.
Contamination of the \citet{harikane22} sample, and the impact on the derived bright end of the UV LF, is discussed further in Section \ref{sec:contam_discussion}.

\cite{endsley2021} select galaxies at $z=6.63-6.83$ using overlapping data from HSC-SSP DR2, VIDEO and the UDS. 
They search in XMM1 and COSMOS, finding 9 objects in XMM1. XMM1-313310 in their study corresponds to VIDEO\_z7\_10 in this work.
We find $z_{\mathrm{phot}} = 6.72$, agreeing with their redshift range. 
XMM1-418672 in their study exists in our initial catalogue, but has $\chi^{2}=15.3$ so is rejected since it fails the $\chi^{2} < 11.3$ cut in XMM-LSS. 
The other 7 objects were selected by \citet{endsley2021} in the deeper UDS data, so are not recovered in this work.

\subsection{Physical properties with \textsc{BAGPIPES}}
\label{sec: bagpipes}

Table \ref{XMM_final} shows the results of our \textsc{BAGPIPES} fitting. 
All photometric redshifts measured by \textsc{BAGPIPES} agree with those measured by \textsc{LePhare} apart from objects VIDEO\_z7\_25 and VIDEO\_z7\_26 in ECDF-S, which agree within $2\sigma$, due to degeneracies in best-fitting redshift because of the gap between the $Z$ and $Y$ bands. 
IRAC detections probe the rest-frame optical, providing improved estimates of stellar masses. 
Measured stellar masses of the galaxies with unconfused \textit{Spitzer}/IRAC photometry are in the range $9.1 \le \mathrm{log}_{10}(M_{\star}/M_{\sun}) \le 10.9$, with a mean stellar mass of $\mathrm{log}_{10}(M_{\star}/M_{\sun}) = 10.0$. Our stellar mass for VIDEO\_z7\_22 agrees with that found by \citet{rebels}, but our stellar mass for VIDEO\_z7\_21 is an order of magnitude higher than theirs. This is because \textit{Spitzer}/IRAC bands for this object are confused, leading to boosted flux due to the foreground object to the north. \citet{rebels} apply neighbour subtraction to their IRAC photometry, which we do not do. Consequently, we mark objects that are confused in their \textit{Spitzer}/IRAC photometry with an asterisk in Table \ref{XMM_final}, indicating that these objects may have overestimated stellar masses. We also show confused IRAC photometry in green and with dashed error bars in the SED fitting plots for all candidates in Appendix \ref{sec:merged}. 
Nearly half (10/23) of our galaxies with unconfused IRAC photometry have stellar masses $\mathrm{log}_{10}(M_{\star}/M_{\sun}) > 10$, suggesting that these galaxies represent some of the most massive galaxies at this epoch \citep[compared to e.g.][]{Bowler14, rebels, labbe22}. The lower end of our derived stellar masses are more uncertain because the rate of \textit{Spitzer}/IRAC non-detections increases for $\mathrm{log}_{10}(M_{\star}/M_{\sun}) \lesssim 9.4$, meaning we cannot directly measure the rest-frame optical emission. Our masses are therefore upper limits.

Further issues arise in estimating stellar masses at these redshifts since it is difficult to tell if there is nebular emission (primarily \textsc{[Oiii]} $+ \ \mathrm{H}\beta$) contaminating the IRAC photometry, outshining the stellar continuum and masking contributions from the older stellar population. Since we have fixed metallicity, this results in a degeneracy between age and dust attenuation.  \citet{whitler23} have shown that properties derived from SED fitting depend on the SFH used. The dependence is strongest for derived age and weakest for stellar mass.  This can be somewhat remedied by restricting the redshift range and using narrowband and intermediate band photometry to compute more precise photometric redshifts \citep{endsley2021}, limiting nebular emission to only one of the two \emph{Spitzer}/IRAC bands and leaving the other free from contamination. Despite this contamination of the IRAC bands, we find that \textsc{BAGPIPES} provides reasonable estimates of the stellar masses in comparison to those derived by excluding contaminated bands entirely \citep[e.g.][]{Bowler_2018}{}{}.

\section{The rest-frame UV Luminosity Function}
\label{sec:uvlf}

We use the LBG candidates found in our selection to determine the UV LF at $z\simeq7$. The $1/V_{\mathrm{max}}$ method is used to compute points on the LF \citep{schmidt}, where $V_{\mathrm{max}}$ is the maximum comoving volume a galaxy can occupy and still be included in our survey. The maximum redshift $z_{\mathrm{max}}$ is determined by redshifting the galaxy in steps of $\Delta z = 0.01$ until it no longer satisfies the $5\sigma$ ($8\sigma$) detection threshold in $Y+J$ in XMM-LSS (ECDF-S). $V_{\mathrm{max}}$ is then the comoving volume between $z=6.5$ and $z=z_{\mathrm{max}}$. An upper limit of $z=7.5$, the maximum considered redshift, is placed on $z_{\mathrm{max}}$. The UV LF is then calculated as

 \begin{equation}
 \label{lf_eqn}
     \Phi(M)d\mathrm{log}(M) = \frac{1}{\Delta M}\sum\limits^{N}_{i}\frac{1}{C(M_{i}, z_{i})}\frac{1}{V_{\mathrm{max},i}} \ ,
 \end{equation}
 where $\Delta M$ is the width of the magnitude bin and $C(M_{i}, z_{i})$ is a completeness correction, which depends on the magnitude and redshift of the object. Poissonian errors are given by

\begin{equation}
\label{lf_error}
     \delta\Phi(M) = \frac{1}{\Delta M}\sqrt{\sum\limits^{N}_{i}\left(\frac{1}{V_{\mathrm{max},i}}\right)^{2}} \ 
 \end{equation}
following \citet{adams22}. Bin widths $\Delta M_{\mathrm{UV}}=0.75$ are chosen for all bins to maximise the number of galaxies in each bin. The bins are centred at $M_{\mathrm{UV}}=-22.175, -22.925$ and $-23.675$. These bins contain 21, 4 and 1 galaxies respectively. The brightest bin only contains VIDEO\_z7\_28, so the binning is also chosen to ensure this object is roughly in the centre of the bin. 
The faintest bin was chosen to ensure we are significantly above where our sample becomes incomplete, as discussed in Section \ref{sec:completeness}.
We compute an upper limit on the number density of galaxies at $M_{\mathrm{UV}} = -24.425$ with bin width $\Delta M = 0.75$. This is calculated by $1/V/\Delta M$, where $V$ is the volume probed by our survey at $6.5 \le z \le 7.5$.

\subsection{Completeness correction}
\label{sec:completeness}

The usable area after accounting for foreground objects and artefacts is 83\% in both fields. This factor is incorporated into the calculation of $V_{\mathrm{max}}$.
 
The focus of this work is the bright end of the rest-frame UV LF. We expect that the brightest bins are close to complete (excluding area lost due to foreground objects) in a $5\sigma \ Y+J$ selection since they are much brighter than the detection threshold ($>8\sigma$ at $M_{\mathrm{UV}} \lesssim -22.4$). We estimate the completeness of our initial selection (see Section \ref{sec:initial selection}) by comparing the number of objects per square degree as a function of apparent magnitude in $Y+J$ in XMM-LSS and ECDF-S with the `ultradeep' stripes from the UltraVISTA Survey \citep[][]{ultravista} in COSMOS, reaching $5\sigma$ depths of 26.3 in $Y+J$ in the ultradeep stripes, $\sim1$ mag deeper than VIDEO. We find, as expected, that our $Y+J$ selection is 95 per cent complete by $M_{\mathrm{UV}} = -21.8$, so we choose our faintest bin to cut off at this absolute UV magnitude. This excludes two candidates from the LF calculation, VIDEO\_z7\_15 and VIDEO\_z7\_16.

SED fitting is in general more complete than a colour-colour selection \citep[e.g.][]{adams20}, motivating its use in this work. 
 To ensure our SED fitting selection is not overly conservative, we test the completeness of our SED fitting by estimating the fraction of genuine high-redshift galaxies incorrectly cut by our selection as a function of absolute UV magnitude and redshift.
The values of $M_{\mathrm{UV}}$ used reaches one magnitude below the $5\sigma$ limit of each of our fields to account for up-scattering of objects into the sample. We use \textsc{BAGPIPES} to generate $z=6-8$ SEDs with ages varying between 50 Myr to 700 Myr (or the age of the Universe at a given redshift). 
We vary the dust attenuation between 0 - 0.5 mag and fix the metallicity at $0.2Z_{\sun}$. 
We generate 26,000 galaxies for ECDF-S and each of XMM-LSS `deep' and `ultradeep'.
Mock photometry in the available bands is generated by perturbing the model photometry using local depths within each field. SED fitting is performed on a catalogue of mock objects, where we apply the $\chi^{2}$ cuts outlined in Sections \ref{sed_preamble} and \ref{sec:contaminants}. The completeness is then the fraction of galaxies recovered in each magnitude-redshift bin. We report the median completeness values for each LF bin in Table \ref{tab:LFpoints}.

\subsection{Cosmic variance}
\label{sec:cosmic_variance}

Galaxy surveys are a discrete sampling of the large scale structure of the Universe.
Substructures such as voids and filaments can thus impact measurements of the LF if the dimensions of the survey are small.
This effect has come to be known as `Cosmic Variance'.
We make use of the \citet{cosmic_variance} calculator to estimate additional uncertainties due to this.
We find that the Poisson errors dominate: cosmic variance contributes just 7.1 per cent to the brightest bin at $M_{\mathrm{UV}} \simeq -23.7$, going down to a 5.7 per cent contribution to our faintest bin at $M_{\mathrm{UV}} \simeq -22.2$. We include these contributions in our LF errors by adding them in quadrature.

\begin{table}
    \centering
    \caption{The rest-UV luminosity function points at $z\simeq7$ derived from the sample presented in this work. The first two columns show the absolute UV magnitude of the bin and the number of galaxies in the bin. All bins have a width of 0.75 mag. The third column shows the comoving number density of galaxies. The final column shows the median completeness in the bin.}
    \begin{tabular}{lccc}

        \hline
         Bin / mag & $n_{\mathrm{gal}}$ & $\phi \ \mathrm{/ \ mag \ / \ Mpc^{3}}$ & Completeness \\[1ex]
         \hline
         -22.175 & 21 &  $2.70\pm0.66\times10^{-6}$ & 0.63 \\[1ex]
         -22.925 & 4 &   $2.81\pm1.54\times10^{-7}$ & 0.76 \\[1ex]
         -23.675 & 1 &   $2.37\pm2.50\times10^{-8}$ & 0.79 \\[1ex]
         -24.425  & 0 &   $<2.07\times10^{-8}$ & - \\[1ex]
         \hline
    \end{tabular}
    
    \label{tab:LFpoints}
\end{table}

\subsection{Results}
\label{sec:results}

\begin{figure}
    \centering
    \includegraphics[width=0.97\columnwidth]{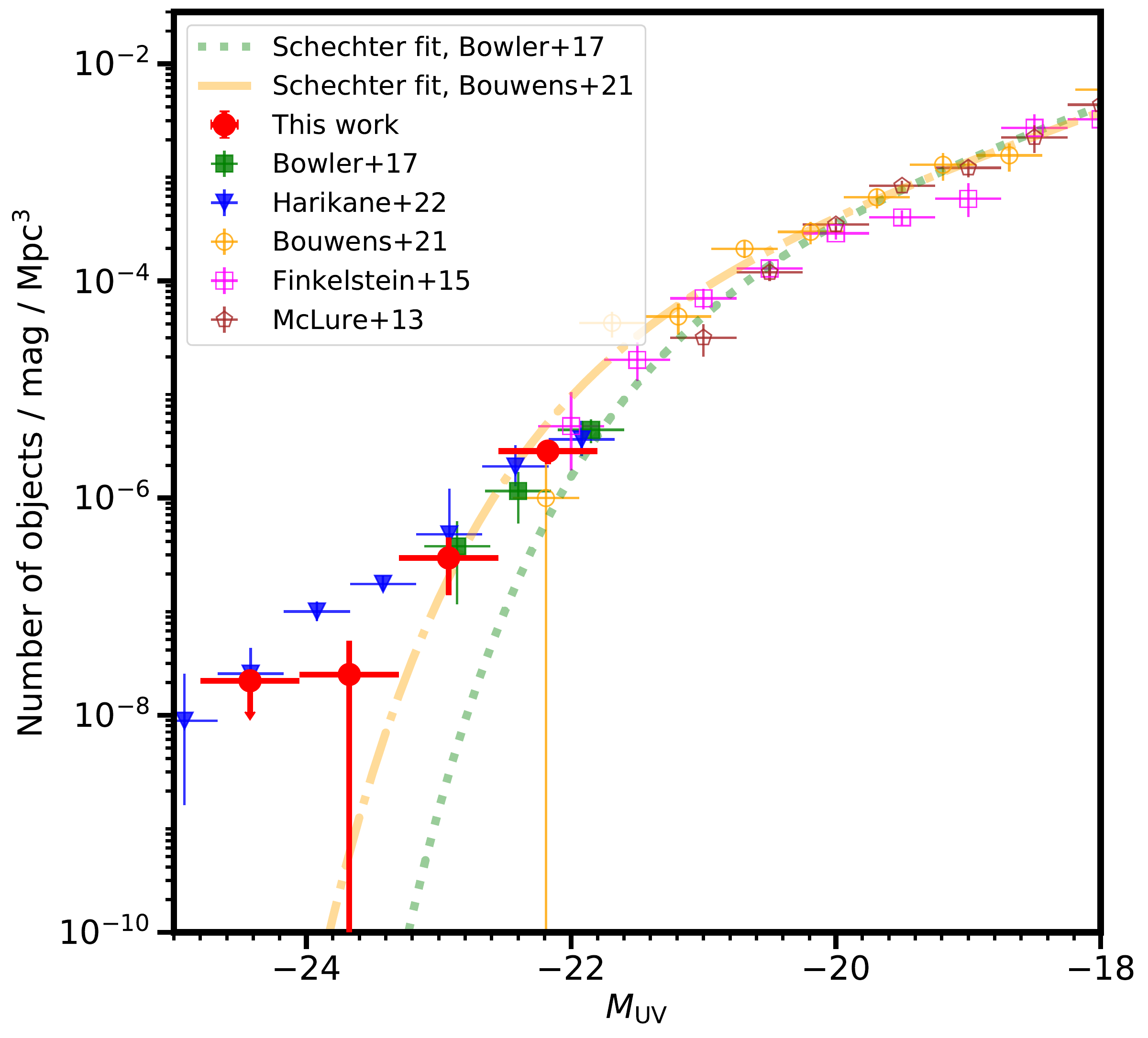}
    \includegraphics[width=0.97\columnwidth]{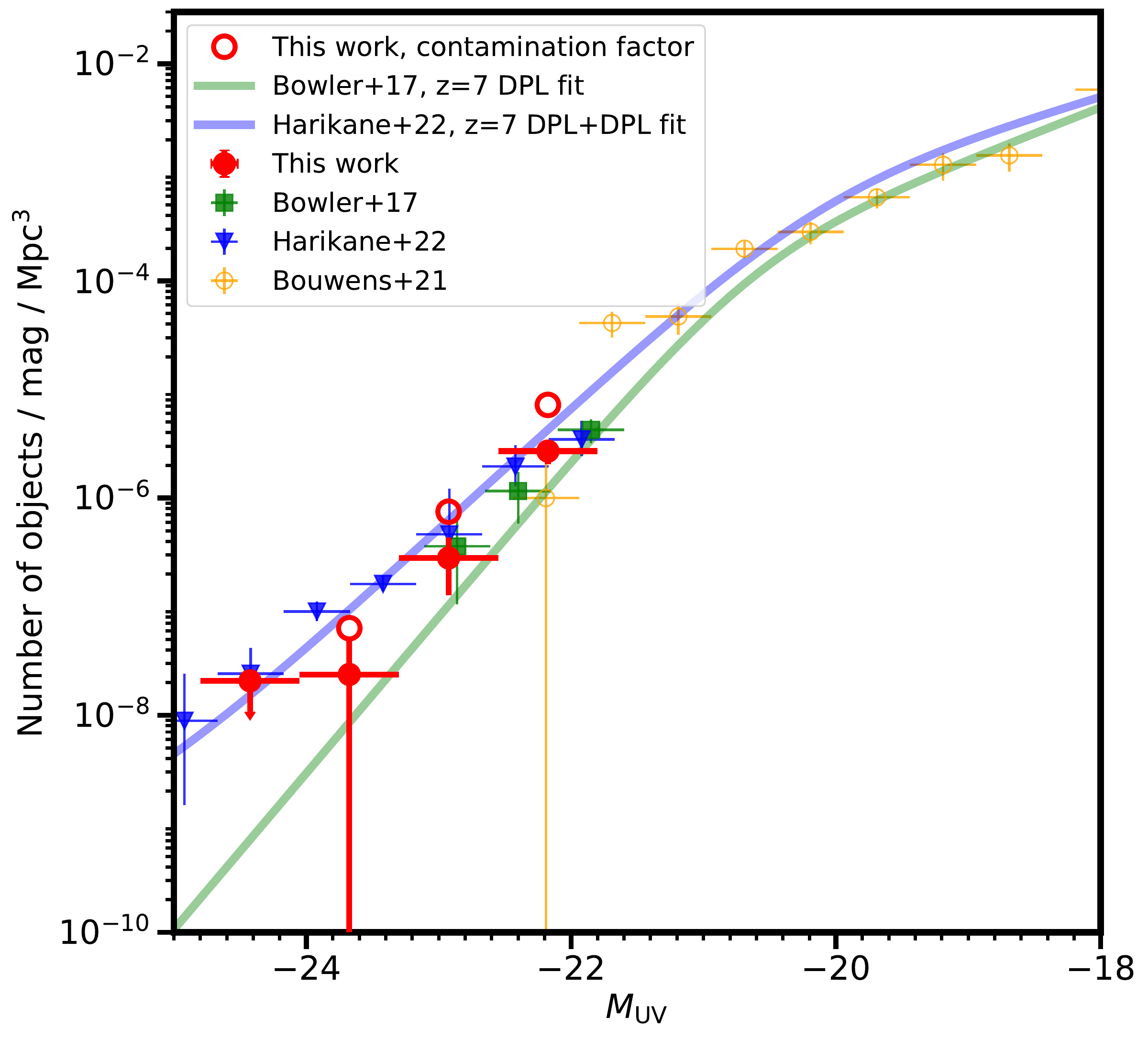}
    \caption{The rest-frame UV luminosity function at $z\simeq7$ derived from 8.2 $\mathrm{deg}^{2}$ of VIDEO data. The red filled circles in both panels show measurements of the number density from this work. We compute an upper limit on the number density of galaxies at $M_{\mathrm{UV}} = -24.425$ as described in Section \ref{sec:uvlf}. Other measurements from \citet{harikane22}, \ \citet{bowler17}, \ \citet{mclure2013}, \ \citet{finkelstein15} and \citet{bouwens21} are shown. The green dotted line and dot-dashed orange line show the best fitting Schechter functions from \citet{bowler17} and \citet{bouwens21} respectively.  The bottom panel shows the DPL fit to the rest-UV LF at $z\simeq7$ derived by \citet{bowler17}, and the DPL+DPL fit derived by \citet{harikane22}, shown as the green and blue lines respectively. The red open circles show our LF points incorporating a contamination factor of 50 per cent, as derived by SED fitting of objects from \citet{harikane22} that overlapped with our catalogues.}
    \label{fig:LF}
\end{figure}

In this section we present our binned rest-frame UV LF measurement at $z\simeq7$ and compare to measurements and best-fitting functions from other studies. In Fig. \ref{fig:LF} we show the measurement of the rest-frame UV LF at $z\simeq7$ from this work. The values of each bin are reported in Table \ref{tab:LFpoints}. 

\subsubsection{Comparison of binned LF points}

We compare our measurement of the $z\simeq7$ UV LF to current wide-area studies from \citet{harikane22} and \citet{bowler17}, and add studies of the faint end using \textit{Hubble} from \citet{mclure2013}, \citet{finkelstein15} and \citet{bouwens21}. 
\citet{bowler17} do not probe as bright as this work because they were limited to a smaller survey area of $1.65 \ \mathrm{deg}^{2}$. 
\citet{harikane22} use $20.2 \ \mathrm{deg}^{2}$ of survey data comparable to the depths of optical bands in this work, allowing them to provide the first constraints on the $z\simeq7$ UV LF beyond $M_{\mathrm{UV}} \lesssim -23$.

%%%%%%%%%%% RESULTS: COMPARING DATA POINTS %%%%%%%%%%%%
Our faintest point at $M_{\mathrm{UV}} \simeq -22.2$ is in agreement with those found by \citet{finkelstein15}, \citet{bowler17}, \citet{bouwens21} and \citet{harikane22}. 
Galaxies at $M_{\mathrm{UV}} \simeq -22$ represent the brightest that can be found by the widest-area \emph{Hubble} surveys, and close to the faintest that can be found from VIDEO. 
The brightest candidate in the current widest area space-based survey \citep{bouwens21} has an absolute UV magnitude $M_{\mathrm{UV}} \simeq -22.1$. 
Low number counts at this magnitude result in large error bars for their brightest bright bin, shown in yellow in Fig. \ref{fig:LF}.
Our point at $M_{\mathrm{UV}} \simeq -22.9$ is in agreement with \citet{bowler17} and \citet{harikane22}, although errors between these three studies span over an order of magnitude. The \citet{bowler17} point lies roughly in the middle of our point and that of \citet{harikane22}. 
Our brightest point in the bin at $M_{\mathrm{UV}} \simeq -23.7$ is in tension with \citet{harikane22} at the $2\sigma$ level, being four times lower than their determination at $M_{\mathrm{UV}} \simeq -23.9$. In Fig. \ref{fig:Muv vs z} we showed that the high-redshift solution of SED fitting for the two likely BDs in their sample (see Fig. \ref{fig:goldrush_dwarfs} and Section \ref{sec:previous work}) puts them at very bright absolute UV magnitudes of $M_{\mathrm{UV}} \sim -24$, explaining this tension. 
We note that our brightest bin contains only one galaxy, VIDEO\_z7\_28.
This source could be a BD, as it is particularly luminous and at a similar redshift to the \citet{harikane22} likely BDs (Fig. \ref{fig:Muv vs z}). If VIDEO\_z7\_28 is indeed shown to be a BD with later follow-up, this would change this point into an upper limit of $2.07 \times 10^{-8}$ objects / mag / $\mathrm{Mpc}^{3}$, leaving our conclusions unchanged.
Finally, we compute an upper limit for the UV LF in a bin at $M_{\mathrm{UV}} \simeq-24.4$ with width $\Delta M = 0.75$, shown in Fig. \ref{fig:LF}. The LF point derived by \citet{harikane22} at $M_{\mathrm{UV}} = -24.42$ lies on our upper limit, within the errors.

In summary, our determination of the UV LF agrees well with previous studies at $M_{\mathrm{UV}} \sim -22$. At brighter magnitudes, however, we find a deficit of sources in comparison to \citet{harikane22}. An analysis of the two brightest objects in their study suggests they are likely to be BDs. We therefore attribute this difference to contamination and emphasise the importance of NIR photometry in the selection of $z\simeq7$ LBGs.

\subsubsection{Schechter vs double power law fitting}

%%%%%%%%%% RESULTS: COMPARING SCHECHTER FUNCTIONS %%%%%%%%%%%%%
In the top panel of Fig. \ref{fig:LF} we show the best-fitting Schechter function fits as derived by \citet{bouwens21} and \citet{bowler17}. Our results appear to agree with the bright-end of the \citet{bouwens21} Schechter fit, although this is due to large errors on our brightest bin. The Schechter function fit derived by \citet{bowler17} uses results from \citet{mclure2013}. 
Similarly to \citet{bowler17} and \citet{harikane22}, we find that the Schechter function does not fully reproduce the number density of bright galaxies, which our results show continues beyond $M_{\mathrm{UV}} \simeq -23$.

%%%%%%%%%% RESULTS: COMPARING DPL FUNCTIONS %%%%%%%%%%%%%
In the bottom panel of Fig. \ref{fig:LF} we show the DPL derived by \citet{bowler17} and the DPL+DPL (galaxy and AGN component) derived by \citet{harikane22}. 
These functions provide a qualitative better fit to the UV LF measurements from our study.
Furthermore, the reduced $\chi^{2}$ value we compute for the DPL of \citet{bowler17} is $\chi^{2}_{\mathrm{red}}=02.3$ which is preferred to their Schechter fit ($\chi^{2}_{\mathrm{red}}=4.7$), as well as the Schechter fit of \citet{bouwens21} ($\chi^{2}_{\mathrm{red}}=4.3$), providing further evidence for a DPL as the form of the rest-UV LF at $z\simeq7$ continuing out to $M_{\mathrm{UV}} \simeq -24$. The DPL+DPL of \citet{harikane22} is in excess of our results, with $\chi^{2}_{\mathrm{red}}=6.2$. Our faintest two bins are in agreement with both \citet{bowler17} and \citet{harikane22}. In particular, the faintest bin is in closer agreement with \citet{harikane22} and the middle bin lies closer to the \citet{bowler17} DPL. The differences are more pronounced at brighter magnitudes, where our results are in tension with the \citet{harikane22} fit at the $2\sigma$ level. The \citet{bowler17} DPL is a factor of 5 lower at $M_{\mathrm{UV}} \simeq -22.9$, and a factor of 10 lower at $M_{\mathrm{UV}} \simeq -23.7$ than \citet{harikane22}. Our upper limit at $M_{\mathrm{UV}} \simeq -24.4$ lies just above the \citet{harikane22} DPL+DPL. Assuming that AGN do not contribute to the bright end of rest-frame UV LF at $z\simeq7$ (see Section \ref{sec:agn_discussion}) such that the number density continues to decline rapidly at brighter magnitudes, we expect that a measurement of the LF at this magnitude would lie closer to the DPL of \citet{bowler17}, which is 20 times lower than predicted by \citet{harikane22}. 

As shown in Section \ref{sec:previous work} and Fig. \ref{fig:goldrush_dwarfs}, the brightest objects from \citet{harikane22} that lie in XMM-LSS are best fit by brown dwarf templates when $YJHK_{s}$ photometry is included. This suggests contamination in their brightest bins. The impact of this contamination on the LF is discussed further in \ref{sec:contam_discussion}. By utilising VIDEO photometry we have thus provided a robust measurement of the rest-frame UV LF at $z\simeq7$ out to $M_{\mathrm{UV}} \simeq -24$.

%%%%%%%%%%%%%%%%% COMPARING TO Z=8 %%%%%%%%%%%%%%%%%%%%%%%

\subsection{Comparison to \texorpdfstring{$\boldsymbol{z}\simeq\boldsymbol{8}$}{z=8} results}
\begin{figure}
    \centering
    \includegraphics[width=0.97\columnwidth]{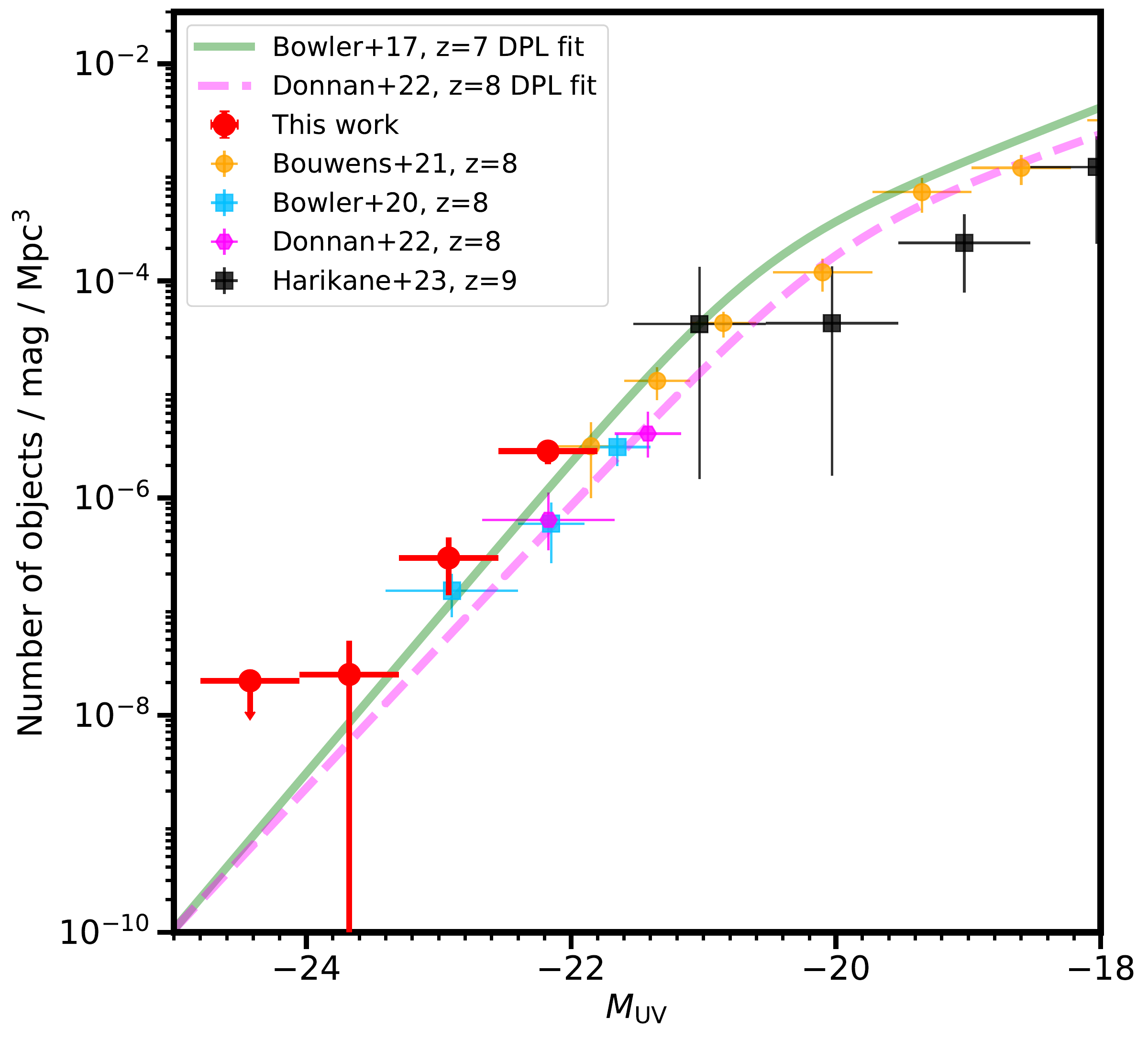}
    \includegraphics[width=0.97\columnwidth]{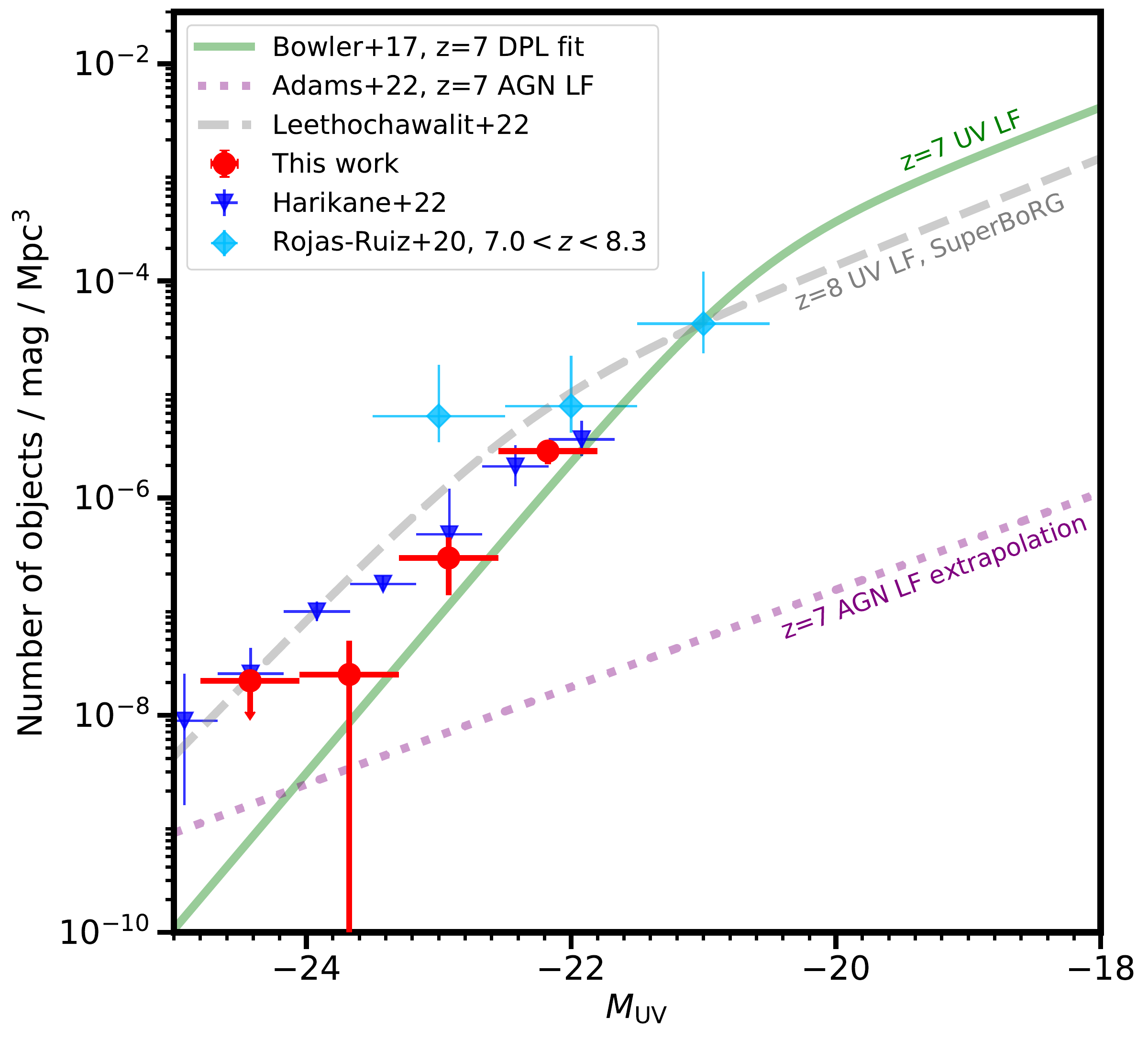}
    \caption{The top panel shows a comparison of our $z\simeq7$ rest-UV LF results (red circles) and the $z\simeq7$ DPL derived by \citet{bowler17} with various $z\simeq8$ results. The blue squares show the $z\simeq8$ rest-UV LF results from \citet{Bowler_2020}. The magenta hexagons show $z\simeq8$ rest-UV LF results from \citet{donnan22}. Both of these studies use deep ground-based optical and NIR imaging. The magenta dashed line is the DPL fit derived by \citet{donnan22}. The yellow points are $z\simeq8$ rest-UV LF results from \citet{bouwens21}, derived from \textit{Hubble} data. The bottom panel again shows our $z\simeq7$ rest-UV LF results (red circles) and $z\simeq7$ DPL derived by \citet{bowler17}. We include the $z\simeq7$ rest-UV LF points found by \citet{harikane22} as dark blue triangles. Here we compare to $z\simeq8$ rest-UV LF results derived from \emph{Hubble} pure-parallel programs, shown as light blue diamonds \citep{rojasruiz20} and the grey dashed DPL fit derived by \citet{leethochawalit}. We also plot the $z\simeq7$ extrapolation of the AGN LF found at $z=3-5$ by \citet{adams22} as the purple dotted line.}
    \label{fig:LF_z8}
\end{figure}

In this section compare our results to various measurements of the rest-frame UV LF at $z\simeq8$ to ascertain whether we find any evidence for evolution in the LF from $z\simeq7-8$. 
In the top panel Fig. \ref{fig:LF_z8} we plot results from other ground-based studies \citep[][]{Bowler_2020, donnan22} who both use deep optical and near-infrared imaging from VIDEO, UDS and UltraVISTA, in a similar manner to this work, to identify robust $z\simeq8$ galaxy candidates. 
We also plot space-based results from \citet{bouwens21} of the faint-end of the $z\simeq8$ LF. 
As expected, our results at $z\simeq7$ are in excess of these results at $z\simeq8$. 
We show the DPLs measured at $z\simeq7$ and $z\simeq8$ by \citet{bowler17} and \citet{donnan22} respectively. 
Little evolution is expected in the bright-end between these epochs \citep[][]{Bowler_2020, harikanez9}{}{}, and this is reflected in the small differences between the DPLs and the slight excess of our results to those at $z\simeq8$. 

In the bottom panel of Fig. \ref{fig:LF_z8} we show recent results from \textit{Hubble} pure-parallel studies. \citet{leethochawalit} measure the $z\simeq8$ LF in 0.41 $\mathrm{deg}^{2}$ of the Search for the Brightest of Reionizing Galaxies and Quasars in HST Parallel Imaging Data (SuperBoRG), and \citet{rojasruiz20} use similar data to derive comparable results. 
Both studies are in tension with this work. The DPL derived by \citet{leethochawalit} is a factor of 2.5-8 higher than our results, with the excess more pronounced at brighter magnitudes. At $M_{\mathrm{UV}} \simeq -24.4$, their DPL lies on the upper limit derived in this work. 
Similarly, the brightest LF point measured by \citet{rojasruiz20} is 20 times higher than our measurement at $M_{\mathrm{UV}} \simeq -23.7$. 

\section{Discussion}
\label{sec:discussion}

Our results agree with other ground-based deep optical+NIR studies at $z\simeq7-8$, but work using deep optical-only ground-based imaging and pure-parallel \emph{Hubble} imaging derive an excess compared to our findings by up to an order of magnitude. In this section we discuss the potential contributions of quasars and brown dwarf contaminants, and implications of our results on the evolution of the bright-end of the rest-frame UV LF. We also consider how upcoming space-based wide-area missions can aid in building a clearer picture of bright LBG abundances at high redshift.

\subsection{Contamination in the bright end of the LF}
\label{sec:contam_discussion}

In Section \ref{sec:results} and Fig. \ref{fig:LF} we showed that our results are in agreement with the steeper DPL found by \citet{bowler17} and are in tension with the flatter decline found by \citet{harikane22}. 
Without the $YJHK_{s}$ bands, most $z\sim7$ candidates found in HSC imaging appear as single band detections, or at most have a detection in the $Z$- and $y$-bands. Thus only a spectral break can be detected, leaving such samples more vulnerable to contamination.
Moreover, by including the HSC narrowbands our objects in XMM-LSS further gain the advantage of the NB921 and $Y$ filters that overlap with the $Z$ and $y$ filters respectively, allowing for better determination of the nature of the break of an object.
 We find this to be powerful in detecting the slope of brown dwarf SEDs.

As discussed in Section \ref{sec:previous work}, when the $YJHK_{s}$ data is included in the SED fitting analysis, we found that 4 out of 8 objects from the GOLDRUSH catalogue that overlaps with our data have poor $z\simeq7$ SED fits. 
The fainter two of these objects have best-fitting SEDs at $z<6.5$, while the brighter two have best-fitting brown dwarf templates, as shown in Fig. \ref{fig:goldrush_dwarfs}. 
This gives us a fiducial contamination rate of 50 per cent. 
We apply this contamination factor to our derived LF points, shown as open red circles in Fig. \ref{fig:LF}, to assess the impact of contamination on our robust sample. 
For our faintest and middle bins, these contaminated points lie slightly above and on the fit derived by \citet{harikane22}, respectively.
%For our two fainter bins, these contaminated points lie on the fit derived by \citet{harikane22}. 
Our upper limit also suggests that their DPL+DPL is overestimated at $M_{\mathrm{UV}} \simeq -24.4$. 
This implies that not using $YJHK_{s}$ results in an overestimation of the bright end of the rest-UV LF at $z\simeq7$. 
Note that our brightest contaminated LF point still lies slightly below the \citet{harikane22} value at $M_{\mathrm{UV}} = -23.92$. 
This suggests that the fiducial contamination rate of 50 per cent could be higher at brighter magnitudes, with brighter bins preferentially affected by brown dwarfs or low-redshift galaxies when only optical data is included. 
This is in tension with the contamination rate estimated by \citet{harikane22}, who claim this is negligible based on results from their spectroscopic catalogue. 
They are also unable to use SED fitting to remove $z\sim2$ interlopers at $z>6$ due to lack of bands redwards of the Lyman break.
This contamination likely causes the unphysical crossing of the UV LFs as derived in their study at $z\simeq6$ and $z\simeq7$ between $-25 \lesssim M_{\mathrm{UV}} \lesssim -23$ (see fig. 8 in \citealp{harikane22}).

\subsection{AGN contribution}
\label{sec:agn_discussion}

In Section \ref{sec:results} we showed that the results of pure-parallel \emph{Hubble} studies \citep{rojasruiz20, leethochawalit} at $z\simeq8$ are significantly in excess of our results at $z\simeq7$. 
\citet{leethochawalit} attribute this excess to a contribution by AGN.
The motivation for this comes from \cite{harikane22} who suggest that the faint-end of the AGN LF dominates at $M_{\mathrm{UV}} \le -24$ at $z=4-7$. 
\citet{adams22} measure the AGN LF in the same fields as this study at $z=3-5$, as well as in the COSMOS field, and find that the AGN LF evolves more rapidly towards higher redshift.
\citet{matsuoka18} find that this evolution continues to $z\simeq6$: their AGN LF crosses the $z\simeq7$ DPL of \citet{bowler17} at $M_{\mathrm{UV}}\simeq-23.6$.  
Thus, if this strong decline in number density of faint UV-selected AGN continues to $z\simeq7$ \citep[AGN are rare at this epoch,][]{mortlock11, banados18, wang21}{}{}, it is unlikely AGN contribute to the $z\simeq7$ galaxy LF at the brighter magnitudes probed by this work and the pure-parallel studies from \emph{HST}. 
Furthermore, on the bottom panel of Fig. \ref{fig:LF_z8} we show the extrapolation of the AGN LF to $z=7$ as estimated by \citet{adams22}. We extrapolate the $z=4.8$ AGN LF to our mean redshift assuming a conservative value of $k = - 0.82$, providing a likely upper limit on the number density of faint $z\simeq7$ AGN. 
Their extrapolation is $\simeq10$ times lower than the double power law computed by \citet{bowler17} at $M_{\mathrm{UV}}=-23$. \citet{adams22} conclude that the AGN number density at $z\simeq7$ is insufficient to contribute to the galaxy LF as found by \citet{bowler17}.
Instead, our results are consistent with little evolution in the bright-end between $z\simeq7-8$, as determined by studies also using deep ground-based optical and NIR imaging \citep[][]{Bowler_2020, donnan22}{}{}.
Contamination in the \cite{leethochawalit} sample could be responsible for the excess of objects found in their work: for two thirds of their sample, only a single filter (F350LP) is available bluewards of the break. 
This is the same filter used in \citet{rojasruiz20} to assert non-detections bluewards of the break, likely leading to imperfect removal of low-redshift galaxies and brown dwarfs.
The true overlap between the galaxy and AGN LFs at $z\simeq7$ therefore likely occurs somewhat brighter, at $M_{\mathrm{UV}}<-24$. 
We note that this discussion pertains to unobscured AGN - highly obscured AGN at $z\simeq7$ have been discovered in the COSMOS field \citep{endsleyAGN}.

\subsection{Astrophysical interpretation}

Robust measurements of the bright end of the rest-UV LF at $z\simeq7$ are vital for understanding the evolution of the most star-forming and arguably massive galaxies between this epoch and the present day. 
The growth of massive galaxies must somehow be suppressed to account for the change in shape of the UV LF over cosmic time, with a DPL observed at high-redshift to the Schechter form seen at low-redshift. 
One mechanism that may govern the shape of the bright end of the LF is quenching by AGN, where feedback is driven by energetic release from their accretion disc \citep[e.g.][]{dave19, lovell22}{}{}, removing gas reservoirs from galaxies and limiting star formation. 
The AGN number density increases rapidly between $z=6-5$, thus we know that AGN activity becomes more common \citep[e.g.][]{matsuoka18, Niida_2020, adams22}. However, our results suggest that the contribution from AGN at $z\simeq7$ to the rest-frame UV LF is minimal.
This indicates that quenching of star formation by AGN is unlikely to impact bright galaxies at $z\simeq7$.

Alternatively, the build-up of dust may lead to significant obscuration of UV light from galaxies, modifying the intrinsic rest-frame UV LF and
causing a steeper decline in the observed bright end of rest-frame UV LF. Star formation enriches the interstellar medium (ISM) with dust, thus preferentially impacting the most massive galaxies. 
For example, \citet{vijayan21} use a simple model linking dust attenuation with ISM metal content and find that dust attenuation becomes important for galaxies with $M_{\mathrm{UV}} \lesssim -21$ at $z\simeq7$. 
Our results suggest dust obscuration does not significantly impact the bright end of the rest-frame UV LF at this epoch.
Dust attenuation values are low on average (see Table \ref{XMM_final}) with no dependence on $M_{\mathrm{UV}}$. 
These results suggest that a lack of dust attenuation at this epoch could explain the observed DPL shape and the almost constant number density of bright $z \simeq 7$ and $z \simeq 8$ sources \citep[e.g. as shown in][]{ferrara22}{}{}.
However, we note that the selection of rest-frame UV-bright galaxies can be blind to dust-obscured star formation, and recent studies are beginning to reveal the importance of this process \citep[e.g.][]{bowler22, inami22, algera23} and constrain dust properties at $z\gtrsim6.5$ \citep[e.g][]{inami22, sommovigo22}.

\subsection{Outlook for upcoming surveys}

At $z\simeq7$ there is a need to bridge the gap between ground-based and space-based studies around the knee of the UV LF. A mismatch can be seen in Fig. \ref{fig:LF} between the brightest bins of \cite{bouwens21} and the faintest bins of \cite{bowler17} and this work. At the knee, the former suffers from low number counts, and the latter suffers from a lack of depth. \textit{Euclid} will be able to address this effectively: it will provide 50 $\mathrm{deg}^{2}$ of $YJH$ imaging down to $5\sigma$ depths of $m_{\mathrm{AB}}\simeq26.4$ in its deep fields \citep{vanmierlo22}, comparable to the depths of UltraVISTA (covering 1.5 $\mathrm{deg}^{2}$). It will also be instrumental in providing much more accurate measurements of the ultra-bright end of the $z\simeq7$ LF. It will provide 15,000 $\mathrm{deg}^{2}$ of $YJH$ imaging down to $m_{\mathrm{AB}}\simeq24.5$ \citep{scaramella22}, slightly deeper than VIDEO. Simply providing thousands more sources will vastly improve number statistics. 
Additionally, with much greater spatial resolution that comes with space-based observatories, blending will be less of an issue, and will allow a more accurate deconfusion of the \emph{Spitzer}/IRAC data.
The morphology of objects may even be used to distinguish between high-redshift galaxies (extended) and brown dwarfs or AGN (point-like). Since AGN are very rare at $z\simeq7$, brown dwarf contaminants that appear as point sources could be efficiently removed from LBG samples via a size or morphology cut. We may also be able to finally probe bright enough to see contributions by the $z\simeq7$ AGN LF, settling the debate about where the overlap between the AGN LF faint end and galaxy LF bright end begins. 

Although $JWST$ is finding many `bright' candidates at ultra-high redshifts \citep[e.g.][]{naidu22}, it may struggle to probe the very bright end ($M_{\mathrm{UV}} < -23$) of the UV LF during the Epoch of Reionization since the area it can survey is more limited than ground-based observations. In the top panel of Fig. \ref{fig:LF_z8} we plot the results of \citet{harikanez9} who measure the $z\simeq9$ UV LF in $\sim90$ square arcminutes of NIRCam imaging. The smaller area compared to this work means they are unable probe brighter than $M_{\mathrm{UV}} \simeq -22$. Wider area studies such as COSMOS-Web \citep{casey22} will be able to cover up to 0.5 $\mathrm{deg}^{2}$, but this is still significantly smaller than what can be achieved from ground-based observatories and with \textit{Euclid} and \textit{Roman}. It is clear that within the next decade, the combination of \emph{JWST} with wide-area space observatories will provide a revolution in the measurement of the $z>6$ LF over a broad magnitude range.

\section{Conclusions}
\label{sec:conclusion}

We have conducted a wide-area search for $z\simeq7$ Lyman break galaxies using deep near-infrared photometry from the VIDEO survey combined with deep optical data and \textit{Spitzer}/IRAC photometry over an area of $8.2$ $\mathrm{deg}^{2}$. Candidates were selected in a $Y+J$ stack reaching $5\sigma$ depths of $m_{\mathrm{AB}} = 25.3$ in XMM-LSS and $8\sigma$ depths of $m_{\mathrm{AB}} = 24.8$ in ECDF-S, with non-detections required in bands bluewards of the Lyman break. 
We show that the inclusion of NIR data from VISTA and \textit{Spitzer} enables a robust removal of low-redshift galaxies and Galactic brown dwarf contaminants.

We found 28 galaxy candidates with a mean redshift $\Bar{z} = 6.74$ and UV absolute magnitudes in the range $-23.5 \le M_{\mathrm{UV}} \le -21.6$. We recovered two spectroscopically confirmed sources from the REBELS Survey \citep[][]{rebels}. 
The \textsc{BAGPIPES} SED fitting code was used to confirm the photometric redshifts and derive galaxy properties.  We derived stellar masses of $9.1 \le \mathrm{log}_{10}(M/M_{\sun}) \le 10.9$ for objects with unconfused \textit{Spitzer}/IRAC photometry, suggesting that these galaxies are some of the most massive at this epoch. 

We measure the UV luminosity function at $z\simeq7$ using our candidates and compare to the current widest area searches: \citet{harikane22} who use optical-only data in an area of $20.2$ $\mathrm{deg}^{2}$, and \citet{bowler17} who use optical and near-infrared data in an area of $1.65$ $\mathrm{deg}^{2}$. 
Our results agree with the DPL fit of \citet{bowler17} to brighter magnitudes, however they lie significantly below the results of \citet{harikane22}. 
Through SED fitting of the \citet{harikane22} candidates, utilizing the VIDEO $YJHK_{s}$ photometry and \emph{Spitzer}/IRAC, we find that the brightest galaxy candidates identified by \citet{harikane22} are likely to be brown dwarfs.
Our results suggest that using optical data alone to select $z\simeq7$ galaxies leads to an overestimate by a factor of 2 at $M_{\mathrm{UV}} \simeq -23$ and by a factor of $10$ at $M_{\mathrm{UV}} < -24$ due to contamination.  

Extrapolating findings from \citet{adams22} predicts a negligible contribution of unobscured AGN at the magnitudes we probe. This is in tension with the conclusions of recent pure-parallel \emph{Hubble} results at $z = 8$ \citep[][]{rojasruiz20, leethochawalit}{}{}, which we find to be in excess of our $z = 7$ LF by a factor of $>10$. Our results provide a robust measure of the bright end of the $z\simeq7$ UV LF, which does not evolve significantly from $z\simeq8$. This suggests a lack of dust attenuation and/or mass quenching between these epochs. 
Upcoming wide-area space missions such as \emph{Euclid} and \emph{Roman} will provide much larger samples extracted from thousands of $\mathrm{deg}^{2}$ of NIR imaging to depths comparable to VIDEO, allowing for determinations of the UV LF beyond $M_{\mathrm{UV}} < -24$ at $z>6$, whilst also providing superior resolution to ground-based imaging, permitting the removal point-like brown dwarf contaminants.

\section*{Acknowledgements}

RGV acknowledges funding from the Science and Technology Facilities Council (STFC) [grant code ST/W507726/1]. 
RAAB acknowledges support from an STFC Ernest Rutherford Fellowship [grant number ST/T003596/1]. 
MJJ acknowledges support of the STFC consolidated grant [ST/S000488/1] and
[ST/W000903/1] and from a UKRI Frontiers Research Grant [EP/X026639/1]. MJJ also acknowledges support from
the Oxford Hintze Centre for Astrophysical Surveys which is funded
through generous support from the Hintze Family Charitable Foundation. 
NJA acknowledges support from the  European Research Council (ERC) Advanced Investigator Grant EPOCHS (788113).

This work is based on data products from observations made with ESO Telescopes at the La Silla Paranal Observatory under ESO programme ID 179.A-2006 and on data products produced by CALET and the Cambridge Astronomy Survey Unit on behalf of the VIDEO consortium.

The Hyper Suprime-Cam (HSC) collaboration includes the astronomical communities of Japan and Taiwan, and Princeton University. The HSC instrumentation and software were developed by the National Astronomical Observatory of Japan (NAOJ), the Kavli Institute for the Physics and Mathematics of the Universe (Kavli IPMU), the University of Tokyo, the High Energy Accelerator Research Organization (KEK), the Academia Sinica Institute for Astronomy and Astrophysics in Taiwan (ASIAA), and Princeton University. Funding was contributed by the FIRST program from the Japanese Cabinet Office, the Ministry of Education, Culture, Sports, Science and Technology (MEXT), the Japan Society for the Promotion of Science (JSPS), Japan Science and Technology Agency (JST), the Toray Science Foundation, NAOJ, Kavli IPMU, KEK, ASIAA, and Princeton University. This paper makes use of software developed for Vera C. Rubin Observatory. We thank the Rubin Observatory for making their code available as free software at http://pipelines.lsst.io/.

This paper is based on data collected at the Subaru Telescope and retrieved from the HSC data archive system, which is operated by the Subaru Telescope and Astronomy Data Center (ADC) at NAOJ. Data analysis was in part carried out with the cooperation of Center for Computational Astrophysics (CfCA), NAOJ. We are honored and grateful for the opportunity of observing the Universe from Maunakea, which has the cultural, historical and natural significance in Hawaii. 

%%%%%%%%%%%%%%%%%%%%%%%%%%%%%%%%%%%%%%%%%%%%%%%%%%
\section*{Data Availability}

All imaging data was obtained from original sources in the public domain. Catalogues of the photometry of samples used in this study can be provided on request.



\bibliographystyle{mnras}
\bibliography{example} % if your bibtex file is called example.bib

%\begin{thebibliography}{99}
%\end{thebibliography}

%%%%%%%%%%%%%%%%%%%%%%%%%%%%%%%%%%%%%%%%%%%%%%%%%%

%%%%%%%%%%%%%%%%% APPENDICES %%%%%%%%%%%%%%%%%%%%%

\appendix

\section{Adding inclusive candidates to the LF calculation}
\label{sec:inclusive calculation}

\begin{table*}
\caption{The photometry of our $z\simeq7$ inclusive candidates. The top section shows objects in XMM-LSS, and the bottom section shows objects in ECDF-S. Objects are ordered by their photometric redshift. The first column shows the object ID, and the next two columns show the coordinates of the candidate. The remaining bands show the photometry in the bands available in each field. We require $<2\sigma$ detections bluewards of and including the $i/I$-bands, so we only present bands redwards of this. The photometry is measured in a 2.0 arcsec diameter circular aperture apart from the \textit{Spitzer}/IRAC bands where 2.8 arcsec diameter apertures are used to account for the broader PSF. The photometry is corrected to a total flux assuming a point-source correction using a $10\times10$ grid of PSFs measured by \textsc{PSFEx} across each VIDEO tile.}
\label{tab:inc_phot}
\begin{tabular}{cccccccccccc}
\hline
ID & RA & DEC & Z & NB921 & y & Y & J & H & Ks & [3.6] & [4.5] \\[1ex]
\hline
170389 & 02:16:38.75 & -04:28:50.98 & >$26.7$ & >$26.4$ & >$25.8$ & $24.8^{+0.3}_{-0.2}$ & $24.6^{+0.3}_{-0.2}$ & $24.3^{+0.4}_{-0.3}$ & >$24.0$ & >$24.5$ & $24.8^{+0.6}_{-0.4}$ \\[1ex]
1415071 & 02:25:49.21 & -04:02:38.00 & >$26.2$ & $25.7^{+0.6}_{-0.4}$ & $24.8^{+0.5}_{-0.3}$ & $24.8^{+0.3}_{-0.2}$ & $24.6^{+0.3}_{-0.2}$ & $24.8^{+0.7}_{-0.4}$ & >$24.0$ & $23.7^{+0.2}_{-0.2}$ & $24.1^{+0.5}_{-0.3}$ \\[1ex]
1017553 & 02:22:52.86 & -05:04:37.97 & >$26.4$ & >$26.0$ & >$24.6$ & $24.7^{+0.3}_{-0.2}$ & $24.7^{+0.4}_{-0.3}$ & $24.5^{+0.4}_{-0.3}$ & >$23.9$ & $23.3^{+0.2}_{-0.1}$ & $24.1^{+0.5}_{-0.4}$ \\[1ex]
914913 & 02:22:07.26 & -05:24:02.32 & >$26.3$ & >$25.5$ & >$24.9$ & $25.1^{+0.2}_{-0.2}$ & $25.2^{+0.5}_{-0.3}$ & $25.1^{+0.7}_{-0.4}$ & $24.8^{+0.7}_{-0.4}$ & $24.0^{+0.2}_{-0.2}$ & $24.5^{+0.7}_{-0.4}$ \\[1ex]
\hline
1495402 & 03:35:25.99 & -28:38:19.13 & >$25.1$ & - & - & $23.9^{+0.1}_{-0.1}$ & $22.9^{+0.1}_{-0.1}$ & $22.6^{+0.1}_{-0.1}$ & $22.4^{+0.1}_{-0.1}$ & $22.5^{+0.2}_{-0.2}$ & $23.0^{+0.2}_{-0.2}$ \\[1ex]
705214 & 03:30:30.53 & -27:23:16.47 & >$25.4$ & - & - & $24.6^{+0.1}_{-0.1}$ & $24.6^{+0.3}_{-0.2}$ & $24.8^{+0.6}_{-0.4}$ & >$24.1$ & $24.1^{+0.2}_{-0.2}$ & >$24.7$ \\[1ex]
1090797 & 03:32:28.28 & -27:48:02.25 & >$25.5$ & - & - & $24.2^{+0.1}_{-0.1}$ & $23.8^{+0.2}_{-0.1}$ & $24.1^{+0.3}_{-0.2}$ & $24.2^{+0.5}_{-0.3}$ & $24.3^{+0.2}_{-0.2}$ & $24.7^{+0.3}_{-0.2}$ \\[1ex]
834371 & 03:31:09.95 & -28:43:55.05 & >$25.0$ & - & - & $25.1^{+0.3}_{-0.2}$ & $23.5^{+0.1}_{-0.1}$ & $22.9^{+0.1}_{-0.1}$ & $22.6^{+0.1}_{-0.1}$ & $21.8^{+0.2}_{-0.2}$ & $21.6^{+0.2}_{-0.2}$ \\[1ex]
\hline
\end{tabular}
\end{table*}

\begin{table*}
\label{tab:inclusive_prop}
\caption{Inclusive candidates in XMM-LSS (top) and ECDF-S (bottom). These are defined as having good brown dwarf fits, $\chi^{2}_{\mathrm{BD}} < 10$, yet have significantly better galaxy fits, $\chi^{2}_{\mathrm{BD}} - \chi^{2} > 4$ when fitted without $GR$ in XMM-LSS and $ugrGR$ in ECDF-S. The first column shows the object ID. The next four columns show properties of the high-z solution: photometric redshift, absolute magnitude in a tophat filter at $1500$\AA \ with width $100$\AA, extinction, and $\chi^{2}$ value. The next two columns show the secondary photometric redshift and $\chi^{2}$. The final two columns show the stellar type for fitting to MLT dwarfs, and $\chi^{2}_{\mathrm{BD}}$.} 
\begin{tabular}{cccccccccc}
        \hline
        ID & $z$ & $M_{\mathrm{UV}}$ & $A_{V}$ & ${\chi}^{2}$ & $z_{\mathrm{gal2}}$ & ${\chi}^{2}_{\mathrm{gal2}}$ & Stellar Type & ${\chi}^{2}_{\mathrm{BD}}$  \\
        & & / mag &/ mag & & & &  &   \\[1ex]
        \hline
        170389 & $7.34^{+0.10}_{-0.14}$ & $-22.6\pm0.2$ & 0.0 & 5.3 & 9.0 & 26.5 & T8 & 9.8\\[1ex]
        914913 & $6.53^{+0.64}_{-0.10}$ & $-21.7\pm0.3$ & 0.0 & 4.3 & 1.2 & 14.4 & T3 & 8.1 \\[1ex]
        1017553 & $6.93^{+0.26}_{-0.34}$ & $-22.2\pm0.2$ & 0.0 & 0.9 & 1.5 & 13.5 & T3 & 6.2\\[1ex]
        1415071 & $6.58^{+0.03}_{-0.10}$ & $-22.1\pm0.1$ & 0.2 & 1.6 & 1.4 & 18.4 & T3 & 9.0\\[1ex]
        \hline
        705214 & $6.61^{+0.66}_{-0.18}$ & $-22.3\pm0.3$ & 0.2 & 3.8 & 1.45 & 11.5 & T8 & 9.8 \\[1ex]
        834371 & $6.99^{+0.56}_{-0.07}$ & $-23.4\pm0.2$ & 0.0 & 1.8 & 1.7 & 12.2 & L7 & 6.4 \\[1ex]
        1090797 & $7.33^{+0.07}_{-0.19}$ & $-23.3\pm0.2$ & 0.0 & 3.2 & 1.5 & 38.4 & T8 & 6.8 \\[1ex]
        1495402 & $7.39^{+0.04}_{-0.46}$ & $-24.1\pm0.2$ & 0.0 & 1.6 & 1.6 & 25.8 & L4 & 5.7 \\[1ex]
        \hline
\end{tabular}
\end{table*}

\begin{table}
    \centering
    \caption{The rest-UV luminosity function points at $z\simeq7$, calculated by incorporating the inclusive candidates. The first two columns show the absolute UV magnitude of the bin and the width of the bin. The third column shows the number of galaxies in the bin. The fourth column shows the comoving number density of galaxies we calculate with Equations \ref{lf_eqn} and \ref{lf_error}. The final column shows the median completeness in each bin.}
    \begin{tabular}{lcccc}

        \hline
         Bin & Bin width& $n_{\mathrm{gal}}$ & $\phi$ & Completeness\\[1ex]
         / mag & / mag & & $\mathrm{/ \ mag \ / \ Mpc^{3}}$ & \\[1ex]
         \hline
         -22.05 & 0.5 & 18 &  $2.57\pm0.97\times10^{-6}$ & 0.68\\[1ex]
         -22.55 & 0.5 & 10 &   $8.40\pm4.51\times10^{-7}$ & 0.74\\[1ex]
         -23.175 & 0.75 & 5 &   $2.02\pm1.19\times10^{-7}$ & 0.79\\[1ex]
         -23.925  & 0.75 & 1 &   $2.37\pm2.50\times10^{-8}$ & 0.82\\[1ex]
         \hline
         & 
    \end{tabular}
    
    \label{tab:LFincpoints}
\end{table}

\begin{figure}
    \centering
    \includegraphics[width=\columnwidth]{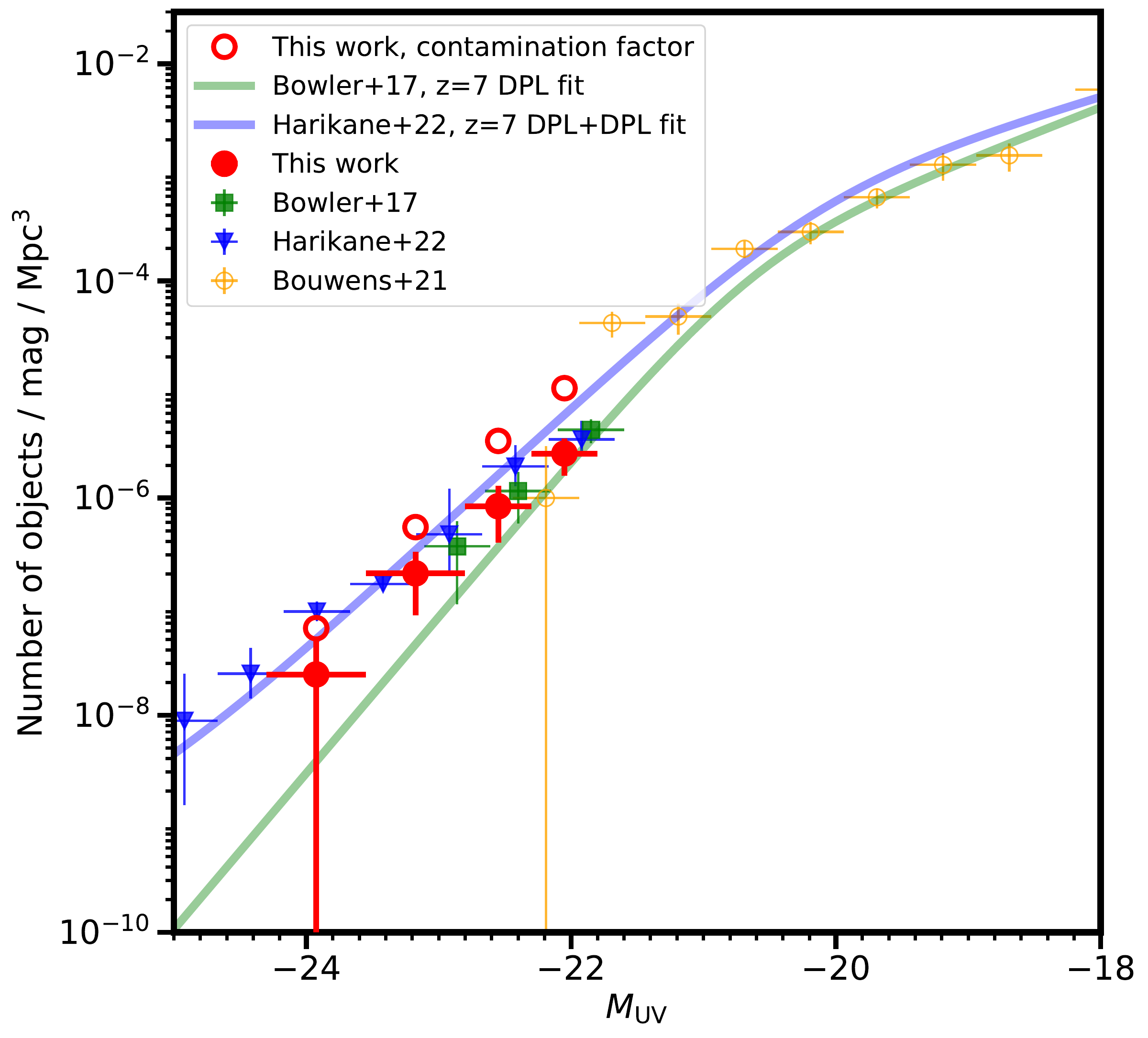}
    \caption{The UV luminosity function at $z=7$, computed including the inclusive sample. The red circles show measurements of the number density from this work. Other measurements from \citet{harikane22}, \citet{bowler17} and \citet{bouwens21} are shown. The solid lines show the best fitting double power laws from \citet{harikane22} (blue) and \citet{bowler17} (green). The red open circles show our LF points incorporating a contamination factor of 50 per cent, as derived by SED fitting of objects from \citet{harikane22} that overlapped with our catalogues.}
    \label{fig:LF_inc}
\end{figure}

In this Appendix we present a more inclusive selection of candidates, which includes sources that have a plausible brown dwarf fit but are still better represented by a high-redshift galaxy model. \citet{Bowler15} state that removing objects with a good brown dwarf fit $\chi^{2} < 10$, likely removes genuine high-redshift galaxies. We retain objects that have good brown dwarf fits and significantly better galaxy fits, $\chi^{2}_{\mathrm{star}} - \chi^{2} > 4$. These candidates are not robust due to the well-fitted brown dwarf SEDs. However, they pass our other selection criteria. We find four inclusive candidates in XMM-LSS, and four in ECDF-S. Their photometry is shown in Table \ref{tab:inc_phot}. The results of SED fitting analysis with \textsc{LePhare} is shown in Table \ref{tab:inclusive_prop}. In ECDF-S the candidates are much brighter than our primary sample: 4 have absolute UV magnitudes $M_{\mathrm{UV}} < -23$, and 1 has $M_{\mathrm{UV}} < -24$. The SED fitting and stamps for the inclusive candidates are shown in Fig. \ref{fig:XMMincl} and Fig. \ref{fig:CDFSincl}. We add these objects to the primary sample and conduct a separate calculation of the rest-UV LF at $z\simeq7$, as outlined in Section \ref{sec:uvlf}. The result is shown in Fig. \ref{fig:LF_inc}. We also plot the DPL fits derived by \cite{bowler17} and \cite{harikane22}, and their LF points. We include \citet{bouwens21} LF points to represent the faint end. The bins are centred at -22.05, -22.55, -23.175 and -23.925, with bin widths of $\Delta M = 0.5$ for the two fainter bins and $\Delta M = 0.75$ for the two brighter bins. The bins contain 18, 10, 5 and 1 candidate galaxies respectively. The LF points are summarised in Table \ref{tab:LFincpoints}. If we include the contamination factor of 50 per cent derived in Section \ref{sec:previous work}, our three faintest bins lie slightly above the DPL+DPL derived by \cite{harikane22}. The brightest bin is consistent with both DPLs, although the Poisson errors are large. The value of the LF in our brightest bin is a factor of four lower than the \citet{harikane22} bin at $M_{\mathrm{UV}} = -23.92$. The contamination factor pushes this bin onto their DPL. 
To summarise, the fiducial UV LF we present in Fig. \ref{fig:LF} is our best estimate of the number density of genuine $z \simeq 7$ sources. However, even if we relax our selection criterion to allow sources which have a plausible brown dwarf fit into the sample, we still find the UV LF to be significantly lower than that found in optical-only surveys for our brightest bins.

\
\
\section{Object Stamps and SED fits}
\label{sec:merged}

In this Appendix we present postage stamps and SED fitting of all candidates in this work. The first two sections are primary candidates in ECDF-S and XMM-LSS. The third section is Lyman-$\alpha$ candidates, all in XMM-LSS. The final two sections are inclusive candidates in ECDF-S and XMM-LSS. 

\subsection{Primary Candidates}
\label{sec: primary stamps}

\subsubsection{XMM-LSS}

%%%%%%%%%%%%%%%%% XMM %%%%%%%%%%%%%%%%%%%%%
\begin{figure*}
     %\centering

         \includegraphics[width=0.75\textwidth]{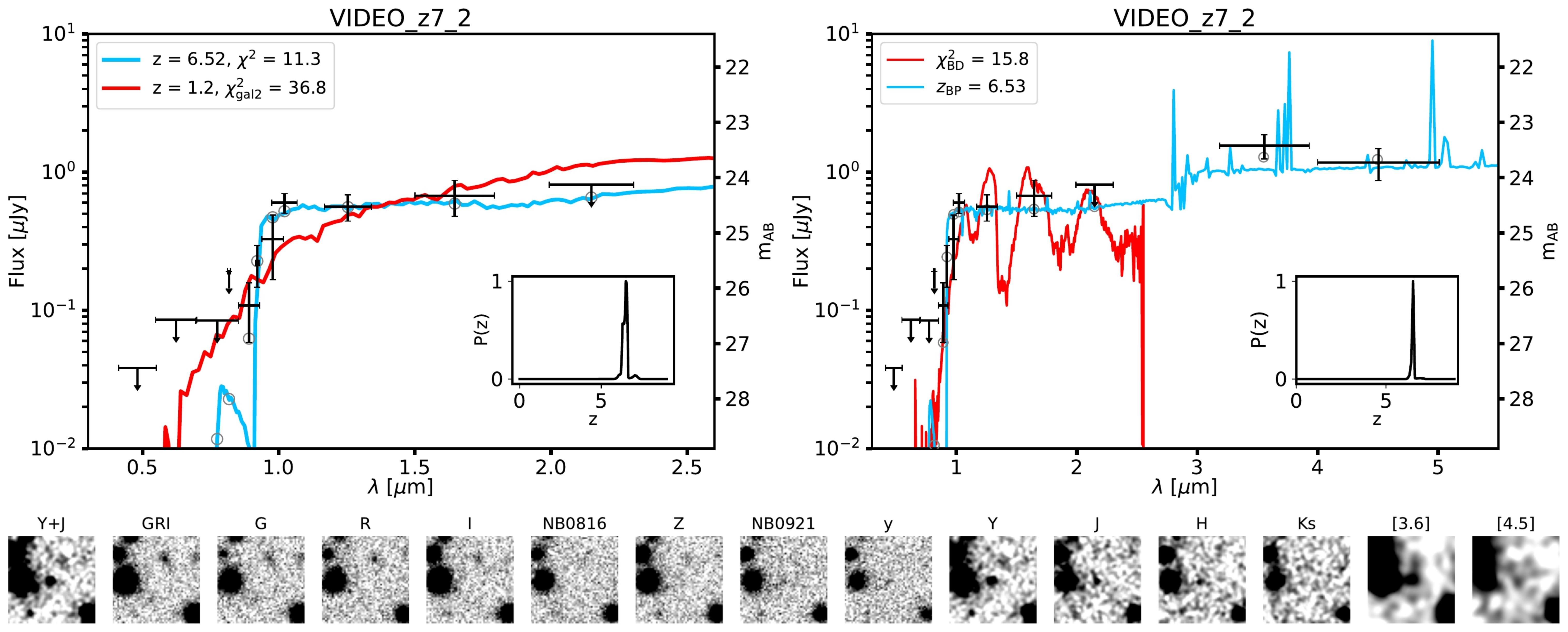}

         \includegraphics[width=0.75\textwidth]{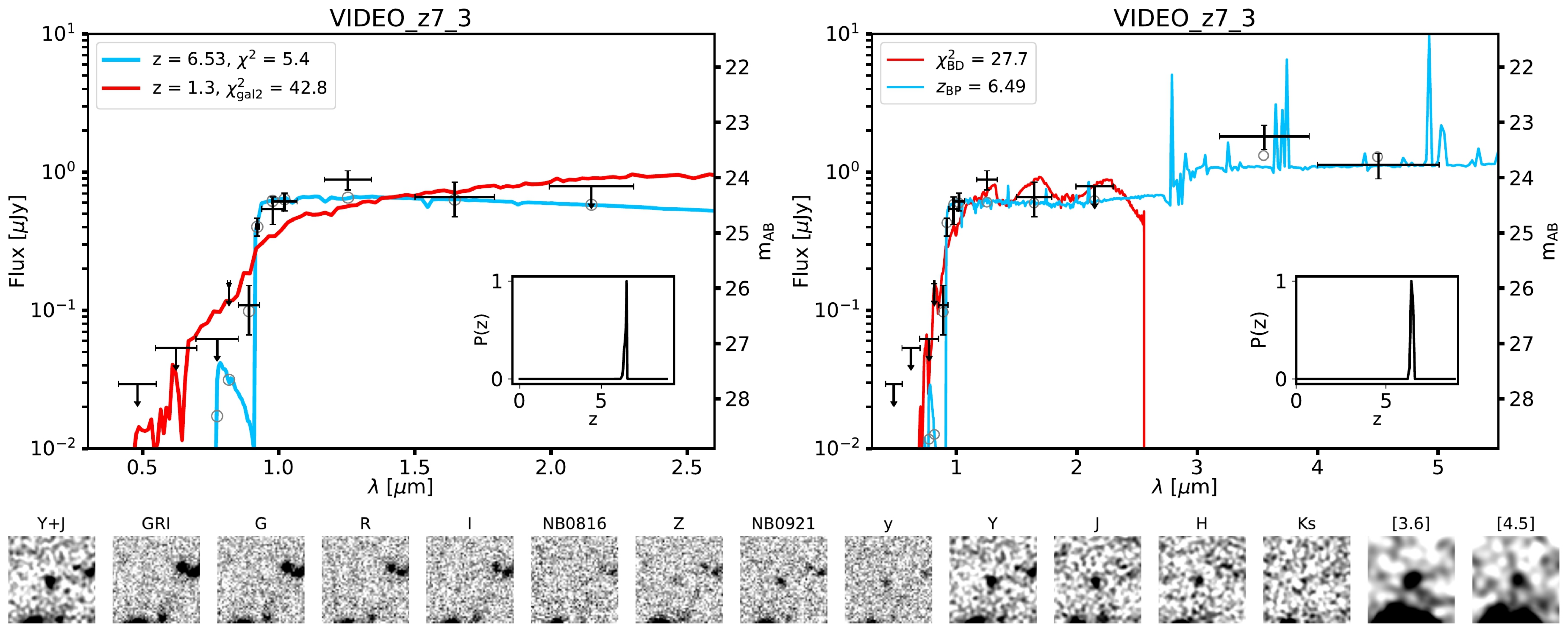}
         
         \includegraphics[width=0.75\textwidth]{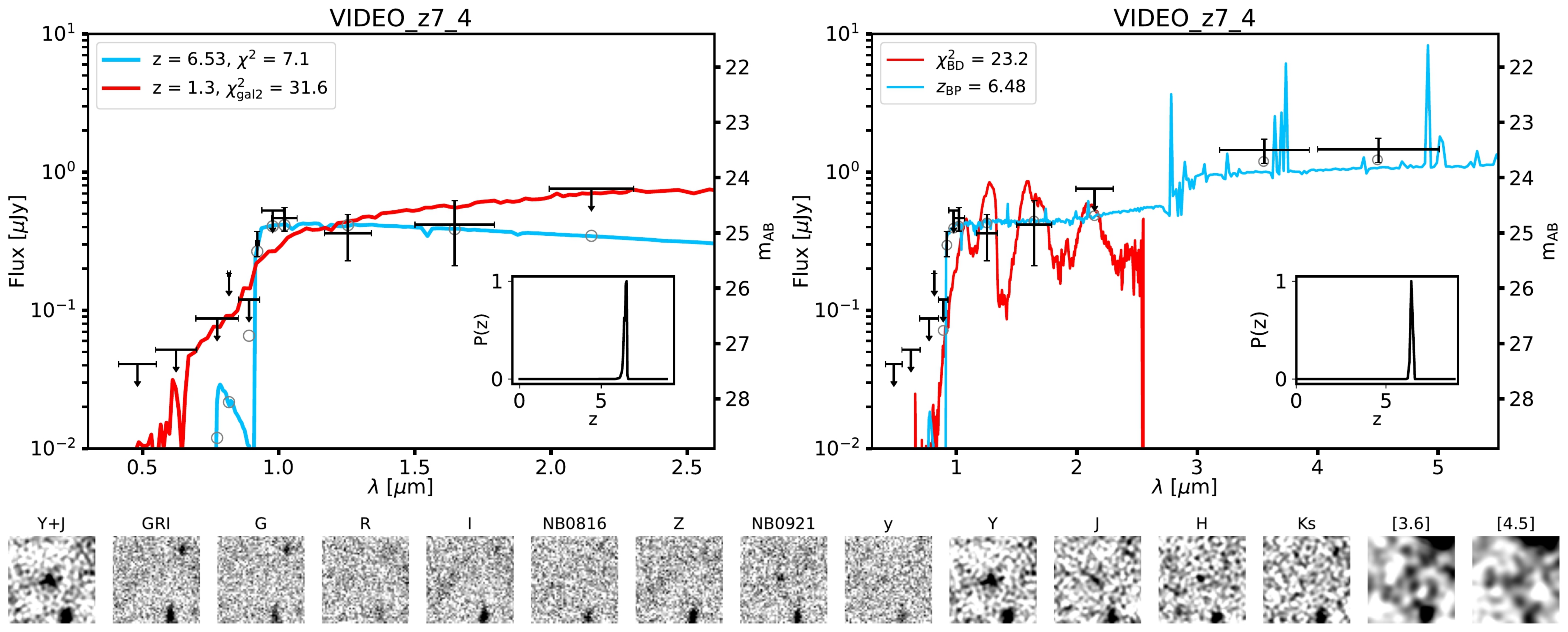}
         
         \includegraphics[width=0.75\textwidth]{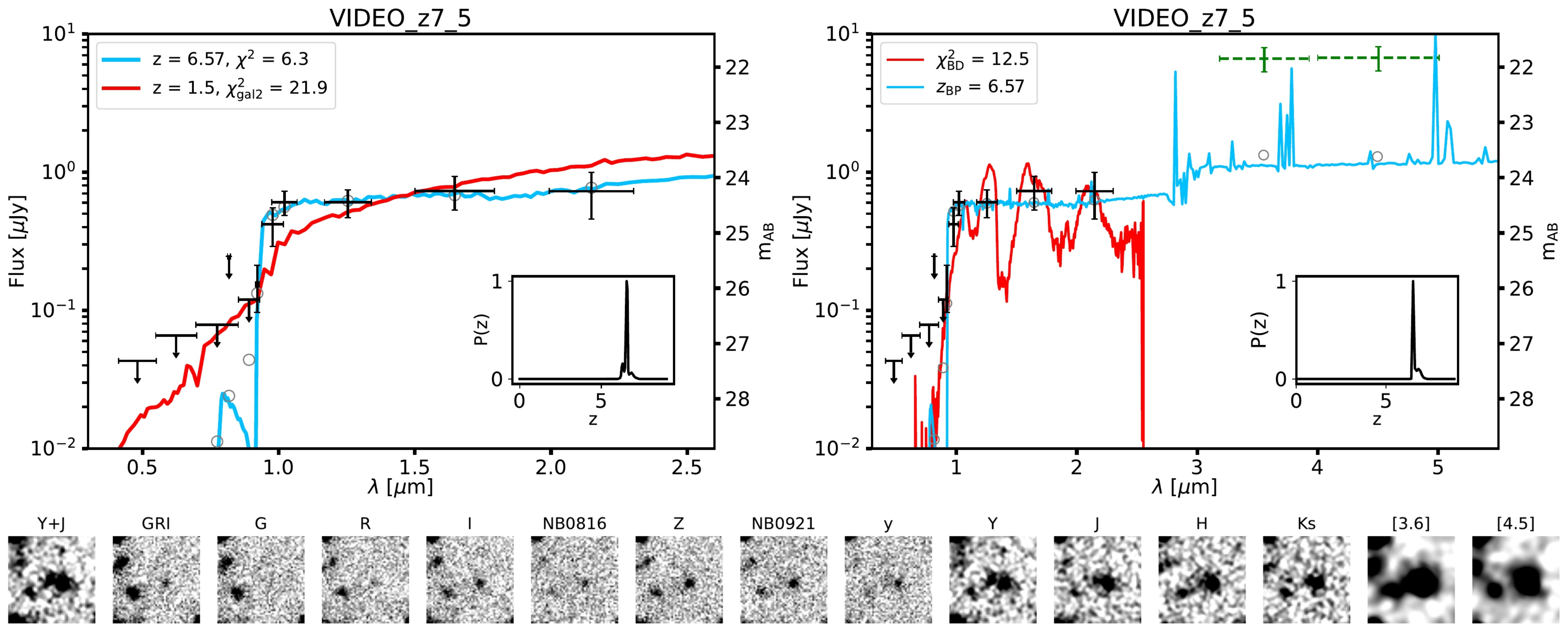}

        \caption{Candidate galaxies in XMM. The plots and stamps are the same as Fig. \ref{fig:candidates}. Confused IRAC photometry is shown in green with dashed wavelength error bars.}
        \label{fig:XMM_objs}
\end{figure*}

\begin{figure*} %\ContinuedFloat
     %\centering

         \includegraphics[width=0.75\textwidth]{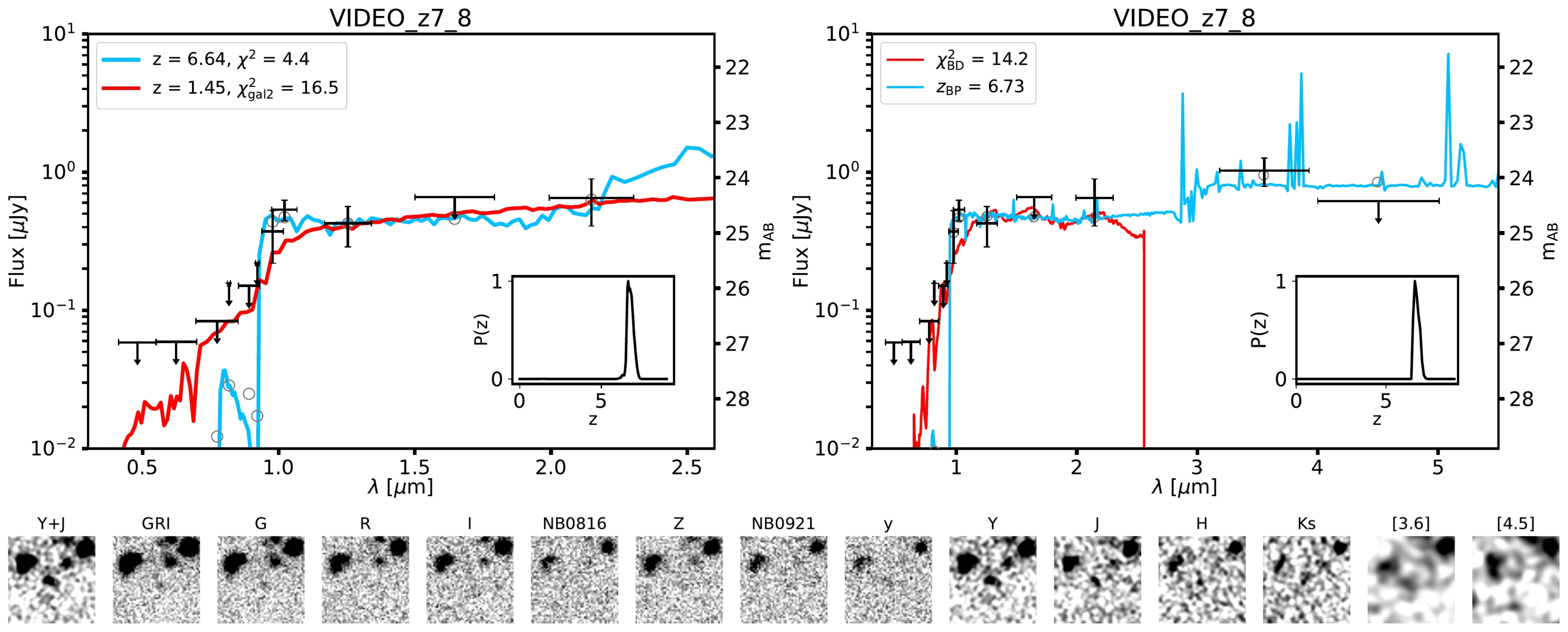}
         
         \includegraphics[width=0.75\textwidth]{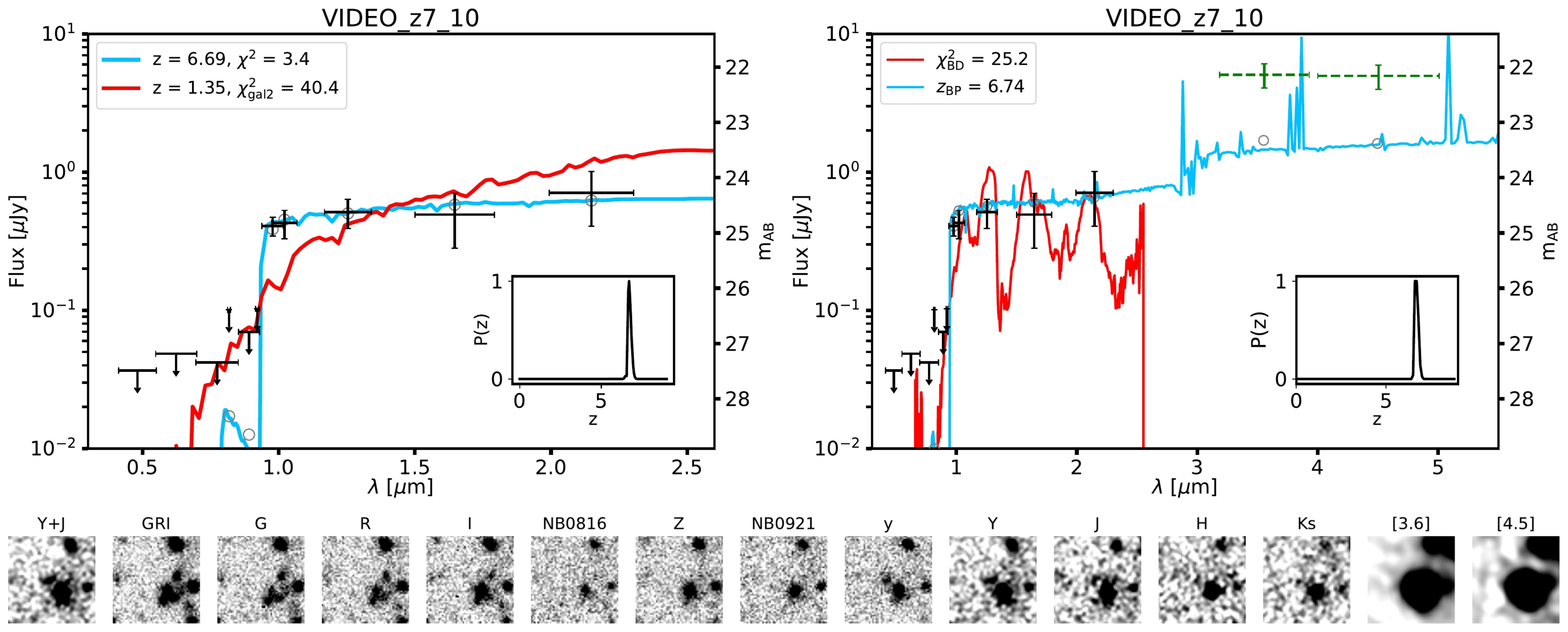}
         
         \includegraphics[width=0.75\textwidth]{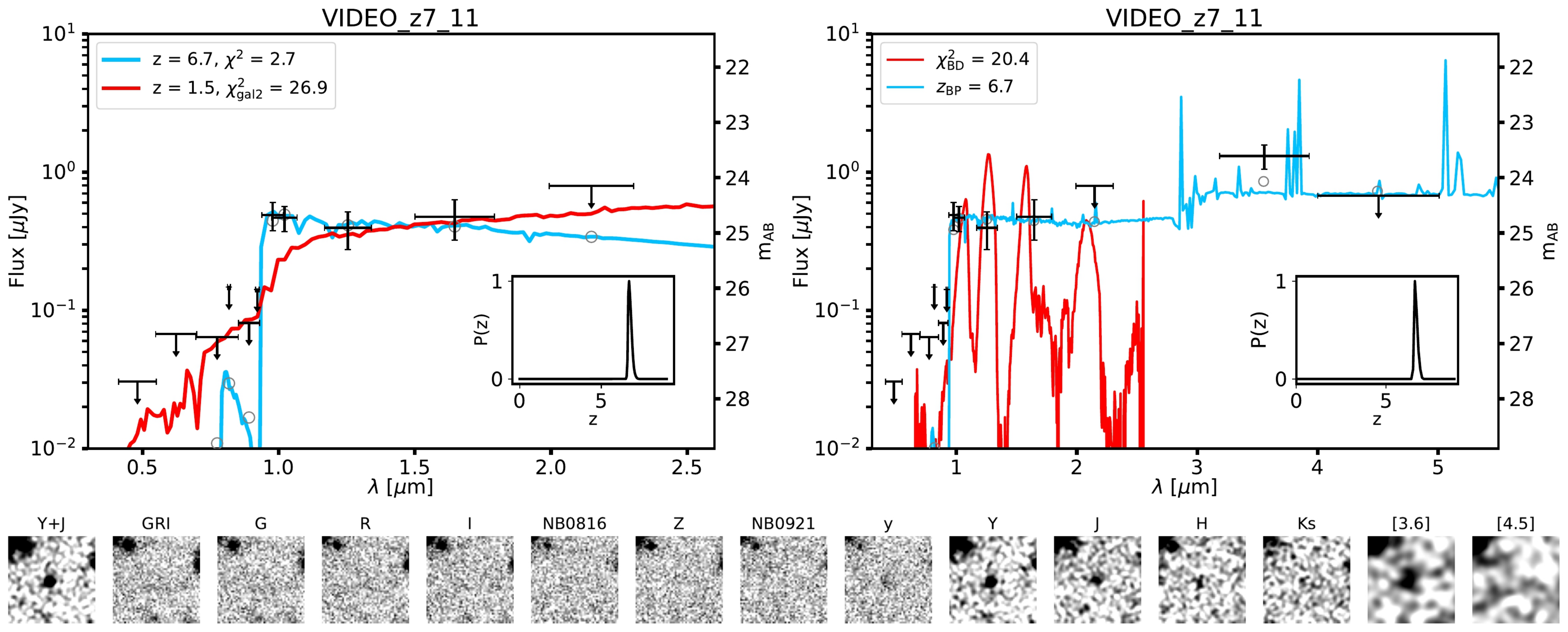}

         \includegraphics[width=0.75\textwidth]{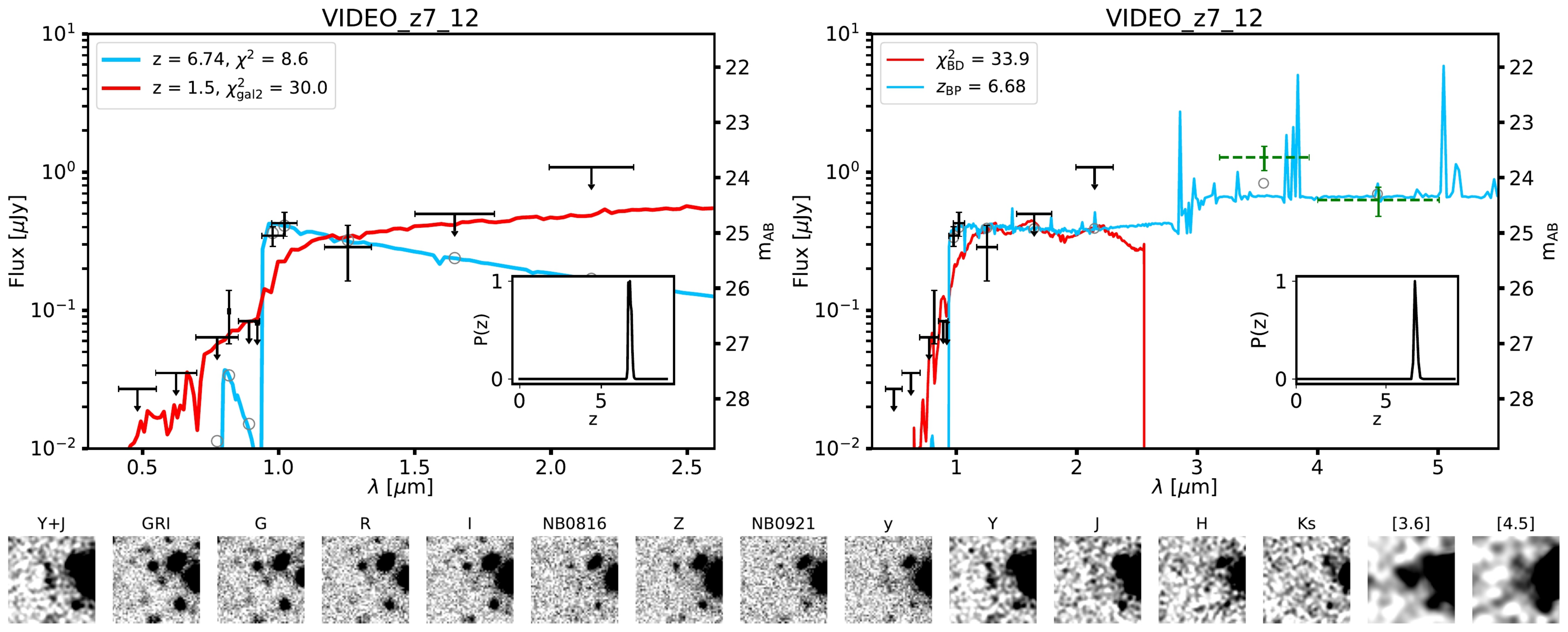}

        \caption{Continued.}
\end{figure*}

\begin{figure*} %\ContinuedFloat
     \centering

         \includegraphics[width=0.75\textwidth]{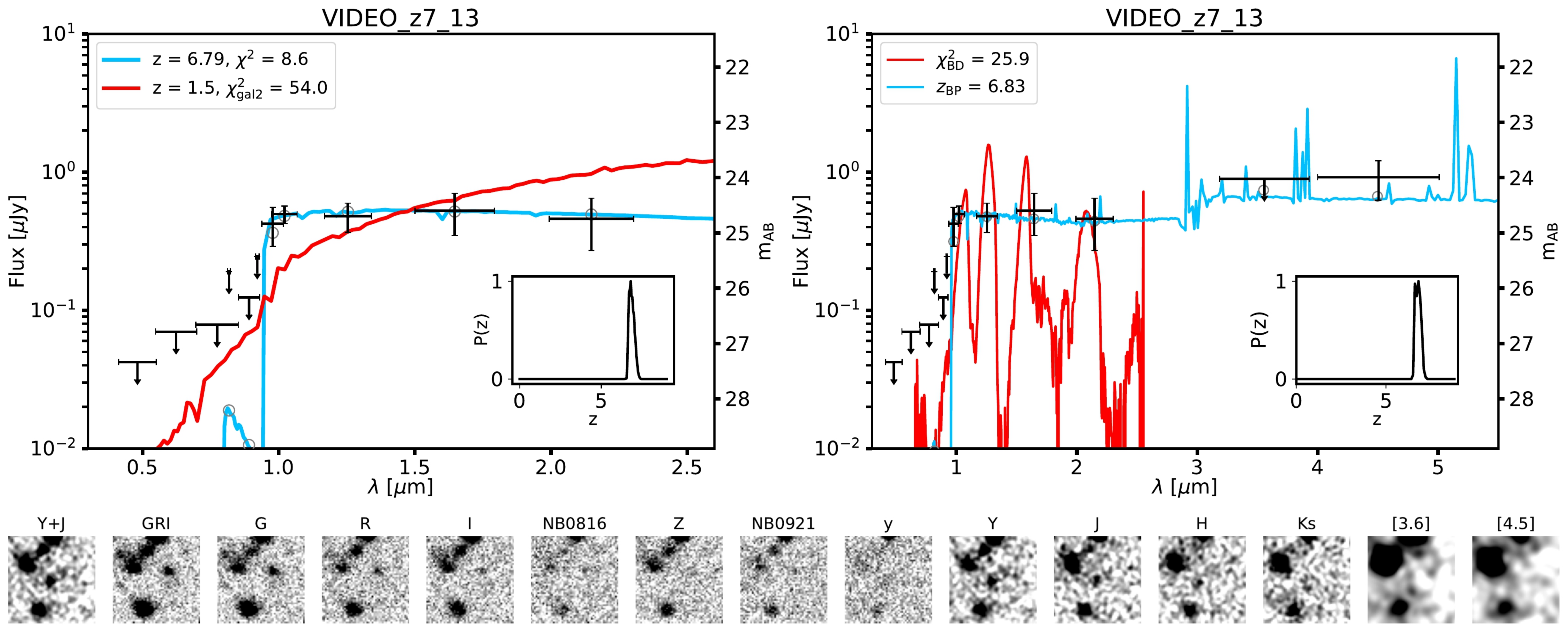}
         
         \includegraphics[width=0.75\textwidth]{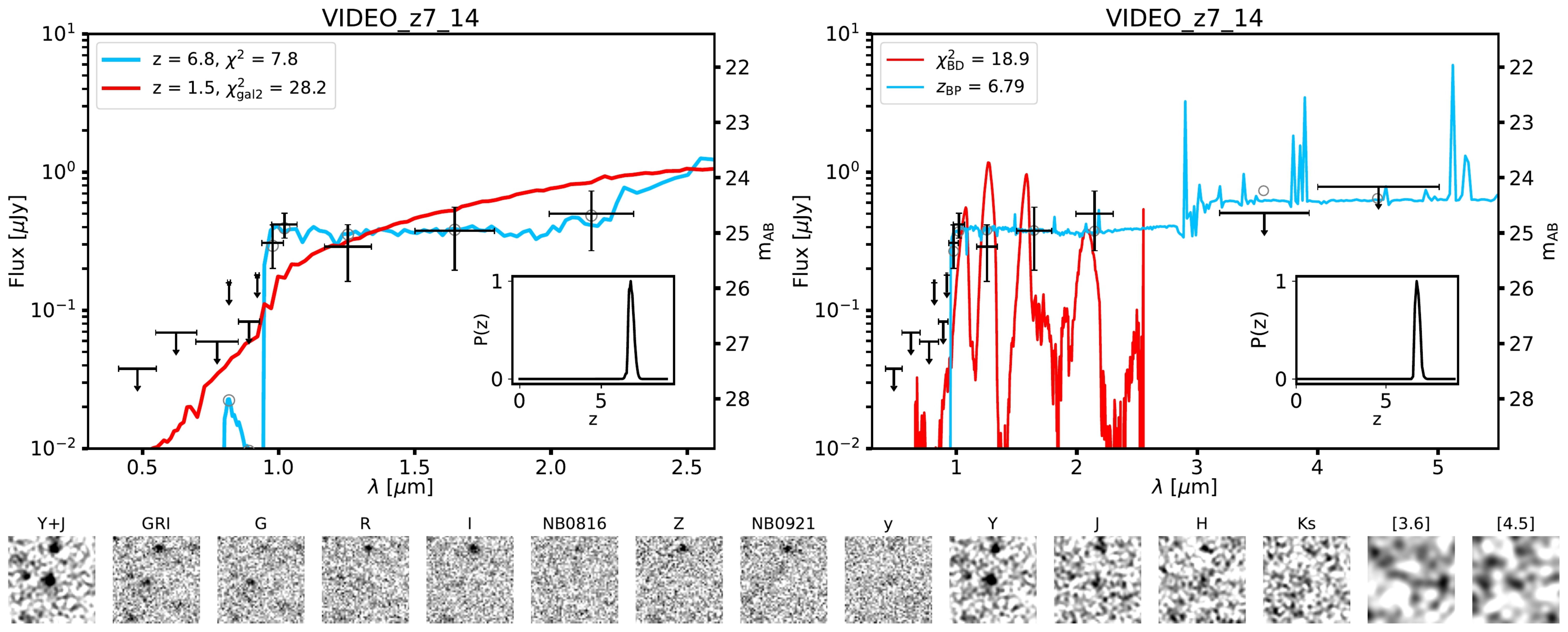}

         \includegraphics[width=0.75\textwidth]{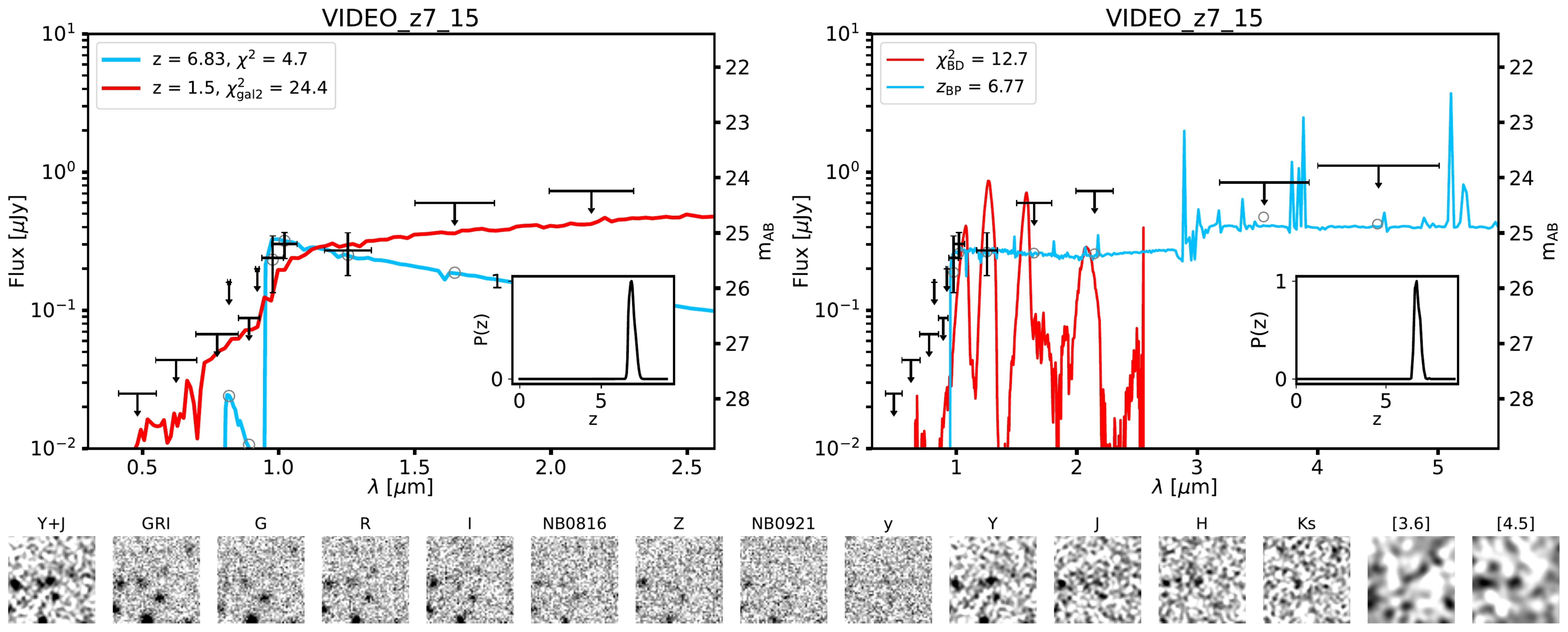}
         
         \includegraphics[width=0.75\textwidth]{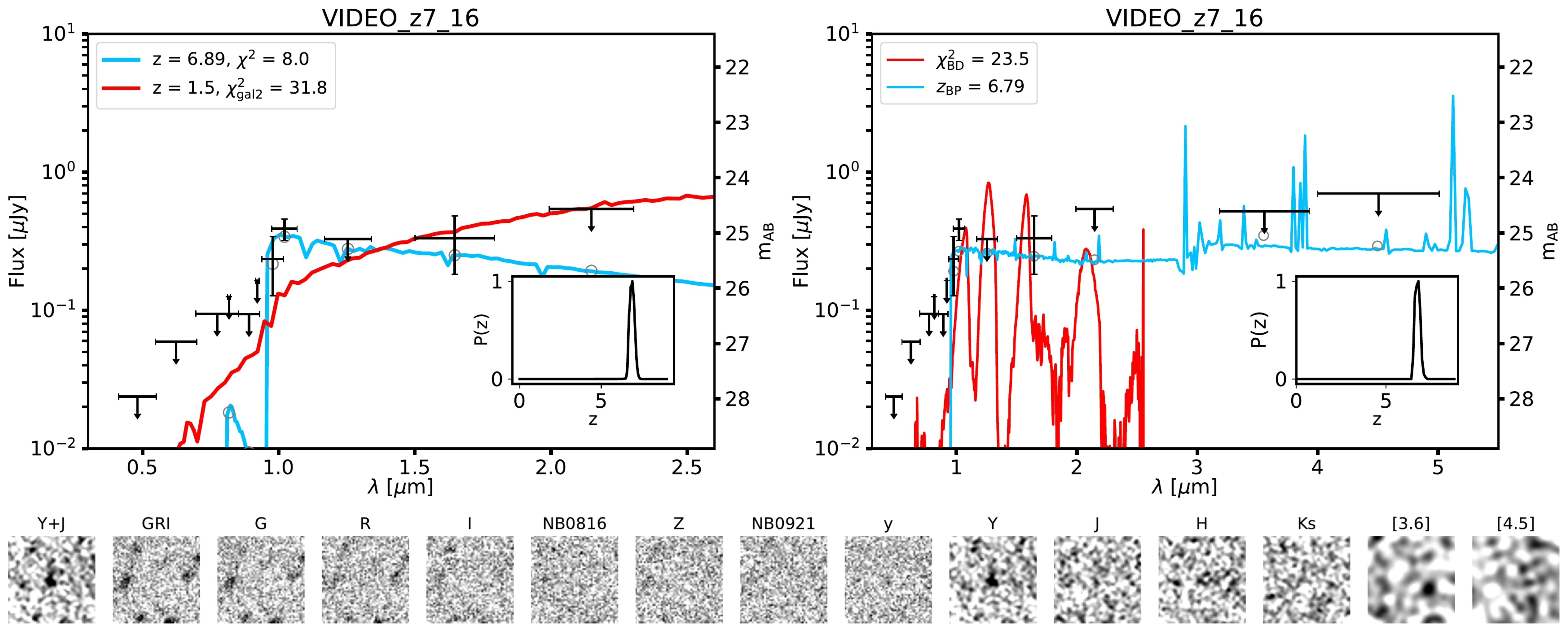}

        \caption{Continued.}
\end{figure*}

\begin{figure*} %\ContinuedFloat
     \centering

         \includegraphics[width=0.75\textwidth]{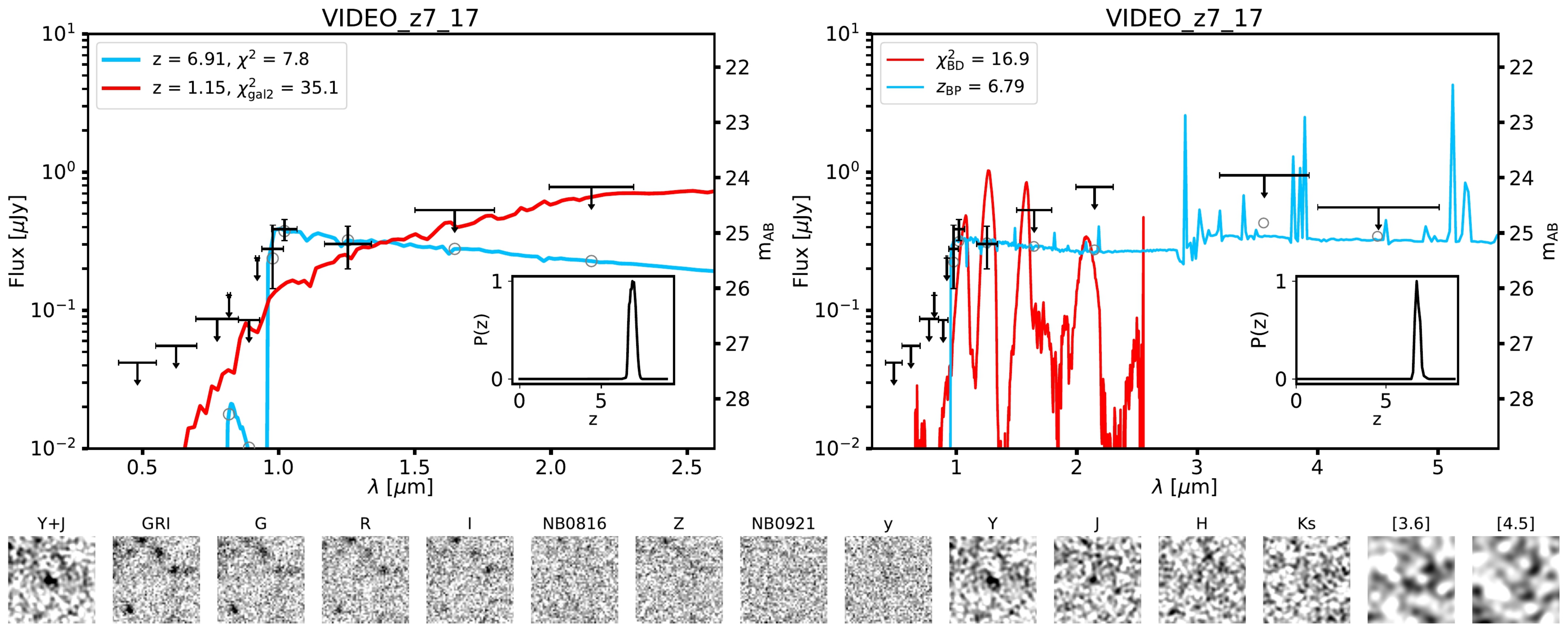}

         \includegraphics[width=0.75\textwidth]{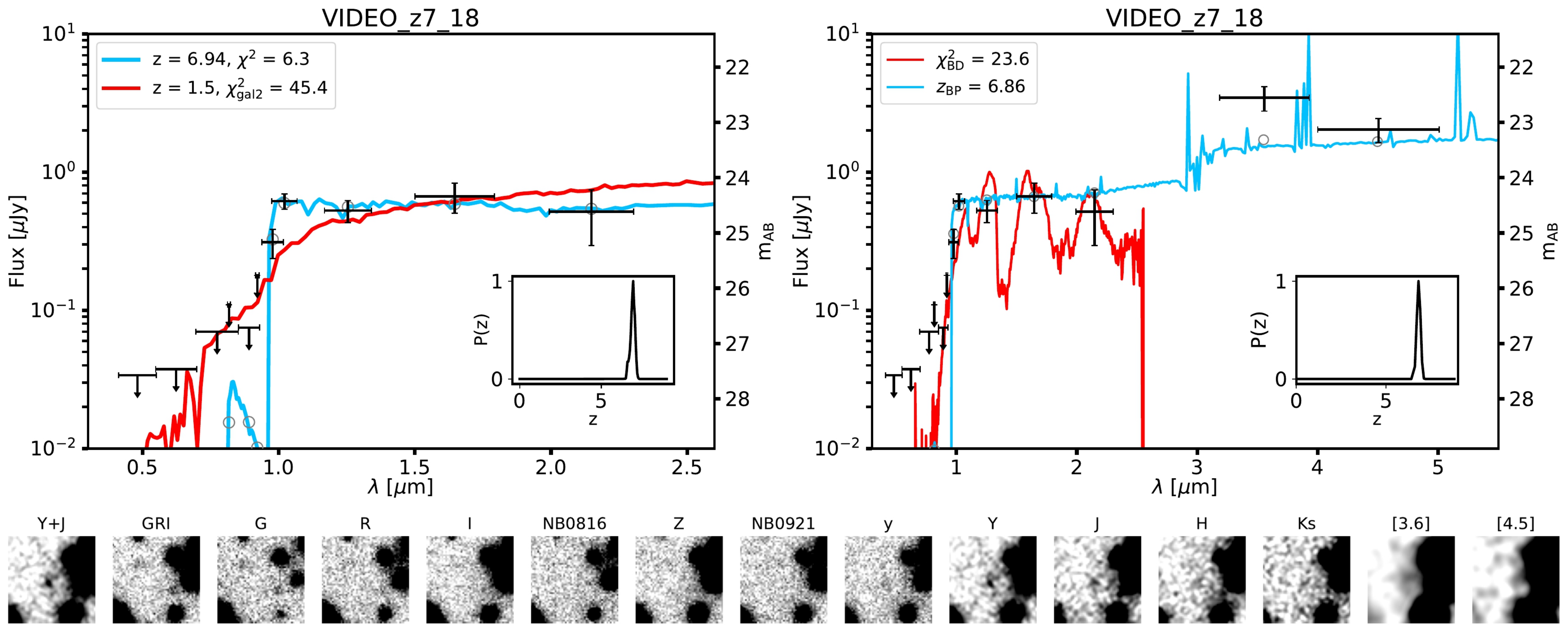}
         
         \includegraphics[width=0.75\textwidth]{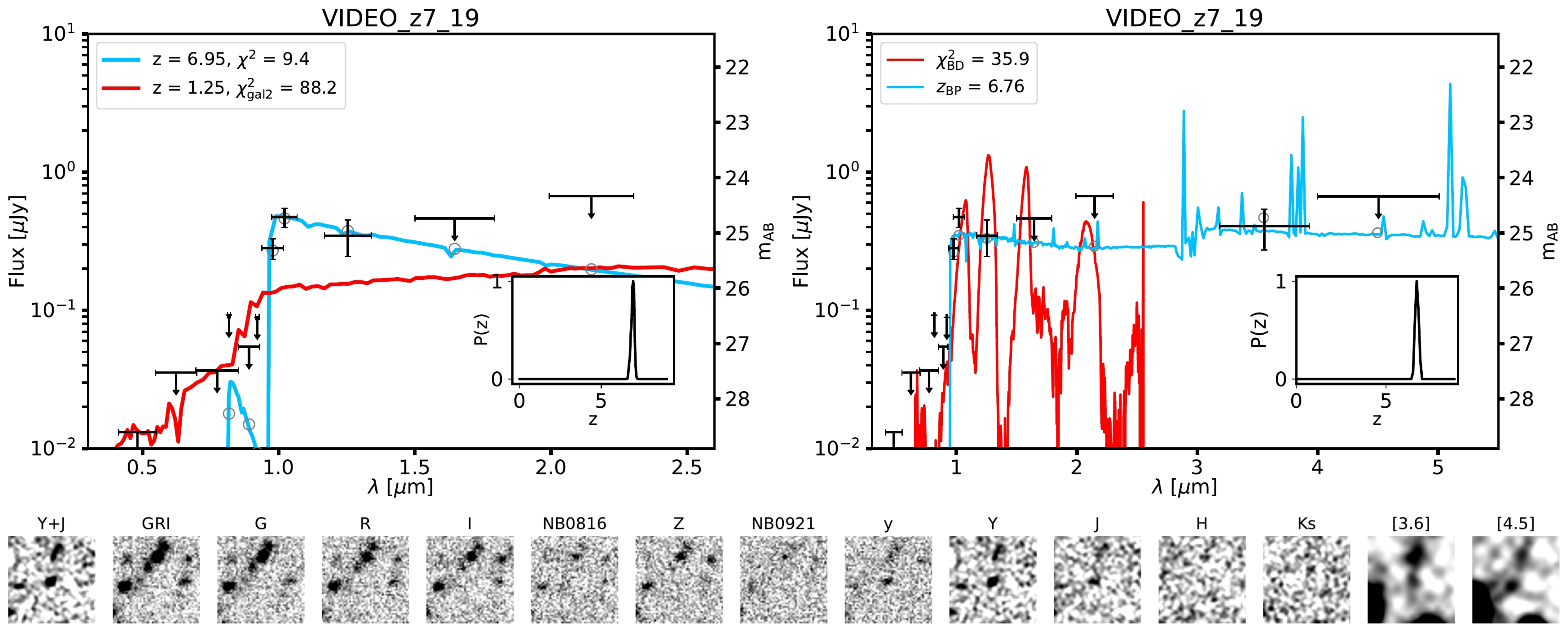}

         \includegraphics[width=0.75\textwidth]{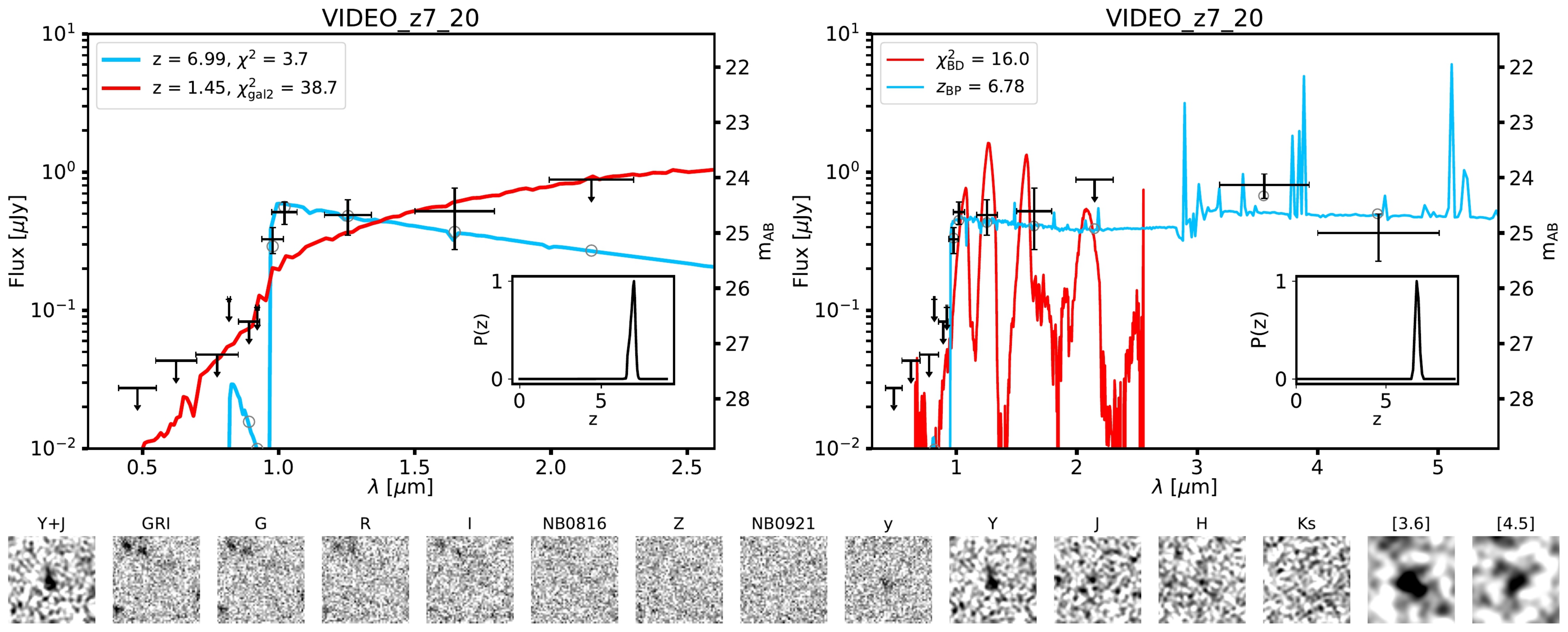}

        \caption{Continued.}

\end{figure*}

\begin{figure*} %\ContinuedFloat
     \centering
         
         \includegraphics[width=0.75\textwidth]{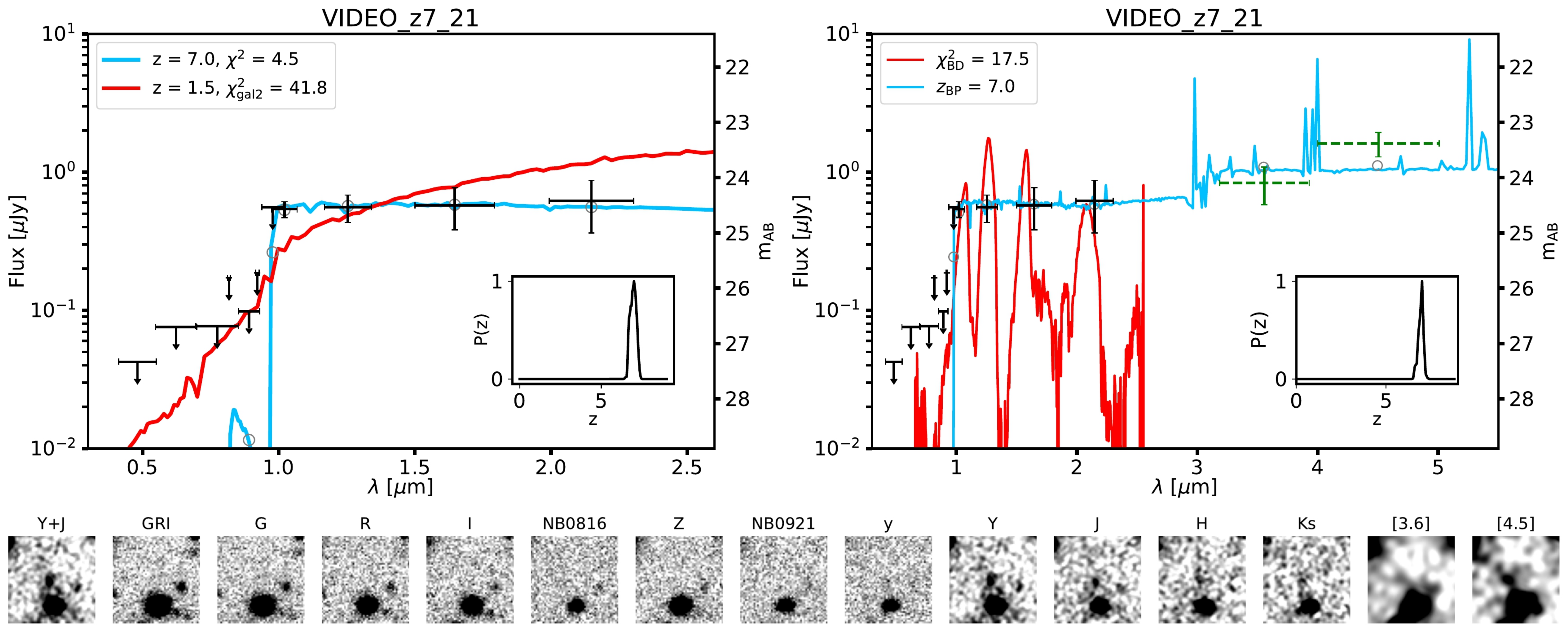}

         \includegraphics[width=0.75\textwidth]{SEDs/VIDEO_z7_22.pdf}

        \caption{Continued.}

\end{figure*}

\subsubsection{ECDF-S}

%%%%%%%%%%%%%%%%% CDFS %%%%%%%%%%%%%%%%%%%%%
\begin{figure*}
     \centering

         \includegraphics[width=0.75\textwidth]{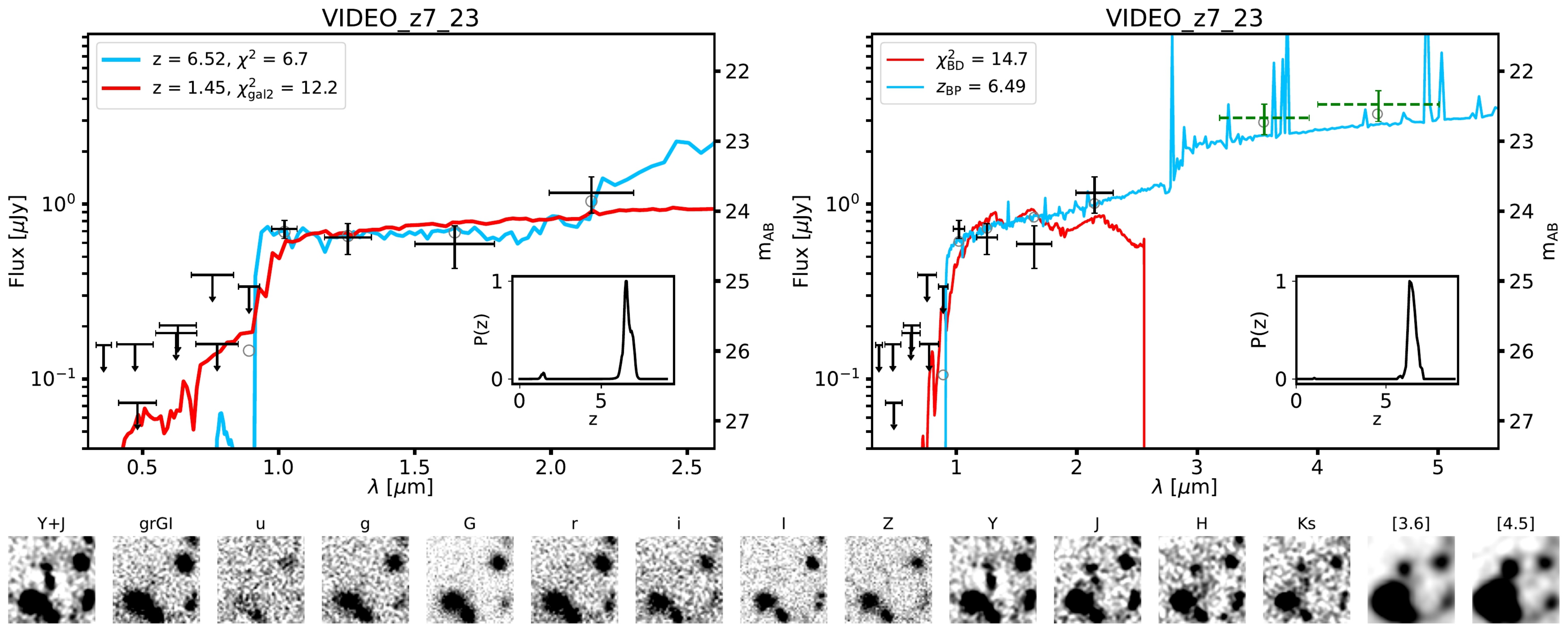}

         \includegraphics[width=0.75\textwidth]{SEDs/VIDEO_z7_24.pdf}
         
         \includegraphics[width=0.75\textwidth]{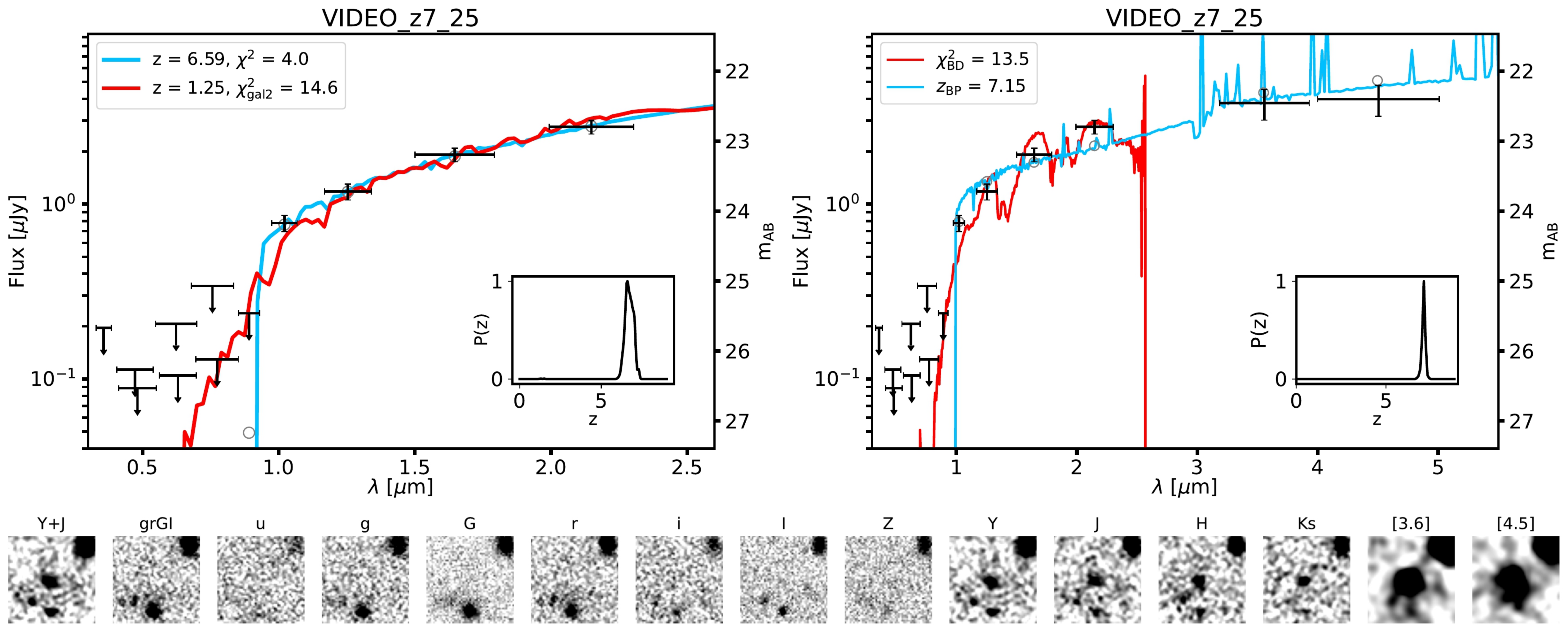}

        \includegraphics[width=0.75\textwidth]{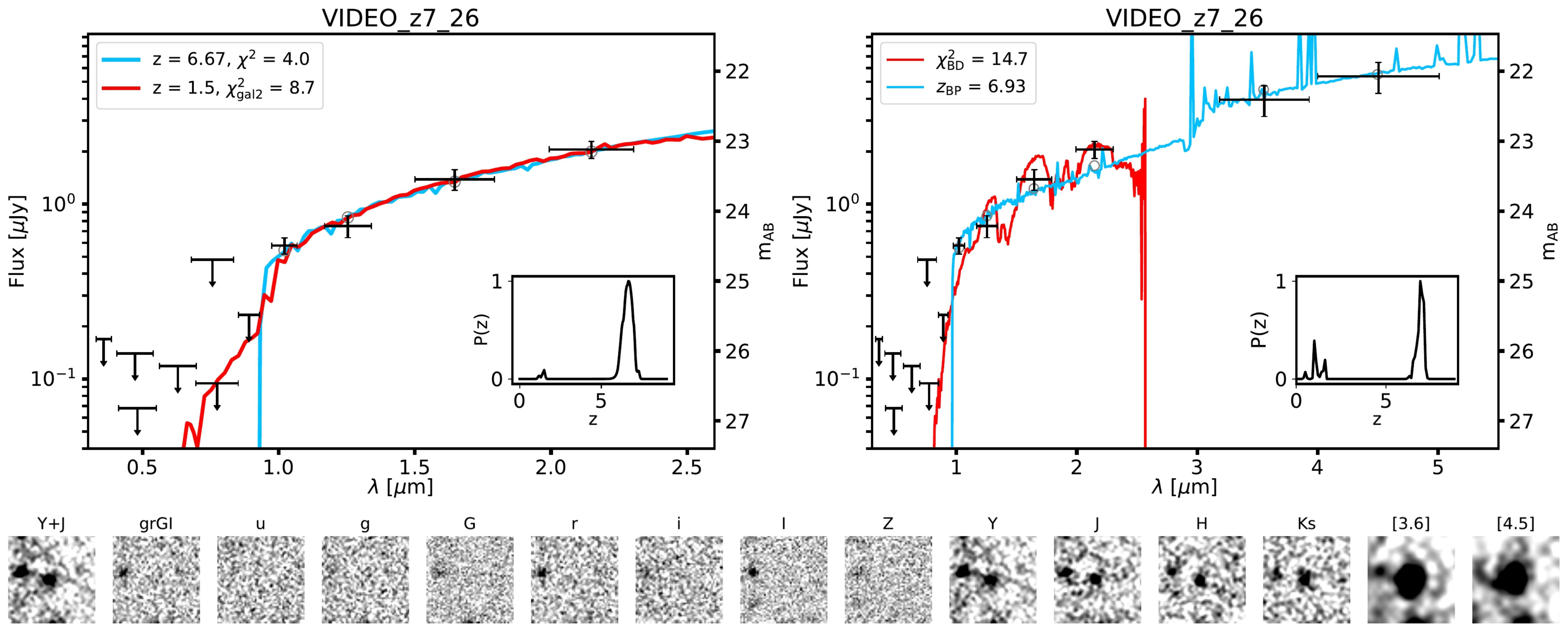}

        \caption{Candidate galaxies in CDFS. The plots and stamps are the same as Fig. \ref{fig:candidates}. Confused IRAC photometry is shown in green with dashed wavelength error bars.}
        \label{fig:CDFS_objs}
\end{figure*}

\begin{figure*} %\ContinuedFloat
     \centering
               
        \includegraphics[width=0.75\textwidth]{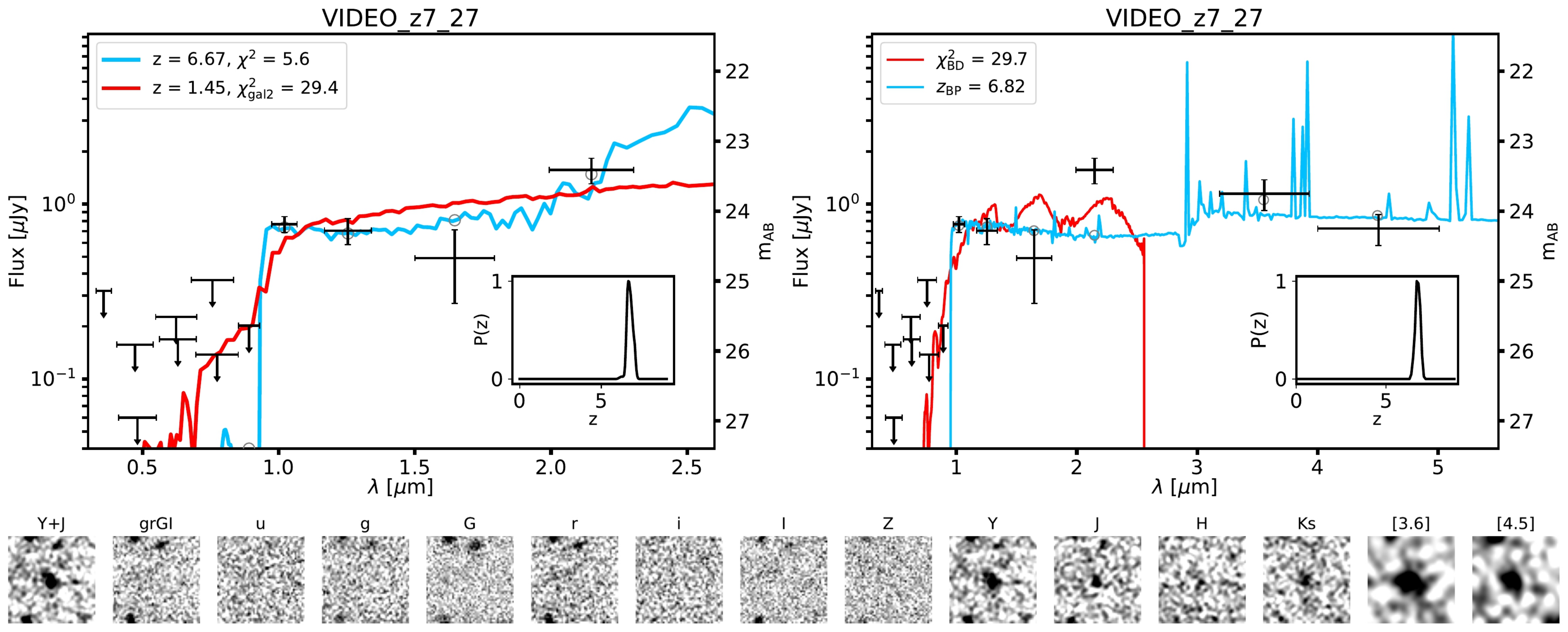}

         \includegraphics[width=0.75\textwidth]{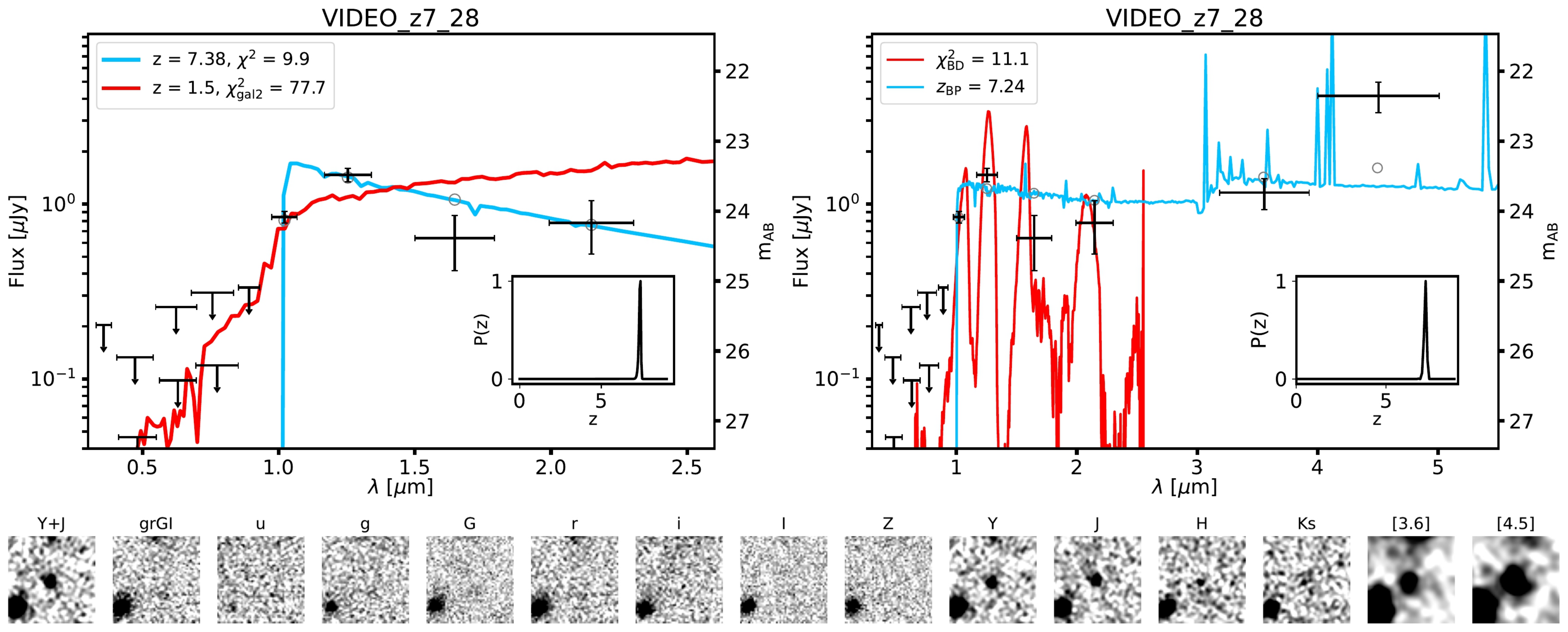}

\caption{Continued.}
\end{figure*}

\subsection{Lyman-\texorpdfstring{$\alpha$}{alpha} Candidates}
\label{sec: lyman stamps}
\subsubsection{XMM-LSS}

%%%%%%%%%%%%%%%%%%%%% XMM LYMAN ALPHA %%%%%%%%%%%%%%%%%%%%%%%%%
\begin{figure*}
     \centering

         \includegraphics[width=0.75\textwidth]{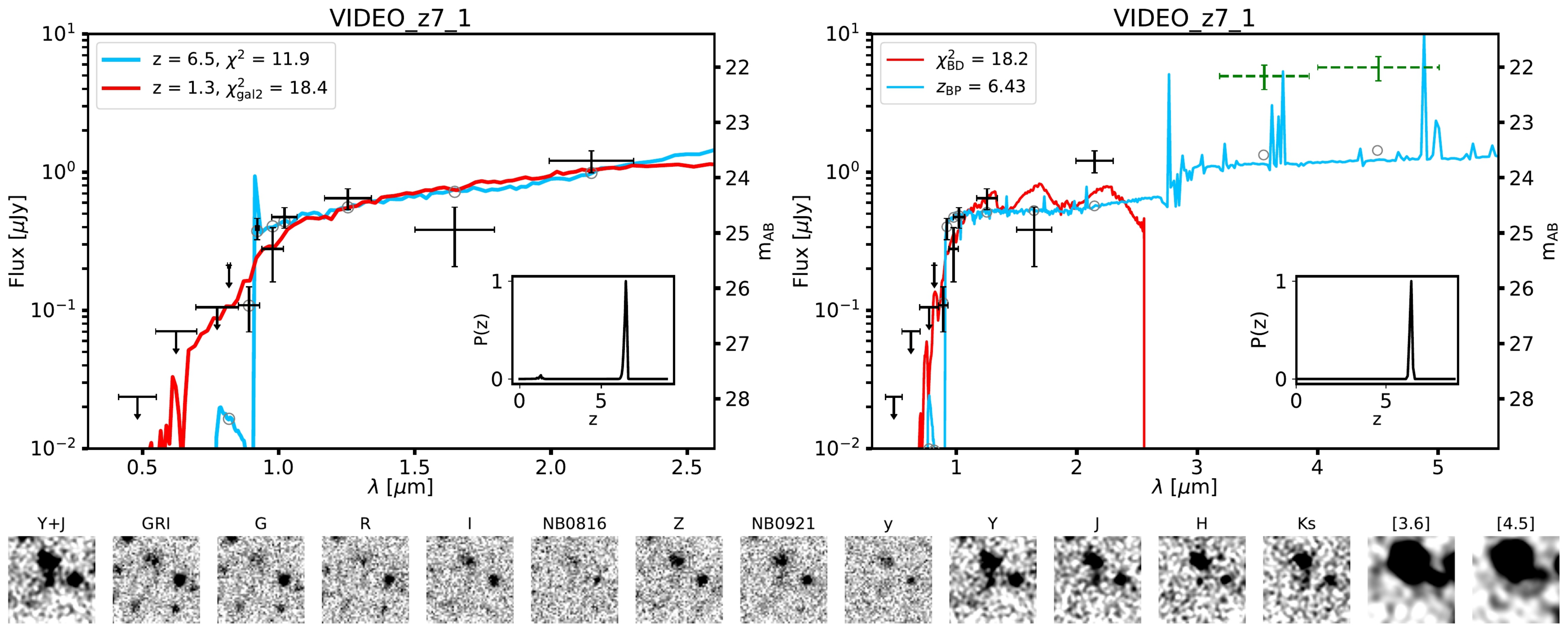}

        \includegraphics[width=0.75\textwidth]{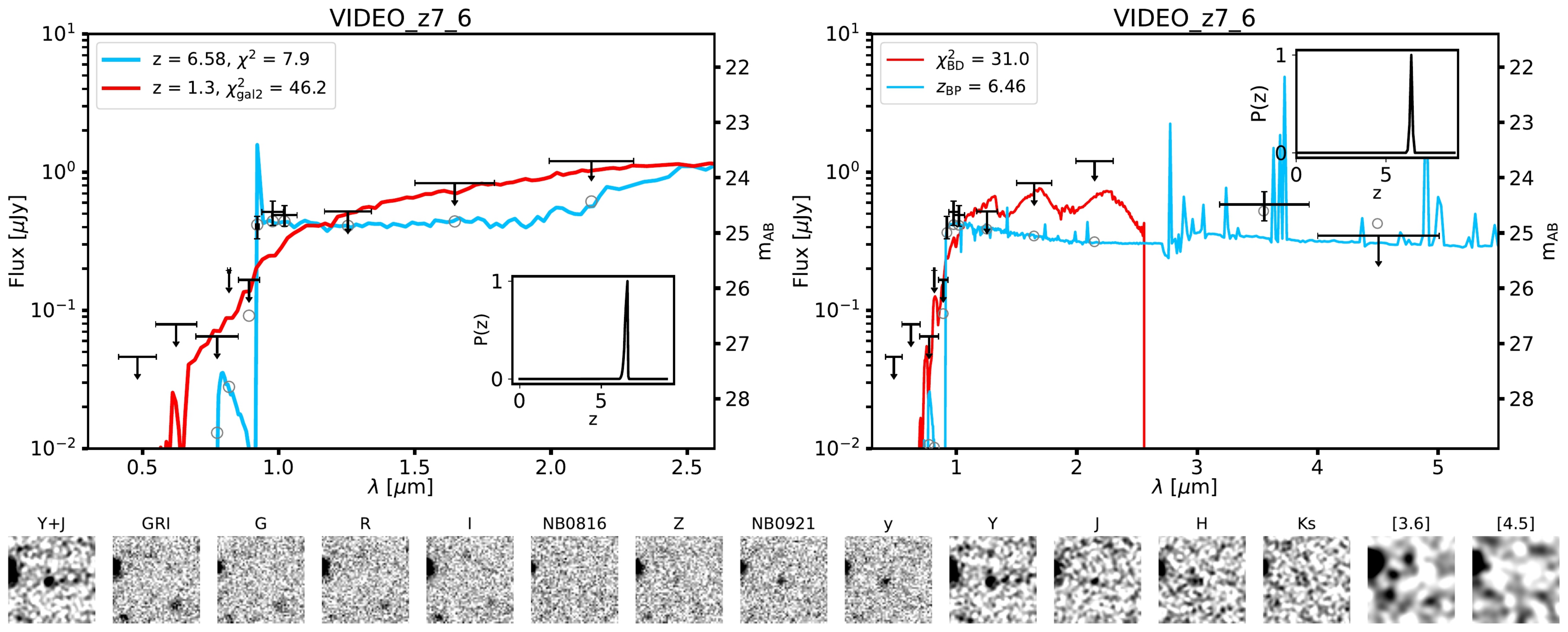}

        \includegraphics[width=0.75\textwidth]{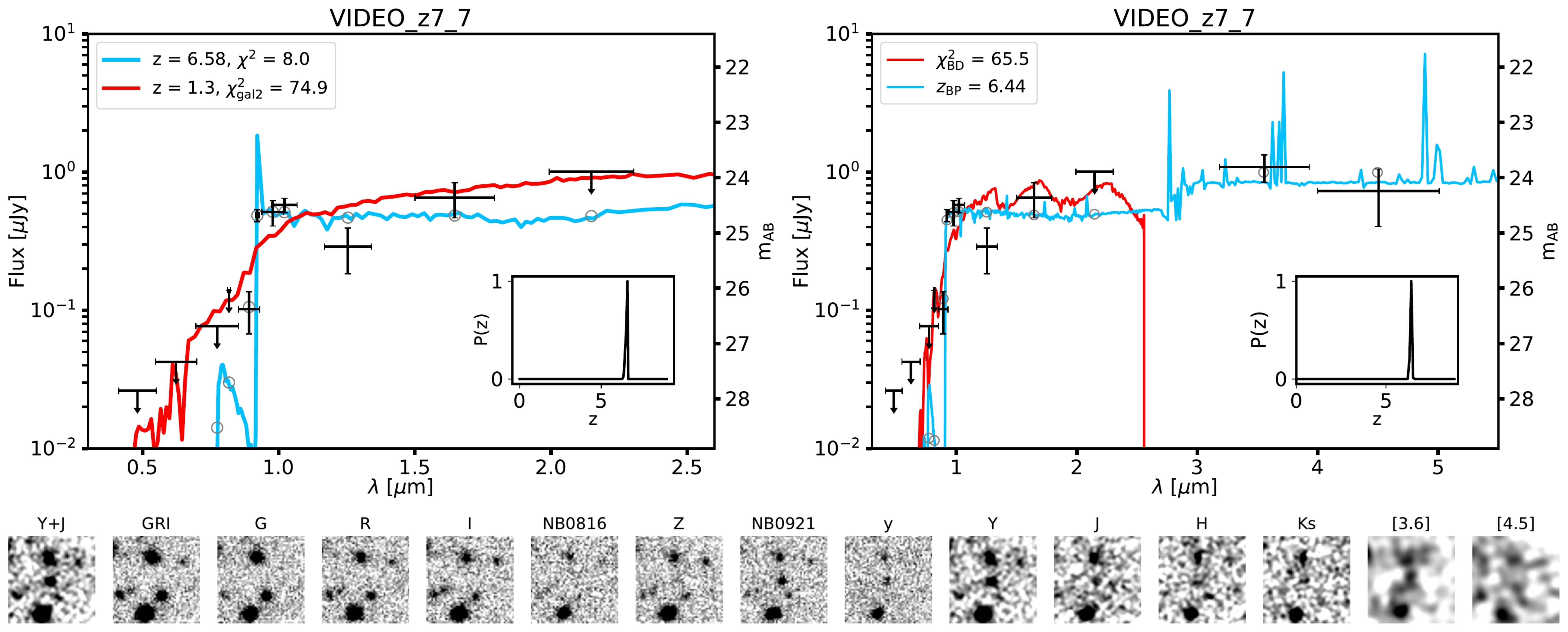}

        \includegraphics[width=0.75\textwidth]{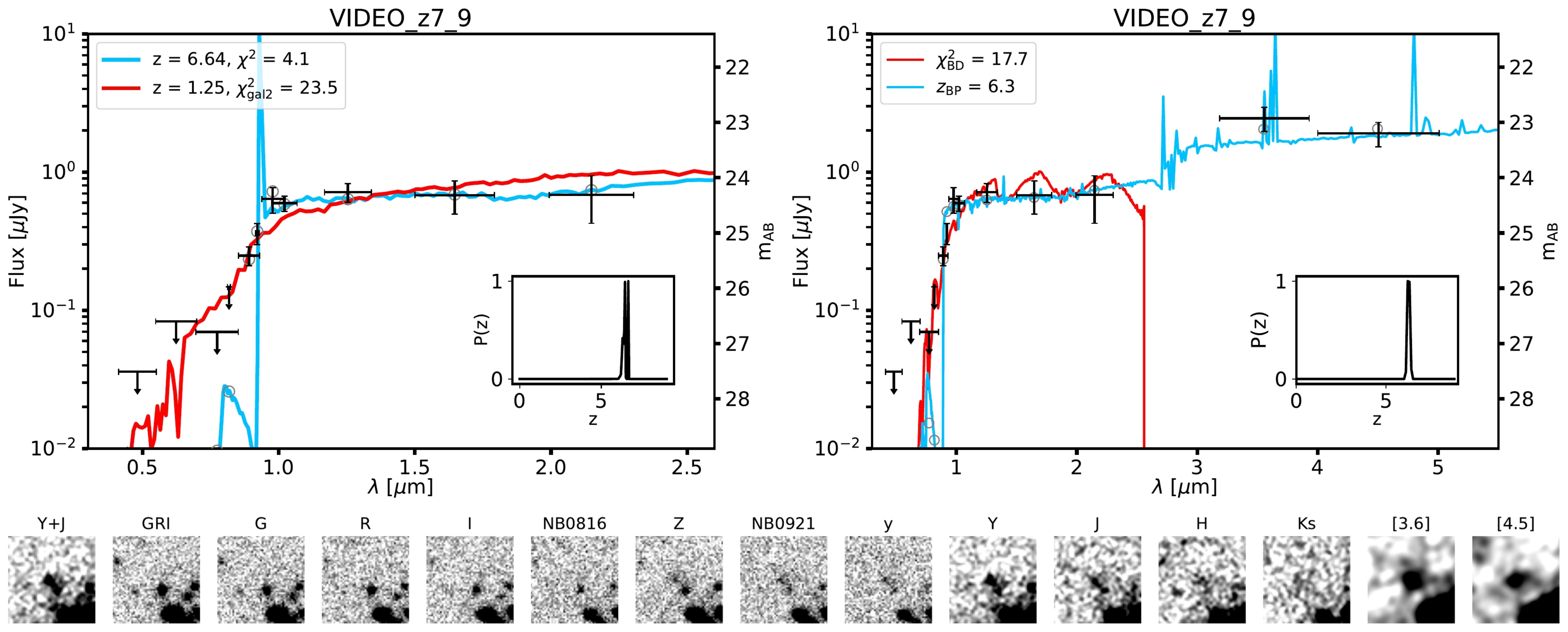}

        \caption{Candidate galaxies in XMM that require a Lyman-$\alpha$ emission line to be included in our sample. The plots and stamps are the same as Fig. \ref{fig:candidates}. Confused IRAC photometry is shown in green with dashed wavelength error bars.}
        \label{fig:XMMlya}
\end{figure*}

%%%%%%%%%%%%%%%%%%%%%%%%%%%%%%%%%%%%%%%%%%%%

%%%%%%%%%%%%%%%%%%%%%%%%%%%%%%%%%%%%%%%%%%%%

%%%%%%%%%%%%%%%%%%%%% XMM INCLUSIVE %%%%%%%%%%%%%%%%%%%%%%%%%

%%%%%%%%%%%%%%%%%%%%%%%%%%%%%%%%%%%%%%%%%%%%%%%%%%

%%%%%%%%%%%%%%%%%%%%% CDFS INCLUSIVE %%%%%%%%%%%%%%%%%%%%%%%%%

\subsection{Inclusive candidates}
\label{sec:inclusive stamps}

\subsubsection{XMM-LSS}
\begin{figure*}
     \centering

         \includegraphics[width=0.75\textwidth]{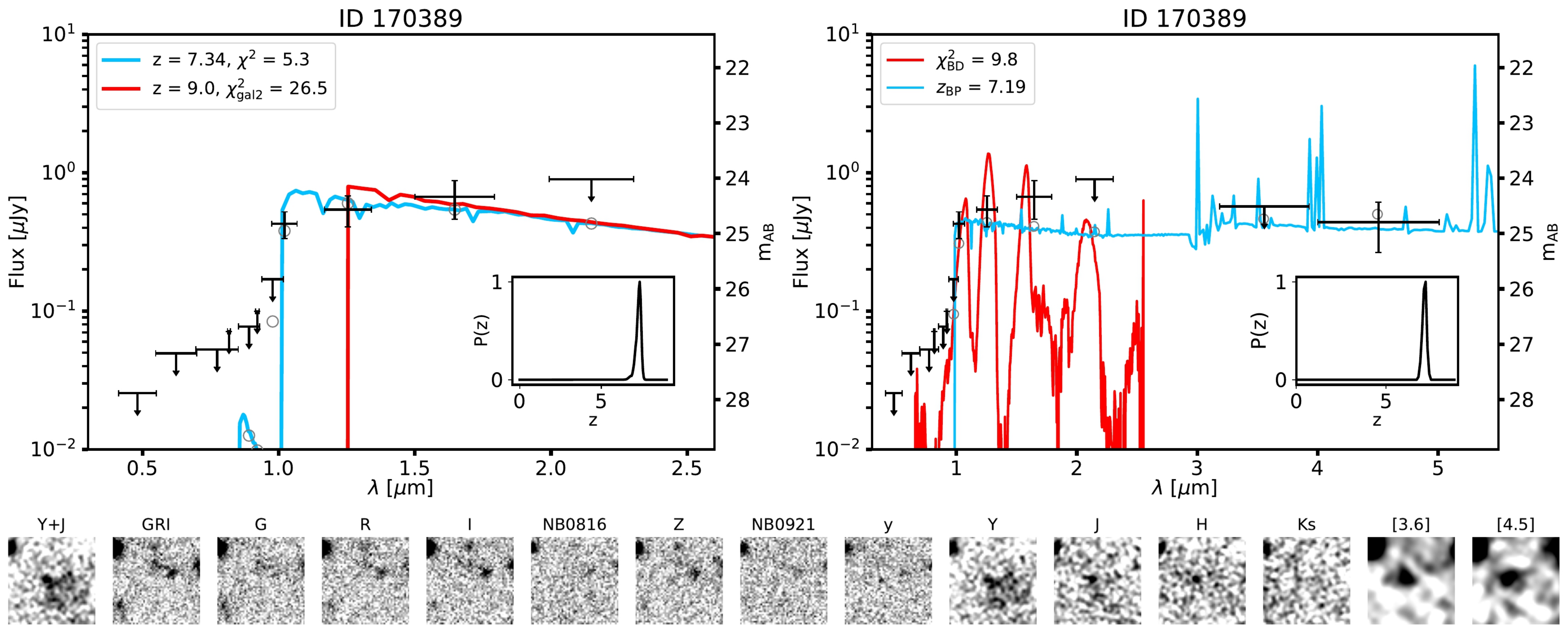}

         \includegraphics[width=0.75\textwidth]{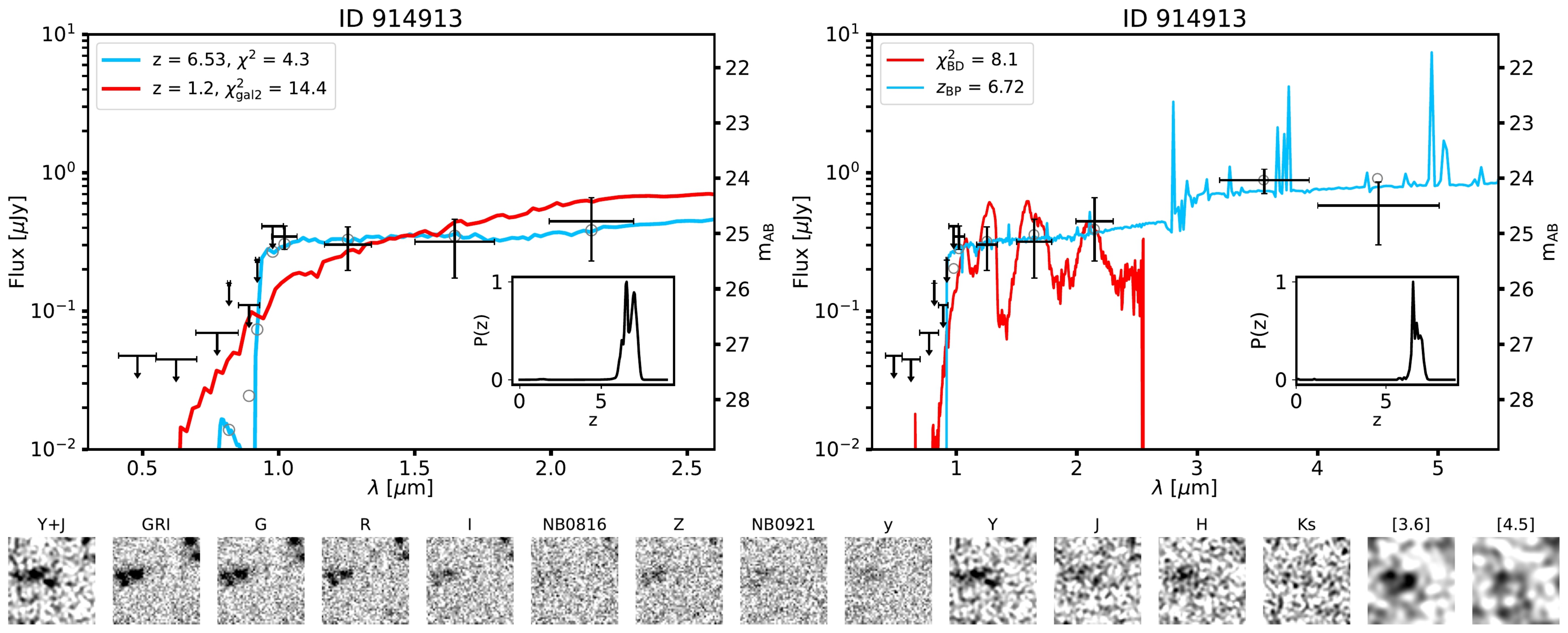}

         \includegraphics[width=0.75\textwidth]{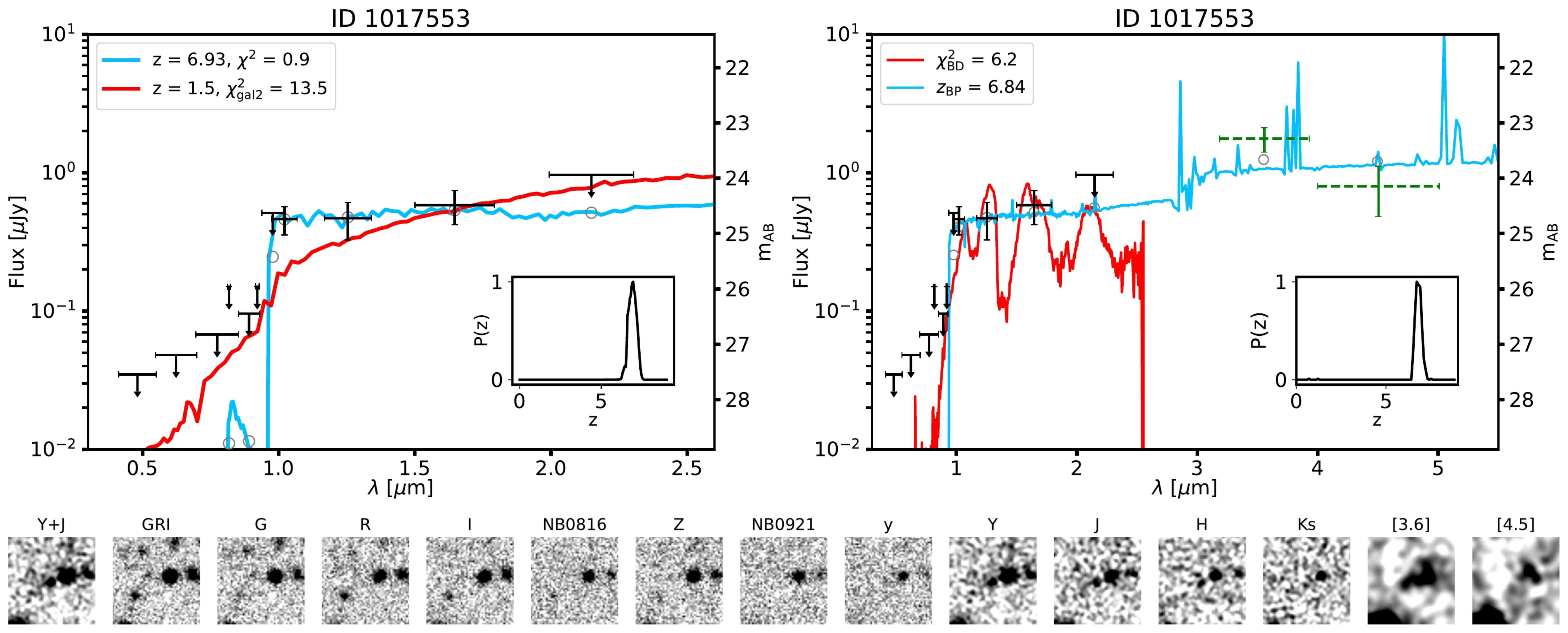}

        \includegraphics[width=0.75\textwidth]{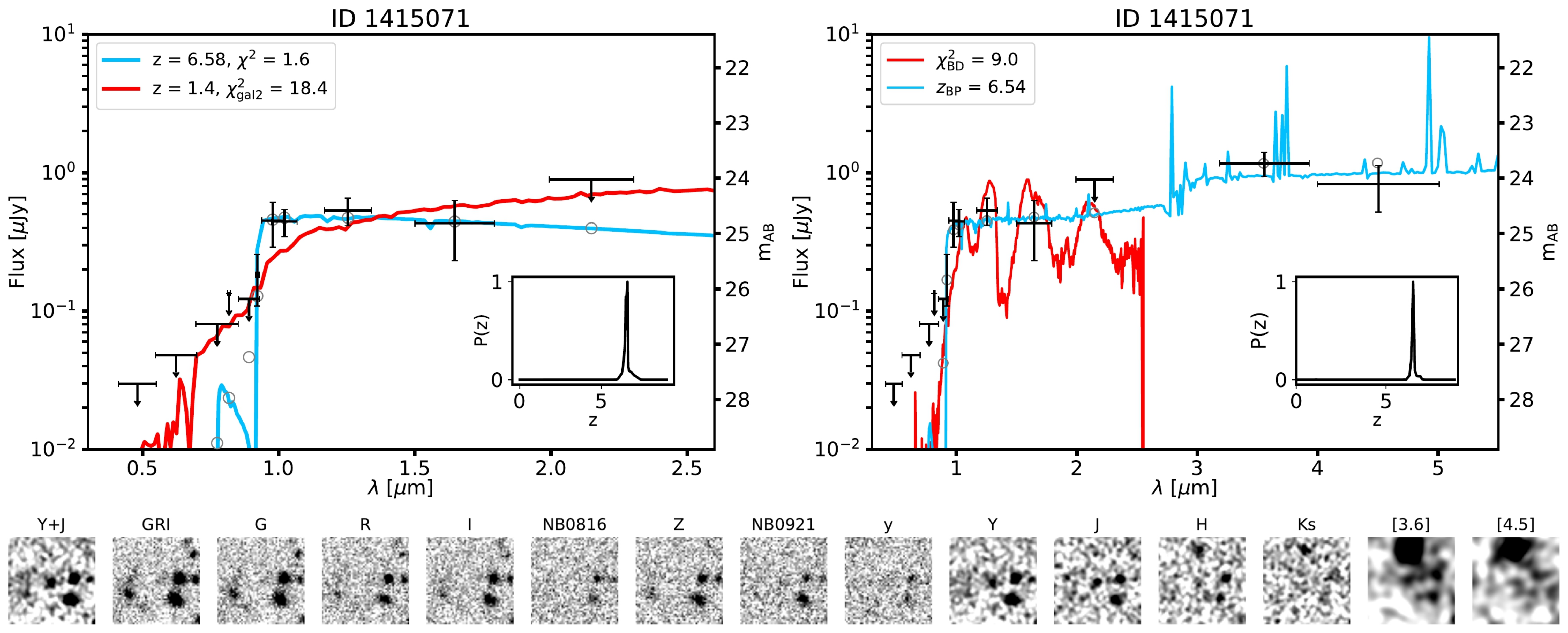}

        \caption{Inclusive candidates in  XMM-LSS. These are defined as having good brown dwarf fits, $\chi^{2}_{\mathrm{star}} < 10$, yet have significantly better galaxy fits, $\chi^{2}_{\mathrm{star}} - \chi^{2} > 4$ when fitted without $GR$. The plots and stamps are the same as Fig. \ref{fig:candidates}. Confused IRAC photometry is shown in green with dashed wavelength error bars.}
        \label{fig:XMMincl}
\end{figure*}

\subsubsection{ECDF-S}
\begin{figure*}
     \centering

        \includegraphics[width=0.75\textwidth]{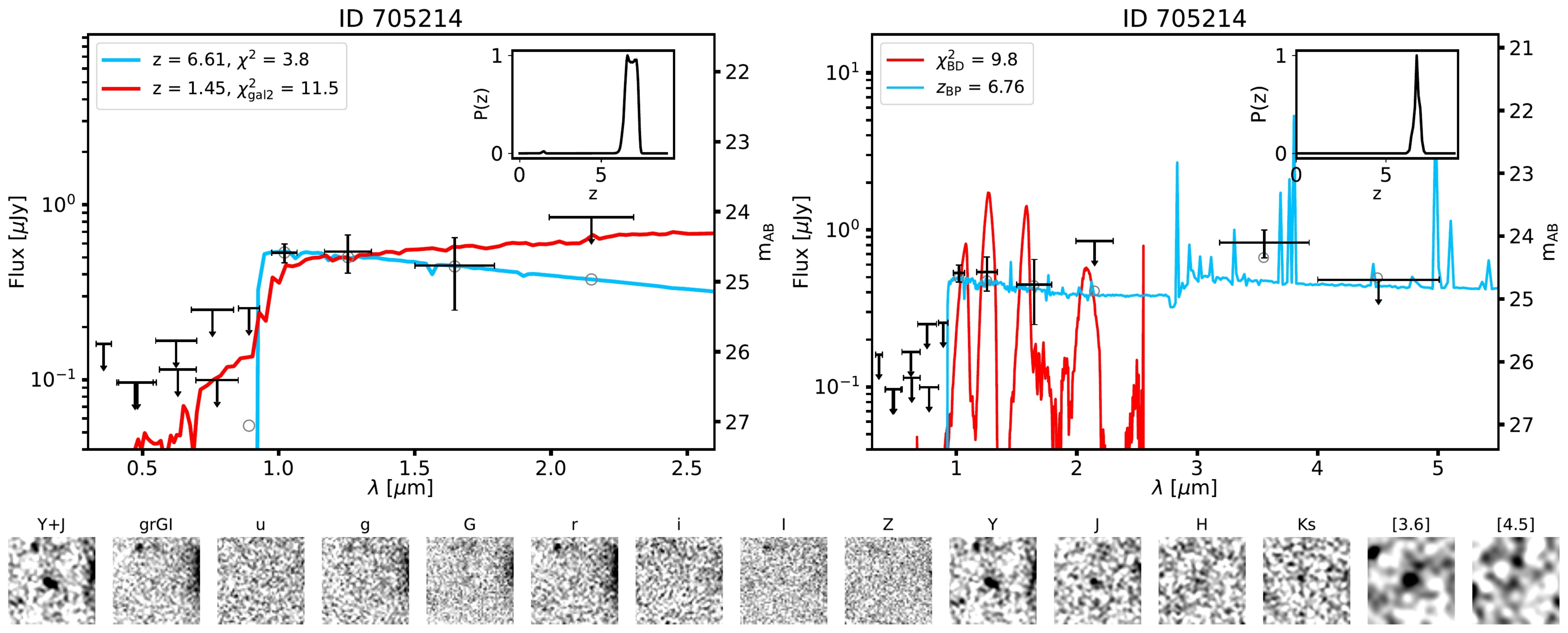}
         
        \includegraphics[width=0.75\textwidth]{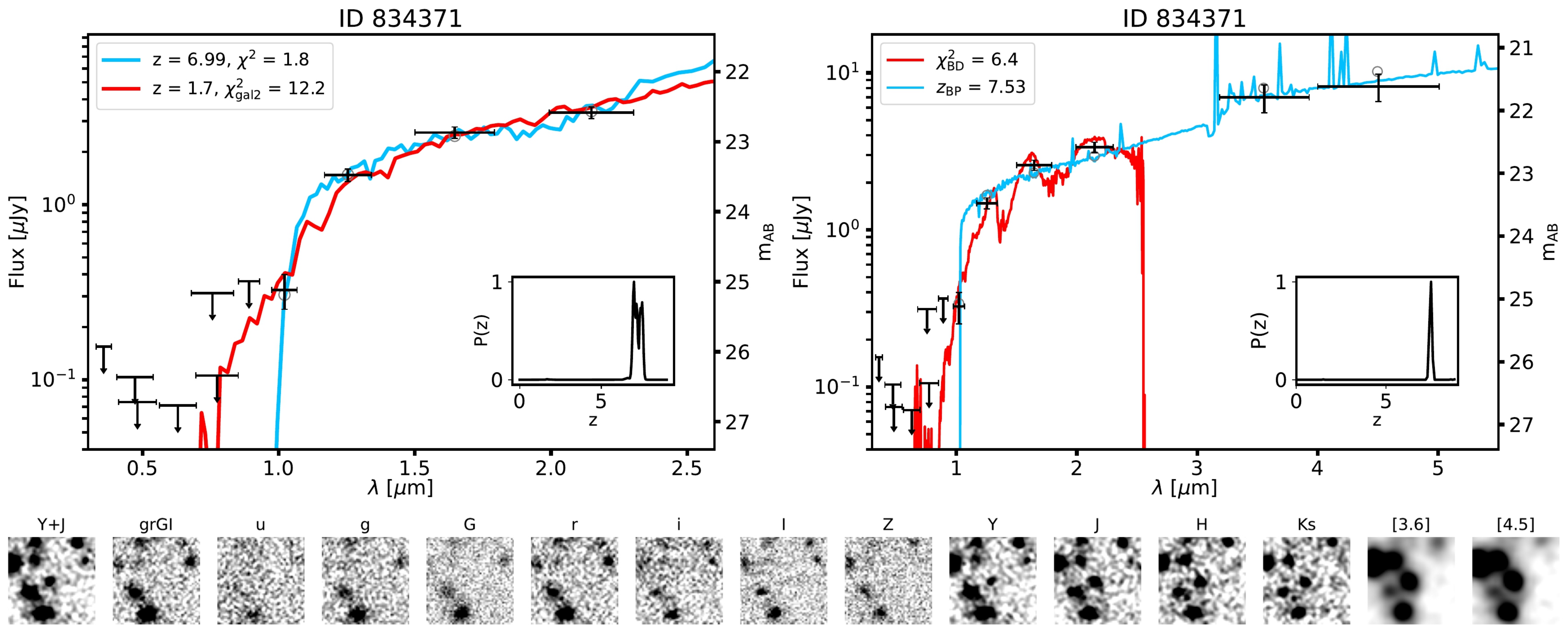}

        \includegraphics[width=0.75\textwidth]{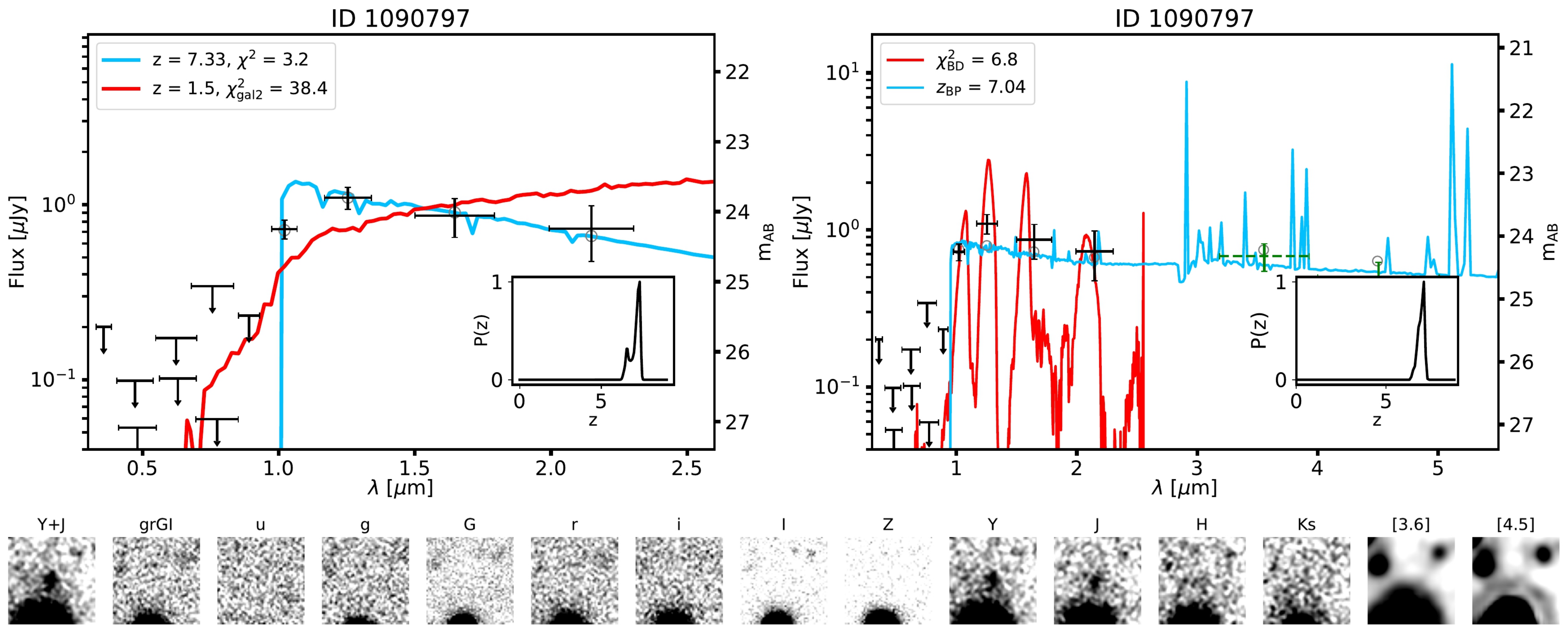}

        \includegraphics[width=0.75\textwidth]{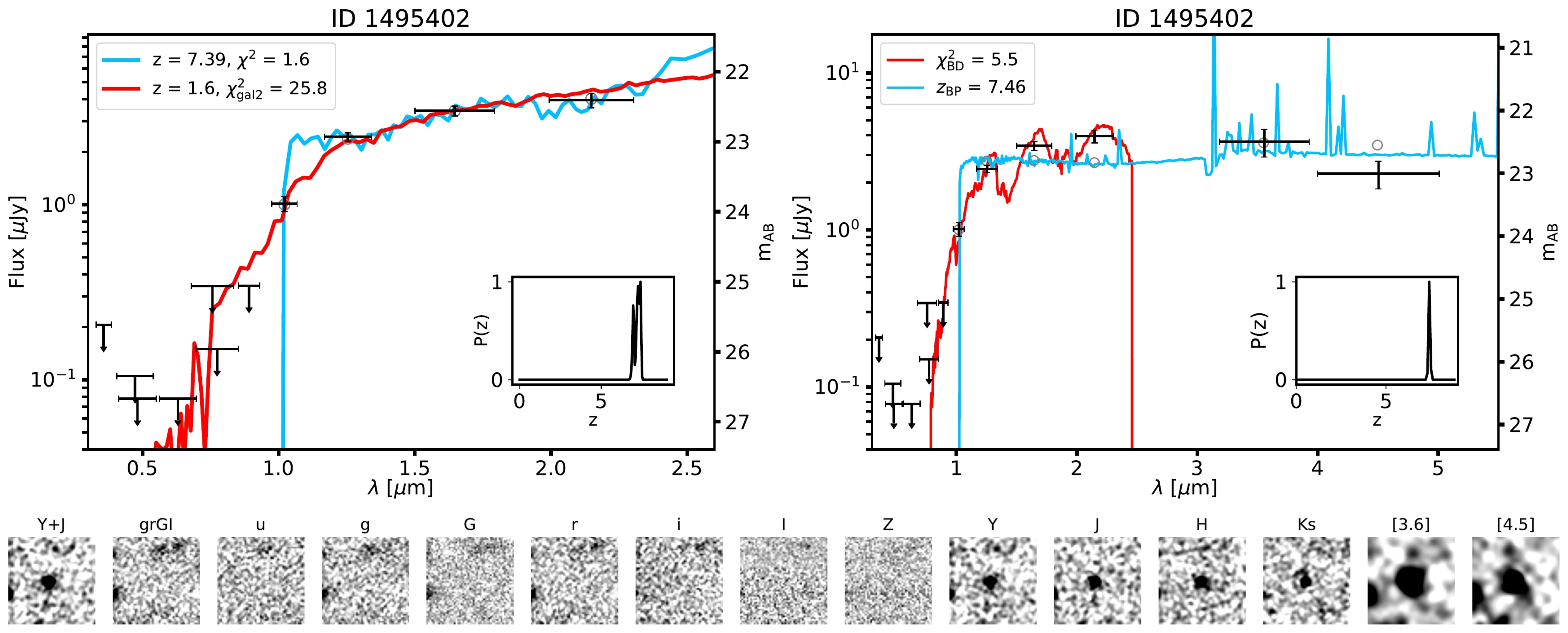}

        \caption{Inclusive candidates in  ECDF-S. These are defined as having good stellar fits, $\chi^{2}_{\mathrm{star}} < 10$, yet have significantly better galaxy fits, $\chi^{2}_{\mathrm{star}} - \chi^{2} > 4$ when fitted without $ugrGR$. The plots and stamps are the same as Fig. \ref{fig:candidates}. Confused IRAC photometry is shown in green with dashed wavelength error bars.}
        \label{fig:CDFSincl}
\end{figure*}

\section{Unusual object in XMM}
\label{sec: 1610530_stamp}

\begin{figure*}
     \centering

        \includegraphics[width=\textwidth]{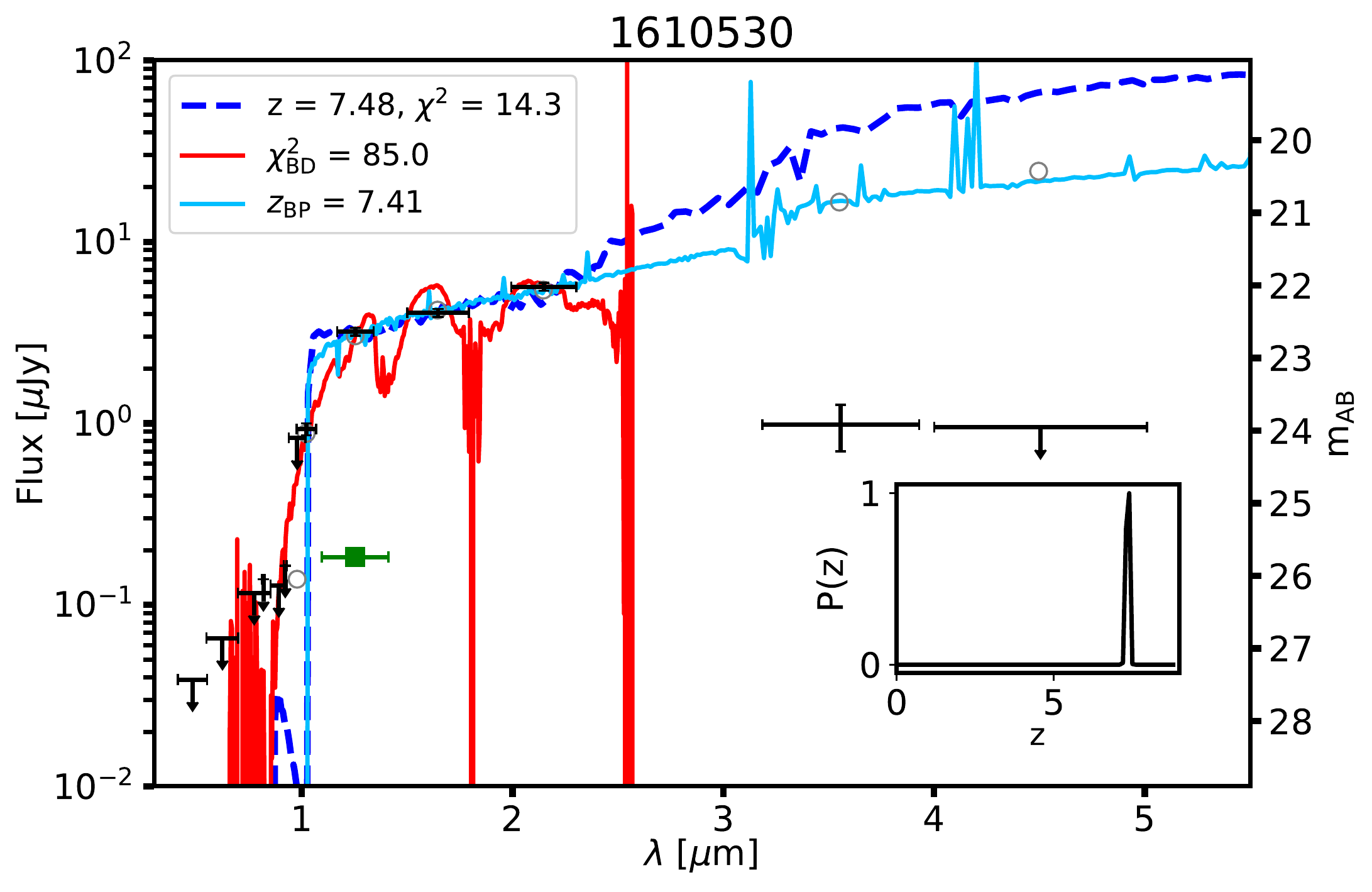}
         
        \includegraphics[width=\textwidth]{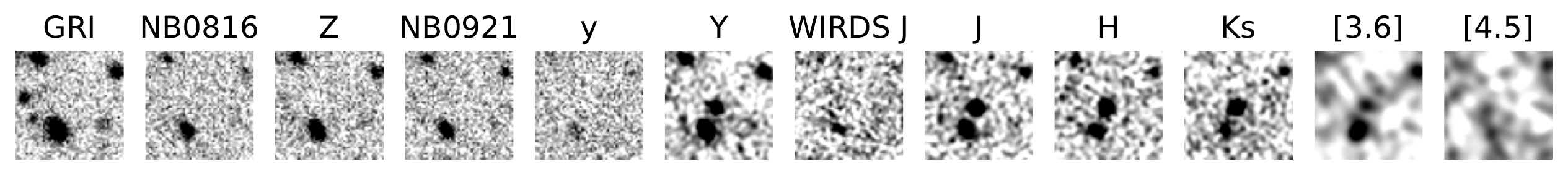}

        \caption{
        SED fitting (top) and postage stamp images (bottom) of ID 1610530, the unusual object discussed in Section \ref{sec:vischeck}. In the top panel, the black points in both plots are the measured photometry, with non-detections replaced by $2\sigma$ upper limits. The blue dashed curve shows the best-fitting \textsc{LePhare} solution. The solid light blue curve shows the best-fit \textsc{BAGPIPES} solution, with its model photometry shown as the grey circles. The red curve shows the best-fitting BD solution. The legend shows the redshift and $\chi^{2}$ of the best-fitting \textsc{LePhare} solution (without IRAC), $\chi^{2}_{\mathrm{BD}}$ (fit without $G$, $R$ and IRAC) and the redshift of the BAGPIPES solution, $z_{\mathrm{BP}}$. The inset black curve is the redshift probability found by \textsc{BAGPIPES}. The bottom panel shows the 10 arcsec $\times$ 10 arcsec postage stamps of the object in the filters used. We also show the WIRDS J-band, where the object is not detected. The green square point in the top panel corresponds to the photometry measured in a 2 arcsec diameter circular aperture in WIRDS-$J$.}
        \label{fig:1610530}
\end{figure*}

In this Appendix we present the SED and postage stamp images of ID 1610530, an unusual object noted in Section \ref{sec:vischeck}. If this object is at high redshift, the rest-frame UV is very bright ($M_{\mathrm{UV}} = -24.4$), but the \textit{Spitzer}/IRAC photometry appears unusually faint, around 3.5-4 mag fainter than suggested by the model. When \textit{Spitzer}/IRAC photometry is included in the SED fitting, the fit becomes very poor with $\chi^{2} = 167.6$. A crosstalk artifact is perhaps ruled out by the $3.5\sigma$ detection in the [3.6] band. Additionally, the object is not detected in WIRDS-$J$. In Fig. \ref{fig:1610530} we present a measurement of the WIRDS-$J$ flux in a 2 arcsec diameter aperture at the position of the object. The measured flux is $\sim3$ mag lower than measured in VIDEO-$J$.

% Don't change these lines
\bsp	% typesetting comment
\label{lastpage}
\end{document}